\documentclass[11pt,a4paper]{article}
\pdfoutput=1
\usepackage{jheppub}
\usepackage{physics}
\usepackage{amsmath}
\usepackage{xcolor}
\usepackage{tensor}
\usepackage{booktabs}
\usepackage{graphicx}
\usepackage{subcaption}

\usepackage{amsthm}
\newtheorem{rem}{Remark}

\DeclareMathOperator{\SO}{SO}
\DeclareMathOperator{\U}{U}
\DeclareMathOperator{\AdS}{AdS}
\DeclareMathOperator{\dS}{dS}
\DeclareMathOperator{\Isom}{Isom}
\DeclareMathOperator{\SU}{SU}

\newcommand*{\dual}{\vb*{\zeta}}
\newcommand*{\bC}{\vb*{C}}
\newcommand*{\Iso}{\Isom_{0}}
\newcommand*{\Poin}{\mathcal{P}}
\newcommand*{\Mink}{\mathbb{M}}

\newcommand*{\half}{\frac{1}{2}}
\newcommand*{\htil}{\tilde{h}}
\newcommand*{\mink}{g^{\Mink}}
\newcommand*{\gsig}{g^{\Sigma}}
\newcommand*{\hperp}{\circledast}
\newcommand{\hnormperp}[1]{\left\lVert #1 \right\rVert^{2}_{\hperp}}
\newcommand*{\hodge}{\star}
\newcommand{\innerp}[2]{\left\langle#1,#2\right\rangle}
\newcommand{\hnorm}[1]{\left\lVert #1 \right\rVert^{2}_{\hodge}}

\newcommand*{\msign}{\mathfrak{s}}
\newcommand*{\that}{\hat{\theta}^{0}}

\newcommand*{\flux}{\mathcal{F}_{m}}

\newcommand*{\nua}{\nu_{\bullet}}

\title{Some Exact Solutions for Maximally Symmetric Topological Defects in Anti de~Sitter Space}
\author{Orlando Alvarez}
\author{and Matthew Haddad}

\affiliation{Department of Physics, University of Miami, 1320 Campo Sano Ave, Coral Gables, FL 33146, USA}

\emailAdd{oalvarez@miami.edu}
\emailAdd{m.haddad@miami.edu}

\abstract{
We obtain exact analytical solutions for a class of $\SO(l)$ Higgs field theories in a non-dynamic background $n$-dimensional anti de~Sitter space. These finite transverse energy solutions are maximally symmetric $p$-dimensional topological defects where $n=(p+1)+l$. The radius of curvature of anti de~Sitter space provides an extra length scale that allows us to study the equations of motion in a limit where the masses of the Higgs field and the massive vector bosons are both vanishing. We call this the double BPS limit. In anti de Sitter space, the equations of motion depend on both $p$ and $l$.  The exact analytical solutions are expressed in terms of standard special functions. The known exact analytical solutions are for kink-like defects ($p=0,1,2,\dotsc;\, l=1$),  vortex-like defects ($p=1,2,3;\, l=2$), and the 'tHooft-Polyakov monopole ($p=0;\, l=3$). A bonus is that the double BPS limit automatically gives a maximally symmetric classical glueball type solution. In certain cases where we did not find an analytic solution, we present numerical solutions to the equations of motion. The asymptotically exponentially increasing volume with distance of anti de~Sitter space imposes different constraints than those found in the study of defects in Minkowski space.}

 \keywords{}
 
\arxivnumber{}

\begin{document}
	\maketitle
	
	\section{Introduction}
	\label{sec:intro}

In this article we obtain exact analytical solutions for a class of $\SO(l)$ Higgs field theories in a non-dynamic background $n$-dimensional anti de~Sitter space $\AdS_{n}$. These field theories admit maximally symmetric $p$-dimensional topological defects. The world brane of a maximally symmetric $p$-defect is  a $q=p+1$ dimensional timelike submanifold $\Sigma^{q} \approx \AdS_{q}$ that is isometrically embedded in $\AdS_{n}$; it is the gauge invariant set corresponding to the zero locus of the Higgs field. The value of $l $ is determined by $n=q+l$.  

The search for a maximally symmetric defect solution to the equations of motion requires the Lorentzian submanifold $\Sigma^{q}$ to admit the largest possible group of isometries. For $q$-dimensional manifolds, this Lie group has dimension $\half q(q+1)$.  The choices of this $q$-dimensional manifold are Minkowski space $\mathbb{M}^{q}$, anti de~Sitter space $\AdS_{q}$, and de~Sitter space $\dS_{q}$. In this article we mostly discuss the anti de~Sitter cases. We show that the Minkowski and de Sitter cases do not give a maximally symmetric solution. The anti de~Sitter case gives a maximally symmetric solution  when the isometric embedding $\AdS_{q} \hookrightarrow \AdS_{n}$ is totally geodesic. In our defect considerations we assume that when we refer any of these maximally symmetric manifolds, we are implicitly considering the simply connected universal covering space.

We need that $p \ge 0$ or equivalently that $q \ge 1$. The mathematical reason is that the formalism we employ requires an Euclidean signature for the metric of the normal tangent space $(T_{\sigma}\Sigma)^{\perp}$ for $\sigma\in\Sigma$. If we allow $q=0$ then $\AdS_{0}$ is a  point and its normal tangent space is the whole tangent space of $\AdS_{n}$ at that point, which has Minkowski signature, and our formulas do not apply directly\footnote{\label{foot:instanton}The $q=0$ case corresponds to instantons, and the embedding manifold is not $\AdS_{n}$ but the Euclidean signature negative constant curvature hyperbolic space $H^{n}$.}.
We are restricted to $\AdS_{n}$ with dimensionality $n = q+l \ge 2$.

We classify AdS topological defects by their degrees of longitudinal and transverse freedom using the tuple notation $(q,l)$. We develop a universal formalism that allows us to study all values of $(q,l)$. We only need to study in detail three types of defects: kink defects, vortices, and monopoles. Our parlance is that a defect with $l=1$ (with one transverse dimension) is kink-like, a defect with $l=2$ is vortex-like, and a defect with $l=3$ is hedgehog-like or monopole-like. For a review of Minkowski space kinks ($l=1$) look in \cite[Chapter~6]{Coleman:symmetry}. The Nielsen-Olesen vortex~\cite{Nielsen:1973cs} has $l=2$, and $p=1$ or equivalently $q=2$. The 'tHooft-Polyakov monopole~\cite{tHooft:1974kcl,Polyakov:1974ek} has $l=3$, and $p=0$ or equivalently $q=1$. For $l \ge 4$, the transverse energy of these spherically symmetric solutions in $\AdS_{n}$ diverges. For a comprehensive review of topological solitons we recommend the book by Manton and Sutcliffe~\cite{Manton:book}. We do not discuss electrically charged defects such as the Julia-Zee dyon~\cite{Julia:1975ff}.

There is a major difference between the study of the equations of motion for maximally symmetric $p$-defects in Minkowski space $\mathbb{M}^{n}$ and in anti de Sitter space $\AdS_{n}$. In Minkowski space, the equations of motion only depend on the transverse dimensionality $l$, and the study of the solutions is independent of the dimensionality $p$ of the defect. This is not the case for  $p$-defects in anti de~Sitter space where the equations of motion depend on both $p$ and $l$. This forces us to do a case by case analysis as we vary $p$ and $l$, \emph{e.g.}, see figure~\ref{fig:allowed}.

Line vortices in $\AdS_{4}$ were discussed in reference \cite{Dehghani:2001ft}, and line vortices in $\dS_{4}$ in \cite{Ghezelbash:2002cc}. In the first reference, approximate analytic solutions for vortices in $\AdS_{4}$ were found after applying some simplifying approximations to the equations of motion. Vortex holography played a strong part of the discussions in these papers.

The study of magnetic monopoles in $\AdS_{4}$, the case with $(q,l)=(1,3)$,  has been around for a while. The earliest work we are familiar with are papers authored by Lugo and  Schaposnik~\cite{Lugo:1999fm}, and Lugo,  Moreno and  Schaposnik~\cite{Lugo:1999ai}. We collectively refer to these two articles as LMS. In appendix~\ref{sec:translation} we discuss how their work is related to ours. Numerical axially symmetric monopole solutions in $\AdS_{4}$ are explored in \cite{Radu:2004ys}. No exact analytic solutions were found in these references.  Approximate analytic and numerical methods are used to discuss multi-monopoles and multi-monopole walls and their importance in the AdS/CFT correspondence in \cite{Bolognesi:2010nb} and in \cite{Sutcliffe:2011sr}. Atiyah~\cite{Atiyah:monopoles} had earlier discussed magnetic monopoles in Euclidean hyperbolic $3$-space $H^{3}$ by exploiting the conformal invariance of the self-dual Yang-Mills equations. His $4$-dimensional manifold was $S^{1}\times H^{3}$ with the Euclidean signature product metric. This manifold is conformal to Euclidean space $\mathbb{E}^{4}$. He used the observation due to Bogomolny that the self-dual Yang-Mills equations applied to a time independent  $\SU(2)$ gauge field are equivalent to  the Bogomolny equations\footnote{The methods that fail in the $\AdS_{4}$ discussion in appendix~\ref{sec:bogomolny} will work positively in Atiyah's scenario.}. The  product manifold Atiyah uses is not the Lorentzian manifold $\AdS_{4}$, and his monopoles are not $\AdS_{4}$ monopoles.

In the 'tHooft-Polyakov monopole in Minkowski space $\mathbb{M}^{4}$ there are two independent length scales determined by the mass of the Higgs scalar $m_{\phi}$ and by the mass of the massive vector meson $m_{A}$. A good way to see this is to observe that after an appropriate rescaling of the fields, the action for the $\SO(3)$ Georgi-Glashow model may be schematically written as
\begin{equation*}
 I = \phi_{0}^{2} \int \dd^{4}x\; \left[ \left(\partial \varphi +A\varphi\right)^{2} - m_{\phi}^{2} \left(\varphi^{2}-1\right)^{2} - \frac{1}{m_{A}^{2}} \left( \partial A + A^{2}\right)^{2} \right]\,,
\end{equation*}
here $\phi_{0}$ is the vacuum expectation value of the scalar field.
For example, you expect solutions of the classical equations of motion for the rescaled dimensionless scalar field $\varphi$  to depend on $m_{A}$ and on the dimensionless ratio $m_{\phi}/m_{A}$: $\varphi=\varphi(m_{A}x,m_{\phi}/m_{A})$. Prasad and Sommerfield~\cite{Prasad:1975kr} discovered an exact solution to the equations of motion by considering a non-trivial limit of the equations of motion in which $m_{\phi} \downarrow 0$, keeping $m_{A}$ fixed, and enforcing an appropriate topological asymptotic boundary condition of the Higgs field $\varphi$ at infinity\footnote{This is the same as letting the $\varphi^{4}$ self coupling $\lambda \downarrow 0$ while maintaining the asymptotic boundary conditions.}. The net effect is that the equations of motion only depend on one length scale $1/m_{A}$ that controls the asymptotic behavior along with the correct boundary conditions imposed manually. The dimensionless parameter $m_{\phi}/m_{A}\downarrow 0$  in the Prasad and Sommerfield limit. The Prasad-Sommerfield solution satisfies the first order equations of Bogomolny~\cite{Bogomolny:1975de} that guarantee a solution with a saturated lower bound on the energy~\cite{Bogomolny:1975de,Coleman:1976uk}. It is known that the mass of the monopole is given by $M = (4\pi\phi_{0}^{2}/m_{A}) \, \mathcal{F}(m_{\phi}^{2}/m_{A}^{2})$ where the function $\mathcal{F}$ satisfies $\mathcal{F}(0) =1$ and $\mathcal{F}(\infty) \approx 1.787$, see \cite[p.~255]{Manton:book}.

In this paper we also discuss the Prasad and Sommerfield limit $m_{\phi}\downarrow 0$ for other values of $q$ and $l$. We refer to this as the limit of Bogomolny, and Prasad and Sommerfield (BPS) even though there may be no Bogomolny equations. The BPS limit was studied in LMS for the case of monopoles in $\AdS_{4}$.

\begin{figure}[tbp]
\centering
\includegraphics[width=0.5\textwidth]{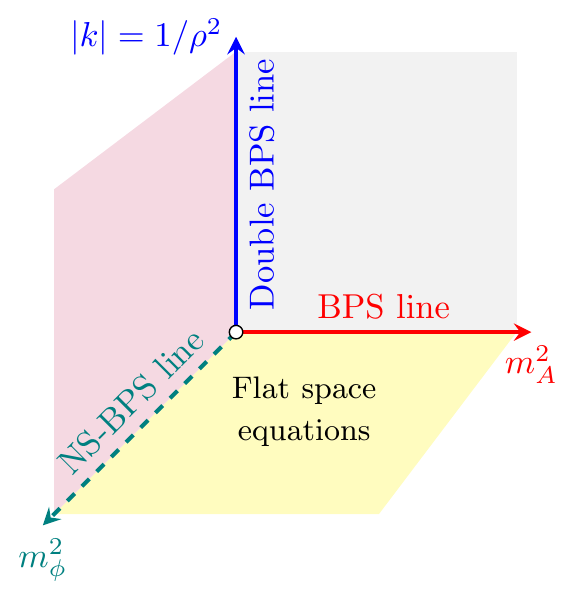}
\caption{\small The  equations of motion for our Higgs model depend on three mass scales $m_{\phi}^{2}$, $m_{A}^{2}$, and $\lvert k \rvert =1/\rho^{2}$ if $l \ge 2$. We explore the parameter space for analytic maximally symmetric solutions to the equations of motion in section~\ref{sec:double}. The Prasad-Sommerfield limiting solutions lie along the BPS line. There are no solutions that satisfy our boundary conditions along the green dashed line labeled NS-BPS. We find new exact spherically symmetric analytic solutions along the double BPS line. In section~\ref{sec:pink-plane} we show that for parameter values in the pink plane $m_{A}^{2} \downarrow 0$ there is an analytic solution for the gauge field, but the scalar field has to be studied numerically. The gray plane $m_{\phi}^{2} \downarrow 0$ is where Lugo,  Moreno and  Schaposnik looked for BPS monopole solutions in $\AdS_{4}$, see appendix~\ref{sec:translation} and figure~\ref{fig:LMS-parameter}.}%
\label{fig:bps-parameter}
\end{figure}
In studying the equations of motion for  defects in $\AdS_{n}$, we encounter an additional length scale $\rho$, the radius of curvature\footnote{The radius of curvature  is defined by $\rho = \lvert k\rvert^{-1/2}$ where $k<0$ is the sectional curvature; $\AdS_{n}$ is a solution of the vacuum Einstein equations $R_{\mu\nu} - \half g_{\mu\nu}R + \Lambda g_{\mu\nu}=0$ with $\Lambda = \half(n-1)(n-2)k$. } of $\AdS_{n}$. Now there are three independent length scales $1/m_{\phi}$, $1/m_{A}$, and $\rho$. This leads to a three dimensional parameter space, see figure~\ref{fig:bps-parameter}, that can be explored for exact solutions of the equations of motion. The appearance of this additional length scale was already noticed in \cite{Lugo:1999fm}. The scalar field solution $\varphi$ of the equations of motion depends on $\rho$ and on two dimensionless quantities  $m_{\phi}\rho$ and $m_{A}\rho$: $\varphi=\varphi(x/\rho,m_{\phi}\rho,m_{A}\rho)$. We attempt to extend the methods of Bogomolny, and Prasad and Sommerfield, and look for regions in the parameter space where we might find exact solutions.  We consider a limit for the equations of motion for a maximally symmetric defect where $\rho\neq 0$ is fixed, but $m_{\phi}\rho \downarrow 0$ and $m_{A}\rho\downarrow 0$. The net effect is that the equations of motion only depend on one length scale $\rho$ that controls the asymptotic behavior along with the correct boundary conditions imposed manually. 
\begin{quote}
\sffamily
\emph{N.B.} This is a very delicate limit because the action is singular in this limit but the equations of motion are not.
\end{quote}
We take the limiting equations of motion as the starting point in our analysis, and we abandon the action. These limiting equations are consistent and do not follow from an action principle. This is analogous to consistent equations of motions such as the self-dual Yang-Mills equations, or the self-dual equations of motion for the $4$-form in type IIB supergravity which are not derivable from an action. The limiting equations of motion partially decouple. The one for the gauge field is completely decoupled from the Higgs field and can be solved independently. The gauge field solution can then be  inserted into the Higgs field equation of motion which is now linear.
In addition, this ``double BPS limit'' preserves the nonlinear interactions of the gauge fields. The vanishing of the dimensionless parameters, $m_{\phi}\rho$ and $m_{A}\rho$,  leads to   exact analytic solutions in a variety of cases, see figure~\ref{fig:allowed-dbps} for admissible pairs $(q,l)$. For example, there are exact analytic solutions for kink defects $(q,l)=(q,1)$ given by \eqref{eq:leq1}, for the Nielsen-Olesen vortex line $(q,l)=(2,2)$ given by \eqref{eq:exact-q2-l2}, and the 'tHooft-Polyakov monopole $(q,l)=(1,3)$ given by \eqref{eq:sol-q1-l3-h} and \eqref{eq:phi-q1-l3-exact}. We note that the transverse size of these $p$-defects in the double BPS limit is comparable to the radius of curvature of $\AdS_{n}$.

These double BPS equations of motion are a first step in a perturbative expansion of the full equations of motion where the small parameters are $m_{\phi}\rho$ and $m_{A}\rho$. In this sense, we can make contact with the action again.

We have not studied the stability of these double BPS solutions in anti de~Sitter space. Our only attempt at trying to prove stability was to look for a Bogomolny type bound. This bound is used to establish the stability of the Prasad-Sommerfield solution~\cite{Bogomolny:1975de,Coleman:1976uk} in Minkowski space. The curvature of $\AdS_{n}$ invalidates some of the Minkowski space arguments as explained in appendix~\ref{sec:bogomolny}. We show that in the monopole case there are no first order Bogomolny type equations that imply the equations of motion. Additionally, there is a partial Bogomolny bound relating the energy to the magnetic charge when the magnetic charge density is non-negative.  In principle, the underlying symmetries of $\AdS_{q}$ and $\AdS_{n}$ should greatly aid in the stability analysis of linear perturbations of the equations of motion around these double BPS solutions.

The organization of this article is the following: A formalism is developed to study  maximally symmetric $p$-defects  in $\AdS_{n}$ in sections \ref{sec:defects} and \ref{sec:darboux}, leading to the general maximally symmetric equations of motion~\eqref{eq:eom-gen} for all values of $(q,l)$. One useful feature of our formalism is that the normal radial coordinate $\nu$ we employ is always the physical distance from the defect world brane no matter what $(q,l)$ pair we are studying. We found it convenient to use an orthonormal frame because the Pythagorean theorem automatically organized the calculation for us. For example, the transverse energy functional consists of five positive semi-definite summands, therefore finiteness of transverse energy follows if each summand is finite. The theorem that requires $\Sigma^{q}$ to be totally geodesic is relegated to  mathematical appendix~\ref{sec:max-sym}.  The finite transverse energy constraints are discussed in section~\ref{sec:finite-E}, and summarized in figure~\ref{fig:allowed}. In section~\ref{sec:double} we discuss the double BPS limit and the exact solutions we found. In section~\ref{sec:pink-plane}, we briefly explore numerically a portion of the parameter space $m_{A} \downarrow 0$ where the gauge field is explicitly known but the scalar field has to be studied numerically. In appendix~\ref{sec:translation} we relate our work to the work of Lugo, Moreno and Schaposnik. Finally in appendix~\ref{sec:bogomolny} we restrict to magnetic monopoles in $\AdS_{4}$ in the BPS limit. We discuss that the BPS equations do not imply the equations of motion. We show the existence of a partial bound on the mass if the magnetic charge density is non-negative.

	\section{Defects in constant curvature spaces}
	\label{sec:defects}
	
We use a $\SO(l)$ gauged Higgs type field theory as the model for a topological defect. We first discuss the definition of a maximally symmetric $p$-dimensional defect in Minkowski space and subsequently generalize the notion to a Lorentzian constant curvature space. A $p$-dimensional \emph{maximally symmetric defect} in $\mathbb{M}^{n}$ is a topologically
stable solution to the equations of motion that is invariant with respect to the action of the subgroup $\Poin(q) \times \SO(l)\subset \Poin(n)$ where $q=p+1$ and $q+l = n$.  Here $\Poin(n)$ is the Poincar\'{e} group, the isometry group of Minkowski space $\Mink^{n}$, and  $\Mink^{n} \approx \Poin(n)/\SO(1,n-1)$.  The world brane (time evolution of the defect brane) for the defect  is the $q$-dimensional manifold $\Sigma^{q}$. Since the symmetry group of the solution is $\Poin(q) \times \SO(l)$ we know that the world brane $\Sigma^{q}$ for the defect  is a timelike $q$-plane. Notice that $\Sigma^{q}$ is intrinsically flat, and that the invariance group of the solution implies that the defect is static for any choice of time direction in $\Sigma^{q}$.

	Let $M^{n}$ be a Lorentzian manifold with constant sectional curvature $k$ and with isometry group\footnote{$\Iso(N)$ is the connected component to the identity of the isometry group of the manifold $N$.} $\Iso(M^{n})$ of dimension $\frac{1}{2} n(n+1)$. A $p$-dimensional \emph{maximally symmetric defect} in $M^{n}$ is a topologically stable solution to the equations of motion that is invariant with respect to the action of the subgroup $\Iso(\Sigma^{q}) \times \SO(l)\subset \Iso(M^{n})$ where $\Sigma^{q}$ is a maximally symmetric $q$-dimensional Lorentzian submanifold. A maximally symmetric $\Sigma^{q}$ is a constant curvature manifold with $\dim\Iso(\Sigma^{q}) = \frac{1}{2} q(q+1)$.  Here $\Sigma^{q}$ is the world brane of the defect core\footnote{By the core of the defect we mean the region in spacetime where the energy density is concentrated.}. Finding such a $\Sigma^{q} \subset M^{n}$ requires an appropriate generalization of choosing a plane. The correct ``flatness'' notion that leads to a maximally symmetric defect is to require $\Sigma^{q}$ to be a  totally geodesic submanifold. 	
	
An embedded submanifold $\Sigma^{q}$ of a general manifold $M^{n}$ is said to be totally geodesic if every geodesic (with respect to the induced metric) on $\Sigma^{q}$ is also a geodesic on $M^{n}$. Next, we see how this is related to the differential geometric data.  Let $D^{\Sigma}$ and $D^{M}$ be the Levi-Civita connections on the respective manifolds.	
A Darboux frame is an orthonormal frame adapted to the orthogonal decomposition $T_{\sigma}M = T_{\sigma}\Sigma +(T_{\sigma}\Sigma)^{\perp}$ for $\sigma\in \Sigma^{q}$. In a submanifold neighborhood of the point $\sigma$,  we consider an orthonormal framing $(\hat{\vb{e}}_{a}, \hat{\vb{n}}_{i})$, where the $\hat{\vb{e}}_{a}$ are tangential to $\Sigma$ and the $\hat{\vb{n}}_{i}$ are normal to $\Sigma$. We use the  index convention that latin indices from the beginning of the alphabet $a,b,c,d$ run from $1,2,\dotsc,q$ and latin indices from the middle of the alphabet $i,j,k,\dotsc$ take $l=n-q$ values from $q+1,\dotsc,n$. 

Let $u=u^{a}\vb{\hat{e}}_{a}$ and $v=v^{b}\vb{\hat{e}}_{b}$ be vector fields tangent to $\Sigma^{q}$, then the two connections are related by $D_{u}^{M}v = D_{u}^{\Sigma}v  -  u^{a}v^{b}\, K_{ab}{}^{i}\, \vb{\hat{n}}_{i}$. The symmetric tensor $K_{ab}{}^{i}$ is called the second fundamental form or the extrinsic curvatures.
If $u$ is a tangent vector on $\Sigma$ then $D^{M}_{u} u = D^{\Sigma}_{u} u - u^{a}u^{b}\, K_{ab}{}^{i}\hat{\vb{n}}_{i}$. In a totally geodesic submanifold, we would have that $D^{M}_{u} u=0$ and $D^{\Sigma}_{u} u=0$ for all geodesics on $\Sigma$. This is only possible if the extrinsic curvatures $K_{ab}{}^{i}=0$. Summarizing, totally geodesic submanifolds are those where the extrinsic curvatures vanish.
From the viewpoint of standard General Relativity, totally geodesic submanifolds are very desirable because if a test mass in $\Sigma$ is given an initial velocity tangential to $\Sigma$ then its motion will remain in $\Sigma$.

As shown in detail in appendix~\ref{sec:max-sym}, a totally geodesic $q$-dimensional submanifold $\Sigma^{q}$ of $\AdS_{n}$ is a constant curvature Lorentzian submanifold with the same sectional curvature $k$ as $M^{n}$ and with a flat normal bundle $(T\Sigma)^{\perp}$. These are the only submanifolds that admit the possibility of finding a maximally symmetric solution, see remark~\ref{rem:maximal} in the appendix~\ref{sec:flat-submanifolds}.

\section{The Darboux frame and the spherically symmetric ansatz}
\label{sec:darboux}

To work out the equations of motion for our maximally symmetric defect, it is convenient to use a coordinate system adapted to the geometry of the problem, \emph{i.e.}, an analog of spherical coordinates. The construction is based on the method of Cartan discussed in our previous paper \cite{alvarezhaddad:egconstcurvature}. Let $\Sigma^{q}$ be a Lorentzian $q$-dimensional submanifold of $\AdS_{n}$. If $\sigma\in \Sigma^{q}$ then in a submanifold neighborhood of the point $\sigma$   consider a Darboux frame $(\hat{\vb{e}}_{a}, \hat{\vb{n}}_{i})$.  The dual Darboux coframe is denoted by $(\varphi^{a}, \varphi^{i})$. Choose a geodesic of $\AdS_{n}$ starting at $\sigma$ with initial normal velocity $\vb*{\nu}=\nu^{i} \, \hat{\vb{n}}_{i}$, and go a distance $\norm{\vb*{\nu}}$ along the geodesic to a point $x \in \AdS_{n}$. The coordinates of the point $x$ are $(\sigma,\nu^{i})$. Formally, this is the exponential map $\exp_{\sigma}: \vb*{\nu} \in (T_{\sigma}\Sigma)^{\perp} \mapsto x \in \AdS_{n}$. Note that $\varphi^{i}=\dd\nu^{i}$. Cartan's idea is to extend the orthonormal Darboux coframe by parallel transporting it along the normal geodesics; in this way you construct an orthonormal coframe at $(\sigma,\vb*{\nu})$ denoted by $(\vartheta^{a},\vartheta^{i})$.
In our previous paper we showed that
	\[
	\vartheta^a = \cosh(\abs{k}^{1/2}\norm{\nu }) \qty[\tensor{\delta}{^a_b} + \frac{\tanh(\abs{k}^{1/2}\norm{\nu })}{\abs{k}^{1/2}\norm{\nu}}\, \nu ^j K_{abj}]\varphi^b
	\]
The velocity of a geodesic is constant, and in this way we denote the orthogonal projector along the velocity vector by $(P_{L})^{ij} = \nu^{i}\nu^{j}/\norm{\vb*{\nu}}^{2}$, and the orthogonal projector perpendicular to the velocity by $(P_{T})^{ij} = \delta^{ij} -\nu^{i}\nu^{j}/\norm{\vb*{\nu}}^{2}$. Using these, it is easy to write down the  orthogonal decomposition of the extended Darboux frame component $\vartheta^{i}= \vartheta_L^i + \vartheta_T^i$  where
\begin{align*}
	\vartheta_L^i &=  P_L(D\nu )^i \,,  &
	\vartheta_T^i &= 
	\dfrac{\sinh(\abs{k}^{1/2}\norm{\nu })}{\abs{k}^{1/2}\norm{\nu }}P_T(D\nu)^i \,.
\end{align*}
In the above $(D\nu)^{i} = \dd\nu^{i} + \omega^{ij}\nu^{j}$, where $\omega^{ij}$ is the connection on the normal bundle $(T\Sigma)^{\perp}$.

The formulas above are general. We are interested in  the Lorentzian case where $\AdS_{q}\approx\Sigma^{q}\hookrightarrow M^{n}$ is a totally geodesic submanifold.  As discussed previously, we know that $K_{abi}=0$ and that the normal bundle $(T\Sigma)^{\perp}$ is flat, see appendix~\ref{sec:max-sym}. We can always locally trivialize the normal bundle so we set $\omega^{ij}=0$, \emph{i.e.}, the normal part of the Darboux frame $\hat{\vb{n}}_{i}$ is parallel along $\Sigma^{q}$. 
In summary we have
\begin{subequations}\label{eq:darboux}
\begin{align}
    \vartheta^{a} &= \cosh(\abs{k}^{1/2}\norm{\nu }) \varphi^b\\
    \vartheta^{i} &= P_L(\dd\nu )^i +\dfrac{\sinh(\abs{k}^{1/2}\norm{\nu })}{\abs{k}^{1/2}\norm{\nu }}P_T(\dd\nu)^i
\end{align}
\end{subequations}
The metric on $\AdS_{n}$ is given by $\dd s^{2}_{\AdS_{n}}= \mink_{ab}\,\vartheta^{a}\otimes\vartheta^{b} + \delta_{ij}\,\vartheta^{i}\otimes\vartheta^{j}$. Going to spherical coordinates we see that
		\begin{equation}
	\dd s^{2}_{\AdS_{n}}  = \cosh^{2}\left( \lvert k \rvert^{1/2} \lVert\vb*{\nu}\rVert\right) \; \dd s^{2}_{\AdS_{q}} + \left[ \dd\nu^{2} + \left(\frac{\sinh\left( \lvert k \rvert^{1/2} \lVert\vb*{\nu}\rVert\right)}{\lvert k \rvert^{1/2} \lVert\vb*{\nu}\rVert}\right)^{2}\;
	\nu^{2}\; \dd s^{2}_{S^{l-1}} \right],
	\label{eq:met-geod}
	\end{equation}
	where $\dd s^{2}_{\AdS_{q}} = \mink_{ab}\, \varphi^{a}\otimes \varphi^{b}$ is the standard constant curvature metric on $\Sigma^{q} \approx \AdS_{q}$, $\nu = \lVert\vb*{\nu} \rVert$ is the radial distance, and $ds^{2}_{S^{l-1}}$ is the round metric on the unit $(l-1)$-sphere.  The part of the metric in the square brackets is the pullback of the metric on $\AdS_{n}$ to $(T_{\sigma}\Sigma)^{\perp}$ via the exponential map $\exp_{\sigma}$. Said differently, this is the induced metric on $\exp_{\sigma}(T_{\sigma}\Sigma)^{\perp}$, the image of the normal tangent space under the exponential map. This induced metric is isometric to the standard metric on Euclidean hyperbolic space $H^{l}$, see the discussion associated with eq.~\eqref{eq:hyp-met}. We \emph{re-emphasize} that $\nu=\lVert\vb*{\nu} \rVert$ is the physical distance from a point $\sigma\in\Sigma^{q}$ to the point $(\sigma,\vb*{\nu}) \in M^{n}$. We will use this physical distance to measure the behavior of our fields as you move away from the defect world brane.
	
	A simple model that has a topological $p$-defect in $\AdS_{n}$ is a Higgs model with an $\SO(l)$ gauge symmetry, $l \ge 2$, where $n= (p+1)+l =q+l$. The model has a scalar field $\Phi^{I}$ that transforms under the vector representation of $\SO(l)$. The uppercase latin indices $I,J,K,\dotsc$ from the middle of the alphabet will take values from $1$ to $l$. We are looking for a maximally symmetric $p$-defect that is invariant under the action of $\Iso(\AdS_{q}) \times \SO(l)$, and therefore our fields do not depend on the coordinates $\sigma^{a}$ of $\AdS_{q}$ and only depend on the normal coordinates $\nu^{i}$. A connection that is compatible with the symmetries is $A_{a}^{IJ} = -A_{a}^{JI}=0$, and $A_{j}^{IJ} = -A_{j}^{JI}$ with covariant derivative $D_{j}\Phi^{I}= \partial_{j}\Phi^{I}+ A_{j}^{IJ}\Phi^{J}$ and curvature $F_{ij}= \partial_{i}A_{j}-\partial_{j}A_{i} + [A_{i},A_{j}]$.
The equations of motion are obtained by extremizing the action $I = \int_{M^{n}} \mathcal{L}$, where $\mathcal{L}$ is the Higgs model Lagrangian density. We are looking for $p$-defect solutions that are maximally symmetric with symmetry group $\Iso(\Sigma) \times \SO(l)$ and under these conditions the action for the Higgs type model is
	\begin{equation}
	I_{\text{spherically sym}} = - \; E_{\perp} \int_{\Sigma^{q}}\dual_{\Sigma}\;,
	\label{eq:full-action}
	\end{equation}
	where $\dual_{\Sigma}$ is the volume element on $\Sigma$.
	The transverse energy\footnote{$E_{\perp}$ is the tension of the associated $p$-brane.} $E_{\perp}$ in a local orthonormal frame for $(T_{\sigma}M)^{\perp}$ is given by
	\begin{align}
	E_{\perp} &= V_{l-1} \int_{0}^{\infty} \dd\nu \left[\cosh\left( \lvert k \rvert^{1/2} \,\nu\right)\right]^{q}\; \left[\frac{\sinh\left( \lvert k \rvert^{1/2} \,\nu\right)}{\lvert k \rvert^{1/2}} \right]^{l-1} 
	\nonumber \\
	&\quad\times \left[\frac{1}{2}\left( D_{i}\Phi^{I}\right)\left( D_{i}\Phi^{I}\right) + U(\lVert \Phi\rVert^{2})
	+ \frac{1}{8g^{2}}\, F_{ij}^{IJ}F_{ij}^{IJ}\right].
	\label{eq:defect-action} 
	\end{align}
The volume element formula is a special case of a result from our previous paper~\cite{alvarezhaddad:egconstcurvature}.
We implicitly assumed spherical symmetry for the Lagrangian density to do the angular integrals. Here $U$ is the potential, $g$ is the gauge coupling constant, and $V_{l-1}$ is the $(l-1)$-volume of the unit sphere $S^{l-1}$. We emphasize to the reader  the presence of the hyperbolic cosine factor that would not be there in the case of $M^{n}=\mathbb{M}^{n}$. The origin of this hyperbolic cosine factor is the metric \eqref{eq:met-geod}. You can make a mistake influenced by the familiarity of working in Minkowski space where the transverse energy would be obtained by using the pullback metric to $(T_{\sigma}\Sigma)^{\perp}$ via the exponential map. In this case for $p$-defects in $\AdS_{n}$, the metric on the normal tangent space is the metric on Euclidean hyperbolic space $H^{l}$
\begin{equation}
 \dd s^{2}_{(T_{\sigma}\Sigma)^{\perp}} = d\nu^{2} + \left(\frac{\sinh\left( \lvert k \rvert^{1/2} \lVert\vb*{\nu}\rVert\right)}{\lvert k \rvert^{1/2} \lVert\vb*{\nu}\rVert}\right)^{2}\;
	\nu^{2}\; ds^{2}_{S^{l-1}}
	\label{eq:hyp-met}
\end{equation}
Said differently, removing the hyperbolic cosine term in \eqref{eq:defect-action} leads to incorrect equations of motion. The equations of motion arise from varying the action \eqref{eq:full-action}. Note that the $k\to 0$ limit of \eqref{eq:defect-action} is the familiar transverse energy for a spherically symmetric $p$-defect in Minkowski space $\mathbb{M}^{n}$.
It is convenient to define the Jacobian factor
	\begin{figure}
		\centering
		\includegraphics[width=0.5\textwidth]{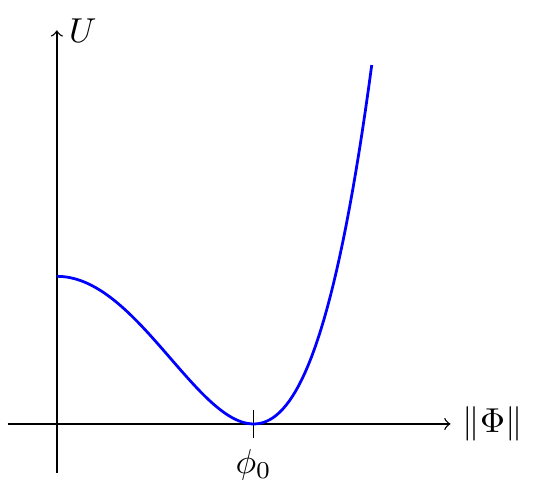}
		\caption{\small Here $U$ is a typical symmetry breaking potential with minima at $\norm{\Phi} = \phi_{0}$. We choose without justification that $U(\phi_{0})=0$.}%
\label{fig:potential}
\end{figure}
	\begin{equation}
	J(\nu) = V_{l-1}\left[\cosh\left( \nu/\rho\right)\right]^{q}\; \left[\rho\, \sinh\left( \nu/\rho\right)\right]^{l-1}\,,
	\end{equation}
	where the ``radius of curvature'' of $\AdS_{n}$ is $\rho = 1/\lvert k \rvert^{1/2}$.
	
	The $\SO(l)$ spherically symmetric ansatz we employ is a generalization and slight variant of the original one used by 'tHooft~\cite{tHooft:1974kcl} and by Polyakov~\cite{Polyakov:1974ek} in the $\SO(3)$ Georgi-Glashow model. The ansatz is
	\begin{equation}
	\Phi^{I}(\vb*{\nu}) = \frac{\nu^{I}}{\lVert \vb*{\nu} \rVert}\,\phi(\nu) \quad\text{and}\quad A^{IJ}(\vb*{\nu}) = \frac{1}{2}\, f(\nu) \left(\nu^{I}\; \dd\nu^{J} - \nu^{J}\; \dd\nu^{I}\right),
	\label{eq:ansatz}
	\end{equation}
	where $\phi$ and $f$ are functions only of the radius $\nu=\lVert \vb*{\nu}\rVert$. We choose the potential $U$ to be of the general symmetry breaking form such as the one shown in figure~\ref{fig:potential}. The form chosen for $\Phi$ is a hedgehog type ansatz with $\phi(\nu) \xrightarrow{\nu\to\infty\,} \pm\phi_{0}$ corresponding to the topological winding number $\pm 1$ solutions. We want the world brane of the $p$-defect to be a gauge invariant set thus we require $\Phi(\vb*{\nu}=0)=0$ which implies $\phi(0)=0$. The form for $A$ is motivated by the abelian constant curvature spherically symmetric ansatz\footnote{The expected behavior near the world brane is constant field strength so we expect $f(0)\neq 0$.} and by the identification of the $\SO(l)$ Lie algebra associated with the gauge group to the one associated with the rotational isometries of $(T_{\sigma}\Sigma)^{\perp}$. In this article we only discuss situations where both the functions $\phi(\nu)$ and $f(\nu)$ are non-trivial functions of $\nu$ that describe localized defects. We note that expressions \eqref{eq:defect-action} and \eqref{eq:ansatz} are actually valid for $l=1$ if you interpret $V_{0}=2$ since $S^{0} \subset \mathbb{E}^{1}$ consists of two points $\{-1,+1\}$. Ansatz \eqref{eq:ansatz} for $l=1$ is the odd parity kink (domain wall) solution with an automatically vanishing gauge field. A brief computation and implementing \eqref{eq:darboux} gives
\begin{align}
	D\Phi^{I} &= \dd\Phi^{I} + A^{IJ}\Phi^{J}
	\nonumber \\
	&= \phi'(\nu) \left(P_{L}\dd\nu\right)^{I} + \left( 1 - \frac{\nu^{2} f(\nu)}{2}\right) \frac{\phi(\nu)}{\nu} \left(P_{T} \dd\nu\right)^{I} 
	\nonumber\\
	& = \phi'(\nu) \left(P_{L}\vartheta\right)^{I} + \left( 1 - \frac{\nu^{2} f(\nu)}{2}\right) \frac{\phi(\nu)}{\nu} \,\frac{\nu/\rho}{\sinh(\nu/\rho)} \left(P_{T} \vartheta\right)^{I}\,.
\end{align}
	To compute $\lVert D\Phi\rVert^{2}$ we exploit that in the orthonormal polar coframe, the longitudinal direction is the radial direction, and  that $P_{T}$ and $P_{L}$ are orthogonal projectors to conclude 
	\begin{equation}
	\lVert D\Phi\rVert^{2}  = \phi'(\nu)^{2} + (l-1) \left( 1 - \frac{\nu^{2} f(\nu)}{2}\right)^{2} \, \frac{\phi(\nu)^{2}}{\nu^{2}} \left(\frac{\nu/\rho}{\sinh(\nu/\rho)}\right)^{2}.
	\end{equation}
	
	Next we compute the curvature by using differential forms $F^{IJ}= \dd A^{IJ} + A^{IK}\wedge A^{KJ}$. The computation is a bit more involved but greatly simplifies by using the orthogonal projectors:
	\begin{gather}
	F^{IJ} = f(\nu) \left( 1 - \frac{\nu^{2} f(\nu)}{4}\right) (\dd\nu)_{T}^{I} \wedge (\dd\nu)_{T}^{J}
	\nonumber \\
	+ \left[ \frac{f'(\nu)}{2\nu} \left( \nu_{j}\nu^{I}\delta^{J}_{l} - \nu_{j}\nu^{J}\delta^{I}_{l}\right) + f(\nu) \left( \delta^{I}_{j} \delta^{J}_{l} - \delta^{J}_{j} \delta^{I}_{l} \right)
	\right] (\dd\nu)_{L}^{j} \wedge (\dd\nu)_{T}^{l}
	\nonumber \\
	= f(\nu) \left( 1 - \frac{\nu^{2} f(\nu)}{4}\right) \left(\frac{\nu/\rho}{\sinh(\nu/\rho)}\right)^{2} \; \vartheta_{T}^{I} \wedge \vartheta_{T}^{J}
	\nonumber \\
	+ \left[ \frac{f'(\nu)}{2\nu} \left( \nu_{j}\nu^{I}\delta^{J}_{l} - \nu_{j}\nu^{J}\delta^{I}_{l}\right) + f(\nu) \left( \delta^{I}_{j} \delta^{J}_{l} - \delta^{J}_{j} \delta^{I}_{l} \right)
	\right] \frac{\nu/\rho}{\sinh(\nu/\rho)} \;\vartheta_{L}^{j} \wedge \vartheta_{T}^{l}
	\label{eq:F}
	\end{gather}
Using the properties of the orthonormal polar coframe we conclude 
	\begin{align}
	\lVert F^{IJ} \rVert^{2} &= (l-1)(l-2) f^{2} \left( 1 - \frac{\nu^{2}f}{4}\right)^{2} 
	\left(\frac{\nu/\rho}{\sinh(\nu/\rho)}\right)^{4}
	\nonumber \\
	&\quad + 2(l-1) \left( \frac{\nu f'}{2} + f \right)^{2} \left(\frac{\nu/\rho}{\sinh(\nu/\rho)}\right)^{2}\,.
	\label{eq:f-sq}
	\end{align}
	The norm on $2$-forms is normalized by observing that in an orthonormal 
	coordinate system in $\mathbb{E}^{2}$, a constant $\U(1)$ field strength is 
	given by $F = f\, \dd\nu^{1}\wedge \dd\nu^{2}$ and $\lVert F \rVert^{2} = 
	f^{2}$. When we write $\lVert F^{IJ}\rVert^{2}$ we mean sum over all $I$ 
	and $J$. For example in an $\SO(2)$ gauge theory, you have $\lVert 
	F^{IJ}\rVert^{2} = \lVert F^{12}\rVert^{2} + \lVert F^{21}\rVert^{2}= 2 
	\lVert F^{12}\rVert^{2}$. In general when summing over the spacetime 
	indices $i$, $j$ you obtain $F_{ij}^{IJ}F_{ij}^{IJ} = 2\,\lVert 
	F^{IJ}\rVert^{2}$. Note that the Yang-Mills lagrangian is quadratic in $f$ 
	in the abelian $\SO(2)$ case.  
	
	Collating all the terms, we have that the transverse energy $E_{\perp}$ of the defect is
	\begin{align}
	E_{\perp} &= V_{l-1} \int_{0}^{\infty} \dd\nu \left[\cosh\left(\nu/\rho\right)\right]^{q}\; \left[\rho\, \sinh\left( \nu/\rho\right)\right]^{l-1} 
	\nonumber \\
	&\times \left\lbrace \left[\half\phi'(\nu)^{2} + \half(l-1) \left( 1 - \frac{\nu^{2} f(\nu)}{2}\right)^{2} \, \frac{\phi(\nu)^{2}}{\nu^{2}} \left(\frac{\nu/\rho}{\sinh(\nu/\rho)}\right)^{2} +U(\phi^{2}) \right]\right.
	\nonumber \\
	&\quad + \frac{1}{g^{2}} \left[ \frac{1}{2}(l-1) \left( \frac{\nu f'(\nu)}{2} + f(\nu) \right)^{2} \left(\frac{\nu/\rho}{\sinh(\nu/\rho)}\right)^{2}\right.
	\nonumber \\
	&\qquad + \left. \left. 
	\frac{1}{4}(l-1)(l-2) f(\nu)^{2} \left( 1 - \frac{\nu^{2}f(\nu)}{4}\right)^{2} 
	\left(\frac{\nu/\rho}{\sinh(\nu/\rho)}\right)^{4}\right] \right\rbrace.
	\label{eq:defect-action-1} 
	\end{align}
	This expression for $E_{\perp}$ is valid for $l \ge 1$.
	\begin{figure}
		\centering
		\includegraphics[width=0.6\textwidth]{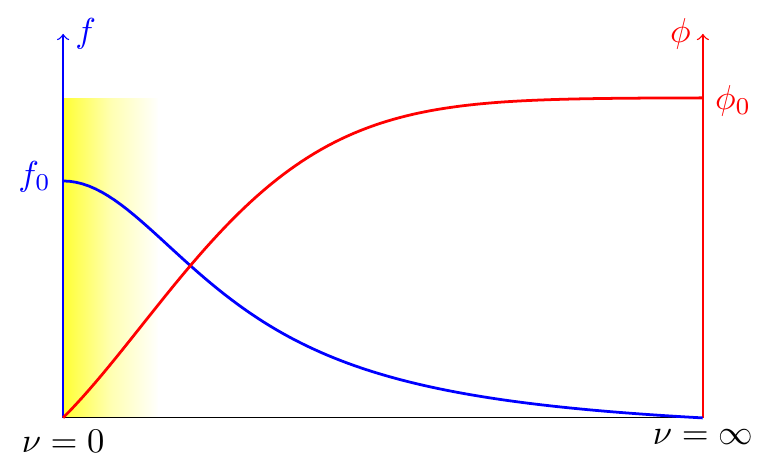}
		\caption{\small The expected behavior of $\phi$ and $f$ as a function of radial distance $\nu$ for a topological defect. The region with the gradient shade in yellow represents the core of the defect.}%
\label{fig:field}
	\end{figure}
	
	Equation~\eqref{eq:defect-action-1} may be simplified by introducing the auxiliary function $h$ defined by
	\begin{equation}
	h(\nu) = 1- \half\, \nu^{2}\, f(\nu)\,.
	\label{eq:def-h}
	\end{equation}
	The expected defect boundary conditions on $h$ are $h(\nu) \xrightarrow{\nu\to 0\;} +1$ and $h(\nu) \xrightarrow{\nu\to\infty\;} 0$, see eq.~\eqref{eq:crit-h}.	In terms of $\phi$ and $h$, the transverse energy $E_{\perp}$ looks like:
	\begin{align}
	E_{\perp} &= V_{l-1} \int_{0}^{\infty} \dd\nu \left[\cosh\left(\nu/\rho\right)\right]^{q}\; \left[\rho\, \sinh\left( \nu/\rho\right)\right]^{l-1} 
	\nonumber \\
	&\times \left\lbrace \left[\half\phi'(\nu)^{2} + \frac{(l-1)}{2} \frac{h(\nu)^{2}\, \phi(\nu)^{2}}{\left[\rho \sinh(\nu/\rho)\right]^{2}} +U(\phi^{2}) \right]\right.
	\nonumber \\
	&\quad + \frac{(l-1)}{g^{2}} \left. \left[ \frac{1}{2}\frac{h'(\nu)^{2}}{\left[\rho \sinh(\nu/\rho)\right]^{2}} +  
	\frac{(l-2)}{4} \frac{\left[h(\nu)^{2}-1\right]^{2}}{\left[\rho \sinh(\nu/\rho)\right]^{4}}\right] \right\rbrace.
	\label{eq:defect-action-2} 
	\end{align}
	
	With the defect boundary conditions our ansatz does not admit pure gauge solutions. To see this we note that if $l \ge 2$ then setting $F^{IJ}=0$ in \eqref{eq:f-sq} leads to $f(\nu)=0$ or $f(\nu)= 4/\nu^{2}$, algebraic forms for $f(\nu)$ that are incompatible with the defect boundary conditions.

If we restrict to the case of $l=2$, then in polar coordinates $(\nu,\varphi)$ for the normal bundle we would have that 
$A^{12}= f(\nu)\cdot\half\,\nu^{2}\;\dd\varphi \xrightarrow{\nu\to\infty\;} \dd\varphi$ and thus we conclude that we have nontrivial holonomy because if we integrate along the circle at infinity we have  $\oint A^{12}=2\pi$. This tells us that the total vortex flux is $2\pi$ in our normalization, that may be related to the conventional normalization by $A = g A_{\text{conv}}$.

	In the above discussion, the action leads to the standard Laplacian in the equations of motion. The situation is the same for the theory with the conformal Laplacian, which is obtained by adding to $I_{\text{spherically sym}}$ a term of the type $c R^{M^{n}} \Phi^{2}$ where $c$ is some fixed constant. Since the scalar curvature $R^{M^{n}}$ is a constant, that term can be absorbed into the potential energy function $U$ as a correction to the quadratic term in $\Phi$. We still choose the potential $U$ to be of the general symmetry breaking form such as the one shown in figure~\ref{fig:potential}.

	Minimizing $E_{\perp}$, we can derive the equations of motion for the defect.
	The  equation of motion for $\phi$ is
	\begin{subequations}\label{eq:eom-gen}
	\begin{align}
	0 &=- \frac{1}{J} \frac{\dd}{\dd\nu}\left( J\, \frac{\dd\phi}{\dd\nu}\right)
	+(l-1)\, h(\nu)^{2} \, \frac{\phi(\nu)}{\left[ \rho \sinh(\nu/\rho)\right]^{2}} + \frac{\dd U}{\dd\phi}\,,
	\nonumber\\
	& = - \frac{\dd^{2}\phi}{\dd\nu^{2}} - \left( \frac{q}{\rho}\,\tanh(\nu/\rho) + \frac{(l-1)}{\rho}\, \coth(\nu/\rho)\right) \frac{\dd\phi}{\dd\nu} 
	\nonumber\\
	&\quad  + \frac{\dd U}{\dd\phi} + \frac{(l-1)}{\left[ \rho \sinh(\nu/\rho)\right]^{2}}\;h(\nu)^{2}\, \phi(\nu)\,. 
	\label{eq:eom1}
	\end{align}
	Similarly, the $h$ equation of motion is
	\begin{align}
	0 &= -\frac{\dd^{2}h}{\dd\nu^{2}} - \left( \frac{q}{\rho}\,\tanh(\nu/\rho) + \frac{(l-3)}{\rho}\, \coth(\nu/\rho)\right) \frac{\dd h}{\dd \nu} 
	\nonumber\\
	&\quad + g^{2} \phi(\nu)^{2}\, h(\nu) + \frac{(l-2)}{\left[ \rho \sinh(\nu/\rho)\right]^{2}}\; \left(h(\nu)^{2}-1\right) h(\nu)\,.
	\label{eq:eom2}
	\end{align}
	\end{subequations}
	
\section{Finite transverse energy constraints}
\label{sec:finite-E}

We would like the $p$-defects in $\AdS_{n}$ to have finite transverse energy \eqref{eq:defect-action-2}. To achieve this, we have to study the convergence of the integral for $E_{\perp}$ in  the two asymptotic limits $\nu\to\infty$ and $\nu\to 0$ where we require that $0<\rho<\infty$. Of these two limits, the $\nu\to\infty$ one is more important. If the transverse energy diverges in the $\nu\to\infty$ limit then there is no hope for a  defect. A divergence in the integral as $\nu\to 0$ may be resolved by considering a different ultraviolet completion of the model. For example, you could add higher derivative terms to the action analogous to what is done in the 4D Skyrme model~\cite{Skyrme} to stabilize the model. In this section, we consider both limits within the context of expression \eqref{eq:defect-action-2} for $E_{\perp}$.

	\subsection{Behavior as $\nu\to\infty$}
	\label{sec:eperp-asymptotics}
	
	The $\nu\to\infty$ behavior is tricky, and has to be carefully analyzed on a case by case basis: $l=1$, $l=2$, and $l \ge 3$. We are interested in solutions that  satisfy the boundary conditions $\phi(\nu) -\phi_{0} \to 0$, and $h(\nu) \to 0$, both exponentially in $\nu$ as $\nu\to\infty$.
	
	The $E_{\perp}$ integrand \eqref{eq:defect-action-2}  consists of five positive semi-definite summands. A finite $E_{\perp}$ solution requires that the integral of each summand converge. These conditions impose growth rates on the fields $h$ and $\phi$. We observe that the growth rates that lead to convergent $E_{\perp}$ are not necessarily the growth rates given by the equations of motion. We are interested in topological defect solutions with  finite transverse  energy so the boundary condition that we choose is $\phi(\nu) \xrightarrow{\nu\to +\infty\,} +\phi_{0}$ for the winding number $+1$ defect. The form of the potential, see figure~\ref{fig:potential}, tells us that as $\nu\to +\infty$ we are near a quadratic minimum and we have that $U(\phi) \approx \frac{1}{2}\, m_{\phi}^{2}\, (\phi-\phi_{0})^{2}$. We would like for the length scale given by the radius of curvature $\rho$ to dominate the Compton wavelength $1/m_{\phi}$ of the scalar field, so in our later applications we require $\rho \ll 1/m_{\phi}$ even though this does not enter into the convergence analysis. Note that the case $\rho \gg 1/m_{\phi}$  essentially reduces to the flat space case. For all $l \ge 1$, we note that  requiring  that the scalar field  decays exponentially
\begin{equation}
	\phi(\nu)-\phi_{0} = O\left(e^{-\half(n-1)\nu/\rho}/\nu^{\half+\epsilon}\right)
	\label{eq:crit-phi}
\end{equation}
as $\nu\to\infty$ for some $\epsilon>0$ guarantees the convergence of the $\phi'(\nu)^{2}$ term and the $U(\phi)$ term\footnote{Convergence of the integral $\int^{+\infty}\psi(\nu)\; \dd\nu$ is guaranteed if there exists $\epsilon >0$ such that $\psi(\nu) = O(1/\nu^{1+\epsilon})$. In fact the integral will converge with the weaker condition $\psi(\nu) = O\left(1/\left[\nu\,(\ln\nu)^{1+\epsilon}\right]\right)$.} in \eqref{eq:defect-action-2}. For the future we note that for topological defects in $\AdS_{n}$, the asymptotic behavior of $\phi$ and $h$ depend not only on the transverse dimensionality $l$ but also on the dimension $p=q-1$ of the defect, see \eqref{eq:def-m-star} and  \eqref{eq:def-mu-star}.  This is different than the situation of the familiar defects in $\mathbb{M}^{n}$.
	
To have a gauge theory, we need $l\ge 2$ which implies that $n = q+l \ge 3$. In the same spirit, we analyze the asymptotic behavior of summands two and four in \eqref{eq:defect-action-2} that are associated with the kinetic energy of the gauge field $h$.  Those integrals will converge if 
\begin{equation}
h(\nu) = O\left(e^{-\half(n-3)\nu/\rho}/\nu^{\half+\epsilon}\right).
\label{eq:crit-h}
\end{equation}

Finiteness of $E_{\perp}$ imposes strong constraints if $l \ge 3$ because the non-abelian gauge fields' self interactions contribute to the energy via the term
	\begin{equation}
	V_{l-1}\; \frac{(l-1)(l-2)}{4 g^{2}} \int^{\infty} \dd\nu\; \left[\cosh\left(\nu/\rho\right)\right]^{q}\; \left[\rho\, \sinh\left( \nu/\rho\right)\right]^{l-1} 
	\frac{\left[h(\nu)^{2}-1\right]^{2}}{\left[\rho \sinh(\nu/\rho)\right]^{4}}
	\label{eq:nonabelian-energy}
	\end{equation}
	in eq.~\eqref{eq:defect-action-2}. We know that $h(\nu) \xrightarrow{\nu\to\infty\;} 0$ and the convergence of the integral may be determined from the asymptotics of the hyperbolic functions
	\begin{equation}
	\int^{\infty}\dd\nu\; e^{(n-5)\nu/\rho}\,.
	\label{eq:nonabelian-div}
	\end{equation}
	This integral will converge if $n<5$. Because $l \ge 3$ and $q \ge 1$, we see that there is exactly one case  where the integral converges, namely $q=1$ with $l=3$. This is the soliton case, a $0$-defect, which is the 'tHooft-Polyakov monopole in $\AdS_{4}$.
	
	The $n=5$ case where the integral is linearly divergent may be salvageable via some unknown method (see the flat space discussion below). There you would have an $\SO(4)$ soliton ($p=0$), and an $\SO(3)$ line defect ($p=1$).
	
	The situation for the existence $\SO(l)$ nonabelian topological defects for $n \ge 6$ appears to be quite dire because of the exponentially divergent energy.
	
	The results are different in the flat space case, \emph{i.e.}, $\rho=\infty$. Here, the asymptotics of \eqref{eq:nonabelian-energy} are given by
	\begin{equation*}
	\int^{\infty} \dd\nu\; \nu^{l-5}
	\end{equation*}
	This integral converges if $l=3$ for all values of $q$. Thus $\SO(3)$ topological $p$-defects are energetically allowed for all values of $p\ge 0$, note that $n \ge 4$. These are 'tHooft-Polyakov $p$-defects that have finite energy. The $l=4$ case has  logarithmically divergent energy and may be salvageable by modifying the model in some way.  An analogy is the XY-model\footnote{The XY model is the $\SO(2)$ nonlinear sigma model in $\mathbb{E}^{2}$, \emph{i.e.}, $n=2$ flat space.}  where the vortex has a logarithmically divergent energy that can be tamed by adding an abelian gauge field. Another approach to finite energy in the XY model is to just have the scalar field but restrict the field configurations to the topological sector with zero net winding number. The consequence of this is that the field decays faster at infinity and leads to finite energy. In this way you can have finite energy vortex anti-vortex pairs. The vortices have an ultraviolet energy divergence near the world brane. This is usually remedied by imposing a finite radius cutoff in the core, or including higher derivative terms such as in the Skyrme model.

	\subsubsection{Asymptotic behavior of $\phi$}
	
	If we look at \eqref{eq:eom1} we note that as $\nu\to\infty$ the term containing $h^{2}$ is very small compared to the other terms because $h$ is exponentially small and $1/\sinh^{2}(\nu/\rho)$ is also exponentially small\footnote{Note that this term is absent in the $l=1$ case.}. In this case, the asymptotic behavior of the equation of motion \eqref{eq:eom1} is
	\begin{equation}
	0= - \frac{\dd^{2}\psi}{\dd\nu^{2}} - \frac{n-1}{\rho}\, \frac{\dd\psi}{\dd\nu} + m_{\phi}^{2}\, \psi\,,\end{equation}
	where $\psi = \phi-\phi_{0}$. The solution to the equation of motion with the correct asymptotic behavior is  $\phi(\nu) - \phi_{0} = -\psi_{*}\; e^{-m_{*} \nu}$ where
	\begin{equation}
	m_{*} =  \frac{n-1}{2\rho} + \sqrt{ m_{\phi}^{2} + \left( \frac{n-1}{2\rho} \right)^{2}\;} \,,
	\label{eq:def-m-star}
	\end{equation}
	and $\psi_{*}>0$ is a constant. This decay behavior leads to convergence in the relevant transverse energy integral summands if $\sqrt{ m_{\phi}^{2} + \left[ (n-1)/2\rho \right]^{2}\;} >0$ by \eqref{eq:crit-phi}. This condition is always satisfied in our models because $n \ge 2$. The contribution to the transverse energy density in the asymptotic region from the purely scalar field part  is
	\begin{equation}
	u_{\phi}(\nu) = \frac{1}{2}\left( \frac{\dd\phi}{\dd\nu}\right)^{2} + U(\phi)
	\approx \frac{1}{2}\left( m_{*}^{2} + m_{\phi}^{2}\right) \psi_{*}^{2}\; e^{-2m_{*}\nu}
	\label{eq:e-dens-defect}
	\end{equation}
	Thus, we  have that
	\begin{gather}
	J(\nu)\,u_{\phi}(\nu) \xrightarrow{\nu\to +\infty\,} V_{l-1}(S^{l-1}) \qty(\frac{1}{2})^{n - 1}\; \rho ^{l-1}\; e^{(n-1)\nu/\rho}\; \frac{1}{2}\left( m_{*}^{2} + m_{\phi}^{2}\right) \psi_{*}^{2}\; e^{-2m_{*}\nu} \nonumber \\
	= V_{l-1}(S^{l-1})\, \frac{m_{*}^{2}+ m_{\phi}^{2}}{2^{n}}\; \rho^{l-1}\; \psi_{*}^{2}\; \exp\left[ -\sqrt{4m_{\phi}^{2}\rho^{2} +(n-1)^{2}\;}\;\bigl(\nu/\rho\bigr)\right] 
	\label{eq:phi-energy}
	\end{gather}
	The important result here is that the pure $\phi$ part contribution to the transverse energy integral converges for all values  of $n \ge 2$.  
The topological kink ($l=1$) exists for all $q\ge 1$. 
	
	\subsubsection{Asymptotic equation of motion for $h$}
	\label{sec:eom-h}
	
	As $\nu\to\infty$ with our boundary conditions, the asymptotic  equation of motion is
\begin{equation}
0 = - \frac{\dd^{2}h}{\dd\nu^{2}} - \frac{n-3}{\rho}\, \frac{\dd h}{\dd\nu} + m_{A}^{2}\, h(\nu)\,,
\end{equation}
where $m_{A}^{2} = g^{2}\phi_{0}^{2}$. The exponentially decaying solution to this equation is $h(\nu) = h_{*}\; e^{-\mu_{*}\nu}$ where
\begin{equation}
\mu_{*} = \frac{n-3}{2\rho} + \sqrt{ \left(\frac{n-3}{2\rho}\right)^{2} + m_{A}^{2}\;\strut}\;,
\label{eq:def-mu-star}
\end{equation}
To have a gauge field, we need $l \ge 2$ which implies that $n\ge 3$. The relevant transverse energy terms will converge if $\sqrt{ [(n-3)/2\rho]^{2} + m_{A}^{2}\;}>0$ according to criterion \eqref{eq:crit-h}.

To compute the contribution of $h$ to the energy density, we consider the second and fourth terms of \eqref{eq:defect-action-2}:	
	\begin{equation}
	u_{h}(\nu) = \frac{l-1}{g^{2}\, [\rho \sinh(\nu/\rho)]^{2}}
	\left( \frac{1}{2}\, h'(\nu)^{2} +  \half\, g^{2}\,\phi(\nu)^{2}\, h(\nu)^{2} \right) .
	\label{eq:h-density-l}
	\end{equation}

	The asymptotic energy density for the gauge field is
	\begin{equation}
	u_{h}(\nu) \approx \frac{4(l-1)\; e^{-2\nu/\rho}}{g^{2}\, \rho^{2}}\;\half\left( \mu_{*}^{2} + m_{A}^{2} \right) h_{*}^{2}\; e^{-2\mu_{*}\nu} \,, 
	\label{eq:uh-l}
	\end{equation}
	and the associated integrand is
	\begin{gather}
	J(\nu)\,u_{h}(\nu) \xrightarrow{\nu\to +\infty\,} V_{l-1}(S^{l-1}) \; \frac{\rho ^{l-1}}{2^{n-1}}\; e^{(n-1)\nu/\rho}\; \frac{2(l-1)\; e^{-2\nu/\rho}}{g^{2}\, \rho^{2}}\; \left( \mu_{*}^{2} + m_{A}^{2} \right) h_{*}^{2}\; e^{-2\mu_{*}\nu}
	\nonumber \\
= V_{l-1}(S^{l-1})  \; \frac{(l-1)\left( \mu_{*}^{2} + m_{A}^{2} \right)}{2^{n-2}\; g^{2}}\; \rho ^{l-3}\; h_{*}^{2}\; \exp\left[- \sqrt{4m_{A}^{2}\rho^{2}+(n-3)^{2}
\;} \bigl(\nu/\rho\bigr)\right]
	\label{eq:h-energy-l}
	\end{gather}

	\subsubsection{The asymptotic behavior of $h$ in the case $l=2$}

	The $l=2$ case is the abelian vortex and we see that in \eqref{eq:defect-action-2} that the non-abelian self-interaction term vanishes identically. 
The asymptotic energy density  integrand for the gauge field is
	\begin{gather}
	J(\nu)\,u_{h}(\nu) \xrightarrow{\nu\to +\infty\,}  \frac{2\pi}{g^{2}\, \rho}\;\frac{\left( \mu_{*}^{2} + m_{A}^{2} \right)}{2^{q}}\; h_{*}^{2}\; \exp\left[ -\sqrt{ 4m_{A}^{2}\rho^{2} + (q-1)^{2}\;}(\nu/\rho)\right] \,.	\label{eq:h-energy}
	\end{gather}
Notice that the  $h$ part contribution to the transverse energy integral converges for all values  of $q \ge 2$. Therefore, the topological abelian vortex localized at a totally geodesic $\AdS_{q} \hookrightarrow \AdS_{q+2}$ is energetically possible, \emph{i.e.},  vortex line, vortex sheet, \emph{etc.}. For $q=1$, the vortex soliton requires $m_{A}>0$.

\subsubsection{The asymptotic behavior of $h$ if $l \ge 3$}
\label{sec:asymp-l3}

\begin{figure}[tbp]
\centering
\includegraphics[width=0.75\textwidth]{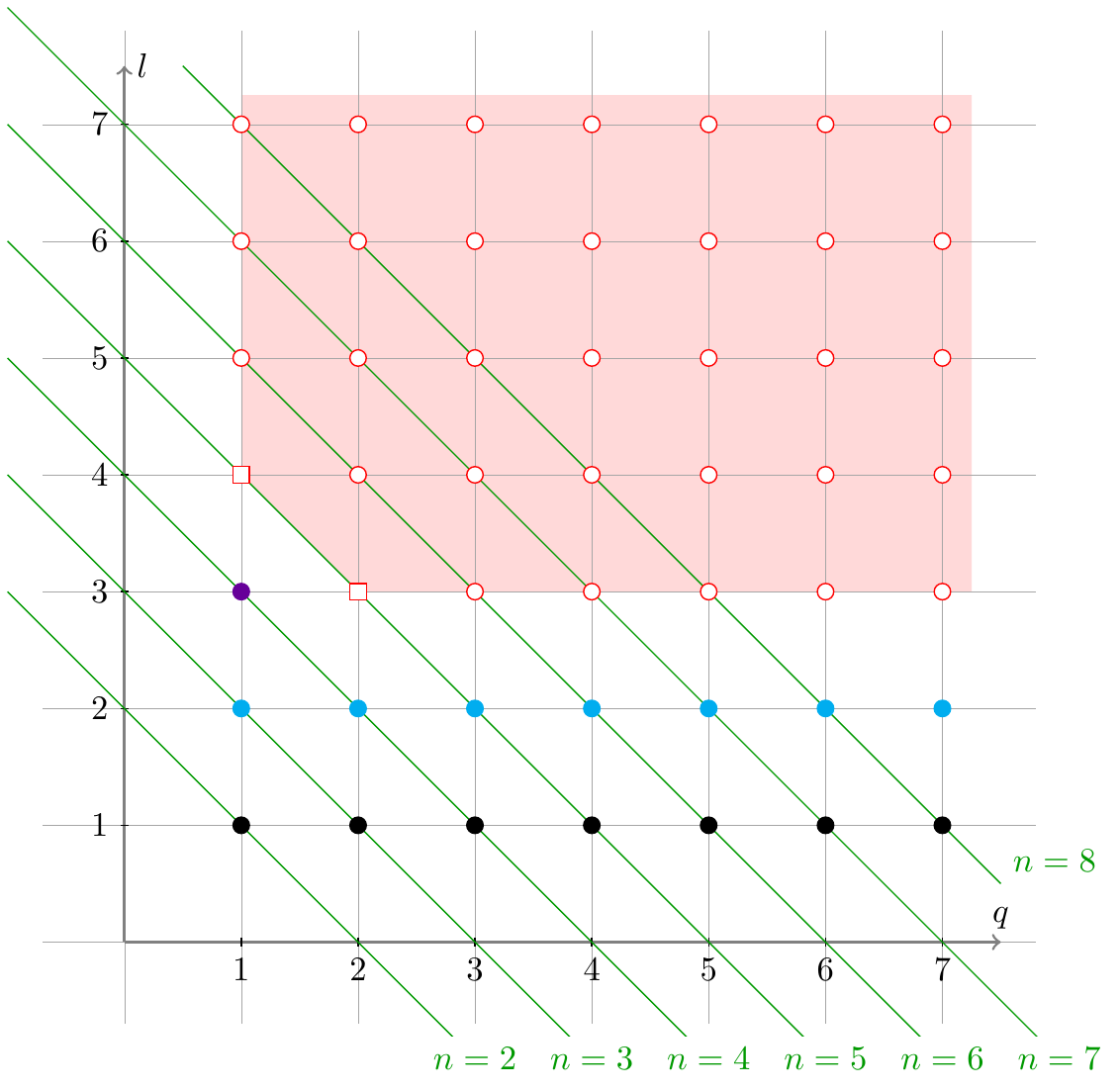}
\caption{\small Allowed values of $(q,l)$ that lead to a finite $E_{\perp}$ as $\nu\to\infty$ are represented by the solid circles. The light red region contains disallowed values of $(q,l)$ due to the divergence of the transverse energy arising from the nonabelian interactions of the gauge field, see eq.~\eqref{eq:nonabelian-div}. The open squares represent values where the $E_{\perp}$ divergence is linear, and the open circles where it is exponential. Remember that $n=q+l$ and that $q \ge 1$ and $l \ge 1$.}%
\label{fig:allowed}
\end{figure}

Expression \eqref{eq:h-energy-l} is the contribution from the second and fourth summands of \eqref{eq:defect-action-2}. We note that since $n \ge 4$ this contribution to the transverse energy converges even if $m_{A}=0$. The only concern is the fifth summand and we have already addressed it in section~\ref{sec:eperp-asymptotics} in the derivation of \eqref{eq:nonabelian-div}. The conclusion is that we do not expect solutions if $n \ge 5$ for $\l \ge 3$. Figure~\ref{fig:allowed} is a summary of values of $(q,l)$ that admit or do not admit solutions.

	\subsection{Behavior as $\nu\to 0$}
	\label{sec:nutozero}
	
	In studying the $\nu\to 0$ behavior of $\phi$ and $h$, we will encounter two second order linear 
ODEs with regular singular points at $\nu=0$. We look for Frobenius solutions of the form $\nu^{\alpha} \left( 1  +O(\nu^{2})\right)$. There are two real Frobenius indices, $\alpha_{+}$ and $\alpha_{-}$, with $\alpha_{+} > \alpha_{-}$. In both the $\phi$ and $h$ cases, $\alpha_{+}-\alpha_{-}=l \in \mathbb{Z}$. From the theorem of Fuchs, we know that there is a solution of the form $\nu^{\alpha_{+}} \left( 1  +O(\nu^{2})\right)$. The solution involving $\alpha_{-}$ may also have a logarithm. This solution does not satisfy the desired boundary conditions or the finite transverse energy constraint. From now on we only consider the $\alpha_{+}$ solution.

	The boundary conditions as $\nu\to 0$ are $\phi(\nu) \to 0$ and $h(\nu) \to +1$.
	First, we examine the $\nu\to 0$ behavior of \eqref{eq:eom1}. For $\nu \ll \rho$, the equation of motion is approximately
	\begin{equation}
	0 = - \frac{\dd ^{2}\phi}{\dd \nu^{2}} - \frac{l-1}{\nu}\, \frac{\dd \phi}{\dd \nu} +(l-1) \frac{\phi}{\nu^{2}} \,.
	\end{equation}
First, we discuss the $l=1$ case where we immediately see that the solution behaves like $\phi(\nu) \sim a + b\nu$. We are interested in setting $a=0$ because that solution  vanishes at $\nu=0$ and has an odd extension to $\nu<0$ corresponding to a kink localized at the origin. Next, we look at $l \ge 2$, where a Frobenius type solution of the form $\phi(\nu) = C \nu^{\alpha} \left( 1 + O(\nu^{2})\right)$ leads to $\alpha_{+}=1$ or $\alpha_{-}= 1-l$.  Thus we conclude that the small $\nu$ behavior of $\phi$ is $\phi(\nu) = \phi'(0)\, \nu + O(\nu^{3})$ for $l \ge 1$.

	Next we look at the $\nu\to 0$ behavior of \eqref{eq:eom2}. From \eqref{eq:defect-action-2} we see that we have to assume that $l \ge 2$ to discuss the $h$ equation of motion. We define $\htil(\nu) = h(\nu) - 1$, then the boundary condition becomes $\htil(0)=0$. For $\nu \ll \rho$ and $\lvert \htil(\nu)\rvert \ll 1$, the equation of motion is approximately
	\begin{equation}
	0 = \nu^{2}\, \frac{\dd^{2}\htil}{\dd\nu^{2}} +(l-3)\,\nu\,\frac{\dd\htil}{\dd\nu} -2(l-2)\, \htil(\nu)\,.
	\end{equation}
The Frobenius  indices are $\alpha_{+}=2$ and $\alpha_{-}= 2-l$. Implementing the boundary condition $\htil(0)=0$  we learn that $\htil(\nu) = -\half\,f(0)\,\nu^{2}+ O(\nu^{4})$, or equivalently $h(\nu) =1 - \half\, f(0)\, \nu^{2} + O(\nu^{4})$.

To summarize, the asymptotic behaviors of the fields for this kind of  topological defect 
is given by
\begin{align}
\intertext{Near region:}
\phi(\nu) &\xrightarrow{\nu\to 0\,} \phi_{\text{near}}(\nu)=\phi'(0)\nu +O\left(\nu^3\right), & l \ge 1 \\
h(\nu)&\xrightarrow{\nu\to 0\,} h_{\text{near}} (\nu)=1-\frac{1}{2} f(0) \nu^2+O\left(\nu^4 \right), & l \ge 2
\intertext{Far region:}
\phi(\nu)&\xrightarrow{\nu\to \infty\,} \phi_{\text{far}} (\nu)=\phi_0-\psi_* e^{-m_* \nu}, & l\ge 1 \label{eq:far-phi} \\
h(\nu)&\xrightarrow{\nu\to \infty\,} h_{\text{far}} (\nu)=h_* e^{-\mu_* \nu}, & l \ge 2
\end{align}
In deriving the asymptotic behavior we explicitly assumed that the radius of curvature satisfied $0<\rho<\infty$. Taking a limit such as $(1/\rho) \downarrow 0$ may be delicate and we have to be very careful. There are no issues with the flat space limit $(1/\rho) \downarrow 0$ if $m_{\phi}>0$ and $m_{A} >0$. On the other hand, you cannot directly apply the asymptotic formulas above when you approach the BPS line in figure~\ref{fig:bps-parameter}. This is a delicate limit in which you have $(1/\rho) \downarrow 0$ and $m_{\phi}\downarrow 0$. The reason is that the assumptions leading to eq.~\eqref{eq:far-phi} are now invalid. The solution~\eqref{eq:bps-soln} of the flat space BPS equations lead to a Coulombic tail asymptotic behavior $\phi(\nu) \to \phi_{0} + O(1/\nu)$ because the flat space $\phi$ field is massless. You have to be equally careful when you approach the NS-BPS line. The behavior along the double BPS line is safe because the radius of curvature remains non-zero and governs the asymptotic behavior.
	
	\section{Double well potential model with the double BPS limit}
	\label{sec:double}

In  this section, we will explore a method of obtaining exact solutions to the equations of motion for some of these defects. We study the equations of motion for a model with potential function
	\begin{equation}
	U(\phi) = \frac{1}{8}\, \lambda \left( \phi^{2} - \phi_{0}^{2}\right)^{2}
	\label{eq:phi-4}
	\end{equation}
It is convenient to rescale to dimensionless variables via $\nu \to \rho \nu$, $\phi \to \phi_{0}\,\phi$. In flat space, the  mass of the Higgs boson is $m_{\phi}^{2}= \lambda \phi_{0}^{2}$, and  $(l-1)$ vector bosons acquire mass via the Higgs mechanism with value $m_{A}^{2}= g^{2}\phi_{0}^{2}$. The transverse energy \eqref{eq:defect-action-2} may be written as
	\begin{align}
	E_{\perp} &= V_{l-1}\,\rho^{l-2}\,\phi_{0}^{2} \int_{0}^{\infty} \dd\nu \left[\cosh\nu\right]^{q}\; \left[\sinh\nu\right]^{l-1} 
	\nonumber \\
	&\times \left\lbrace \left[\half\phi'(\nu)^{2} + \frac{(l-1)}{2} \frac{h(\nu)^{2}\, \phi(\nu)^{2}}{\left[\sinh\nu\right]^{2}} 
	+\frac{1}{8} (m_{\phi}\rho)^{2}\left(\phi^{2}-1\right)^{2} \right]\right.
	\nonumber \\
	&\quad + \frac{(l-1)}{(m_{A}\rho)^{2}} \left. \left[ \frac{1}{2}\frac{h'(\nu)^{2}}{\left[\sinh\nu\right]^{2}} +  
	\frac{(l-2)}{4} \frac{\left[h(\nu)^{2}-1\right]^{2}}{\left[\sinh\nu\right]^{4}}\right] \right\rbrace.
	\label{eq:defect-action-3} 
	\end{align}
We note that $[\rho]=M^{-1}=L$, $[\phi_{0}] = M^{(n-2)/2}$, therefore the prefactor of the integral may we written as $\left(\rho^{n-2}\phi_{0}^{2}\right)/\rho^{q}$, and we see that $[E_{\perp}] = M^{q}= M/L^{p}$ as required.	The equations of motion are
\begin{subequations}\label{eq:eomx}
	\begin{align}
	- \half (m_{\phi}\rho)^{2}(\phi^{2}-1)\phi&= - \frac{\dd^{2}\phi}{\dd\nu^{2}} - \left( q\,\tanh\nu + (l-1)\, \coth\nu\right) \frac{\dd\phi}{\dd\nu} 
	\nonumber\\
	&\quad   + \frac{(l-1)}{\left[ \sinh\nu\right]^{2}}\;h(\nu)^{2}\, \phi(\nu)\,, 
	\label{eq:eom1x}\\
		- (m_{A}\rho)^{2} \phi(\nu)^{2}\, h(\nu) &= -\frac{\dd^{2}h}{\dd\nu^{2}} - \left( q\,\tanh\nu + (l-3)\, \coth\nu\right) \frac{\dd h}{\dd \nu} 
	\nonumber\\
	&\quad  + \frac{(l-2)}{\left[\sinh\nu\right]^{2}}\; \left(h(\nu)^{2}-1\right) h(\nu)\,.
	\label{eq:eom2x}
	\end{align}
\end{subequations}
Note that the equations of motion for a $p$-defect in $\AdS_{n}$ depend on both 
$q$ and $l$. The equations of motion \eqref{eq:eomy} for a $p$-defect in 
$\mathbb{M}^{n}$ depend only on $l$.

In the  Minkowski space $\mathbb{M}^{n}$ version of the Higgs  model, there are two independent length scales $1/m_{\phi}$ and $1/m_{A}$ that enter the equations of motion. We are studying topological defects in $\AdS_{n}$ and there is automatically an independent third length scale $\rho$, the radius of curvature of  $\AdS_{n}$. 
The spherically symmetric equations of motion \eqref{eq:eomx} depend on three mass scales $m_{\phi}^{2}$, $m_{A}^{2}$, and $1/\rho^{2}$ if $l \ge 2$. The radius of curvature $\rho$ is implicit in \eqref{eq:eomx} because the $\nu$ coordinate that appears in those equations of motion is actually the dimensionless radial distance $\nu/\rho$. From the form of the equations of motion there are various parameter limits that can be studied, see figure~\ref{fig:bps-parameter}. The limit $1/\rho^{2}\downarrow 0$ corresponds to the Minkowski flat space equations of motion. This is the yellow planar region in the figure. At a boundary of this region is the BPS line corresponding to solutions where $m_{\phi}^{2}\downarrow 0$ but $m_{A}^{2} \neq 0$. There is also a very delicate non-standard BPS (NS-BPS) like limit in flat space where you consider $m_{A}^{2} \downarrow 0$ with $m_{\phi}^{2}\neq 0$; there are no solutions with our ansatz because because the $h$-field does not satisfy the two point boundary conditions at $\nu=0$ and $\nu=\infty$. This NS-BPS limit is denoted by the dashed green line.  LMS studied the gray planar region given by $m^{2}_{\phi} \downarrow 0$, and they were interested in the the BPS line boundary but did not consider the double BPS line, see appendix~\ref{sec:translation}. We will find explicit analytic solutions with $1/\rho^{2} \neq 0$ but $m_{\phi}^{2} \downarrow 0$  and $m_{A}^{2} \downarrow 0$ along what we call the double BPS limit. Finally in the limit $m_{A}^{2} \downarrow 0$ we  have the pink planar region where we have an analytic form for the $h$-field, and a numerical solution for the $\phi$-field which are briefly discussed in section~\ref{sec:pink-plane}.

At this point, it is worth the effort to be more explicit in the limit we are taking, as it will appear in much of what is to follow. As previously mentioned, this limit is when $m_{\phi}^{2} \downarrow 0$ and $m_{A}^{2} \downarrow 0$, and can only be taken when there is a third length scale for the physics. In our case, we have one given by the radius of curvature of $\AdS_{n}$. It is also important to understand the way that this limit is taken. Up until this point, we have been using the action of a topological defect embedded in $\AdS_{n}$, this is what leads to the transverse energy integral~(\ref{eq:defect-action-3}) and the equations of motion~(\ref{eq:eomx}). What we do is take the double BPS limit in the equations of motion, allowing the masses to fall towards zero and partially decoupling the ODEs, see~\eqref{eq:eom-bps} below. As remarked in the Introduction, the action is singular in this limit but the equations of motion are not. The double BPS equations of motion do not follow from any action.
The solutions that we obtain from this method are a good point to start in a perturbative analysis of the full equations of motion that do follow from the action.

\begin{figure}[tbp]
\centering
\includegraphics[width=0.75\textwidth]{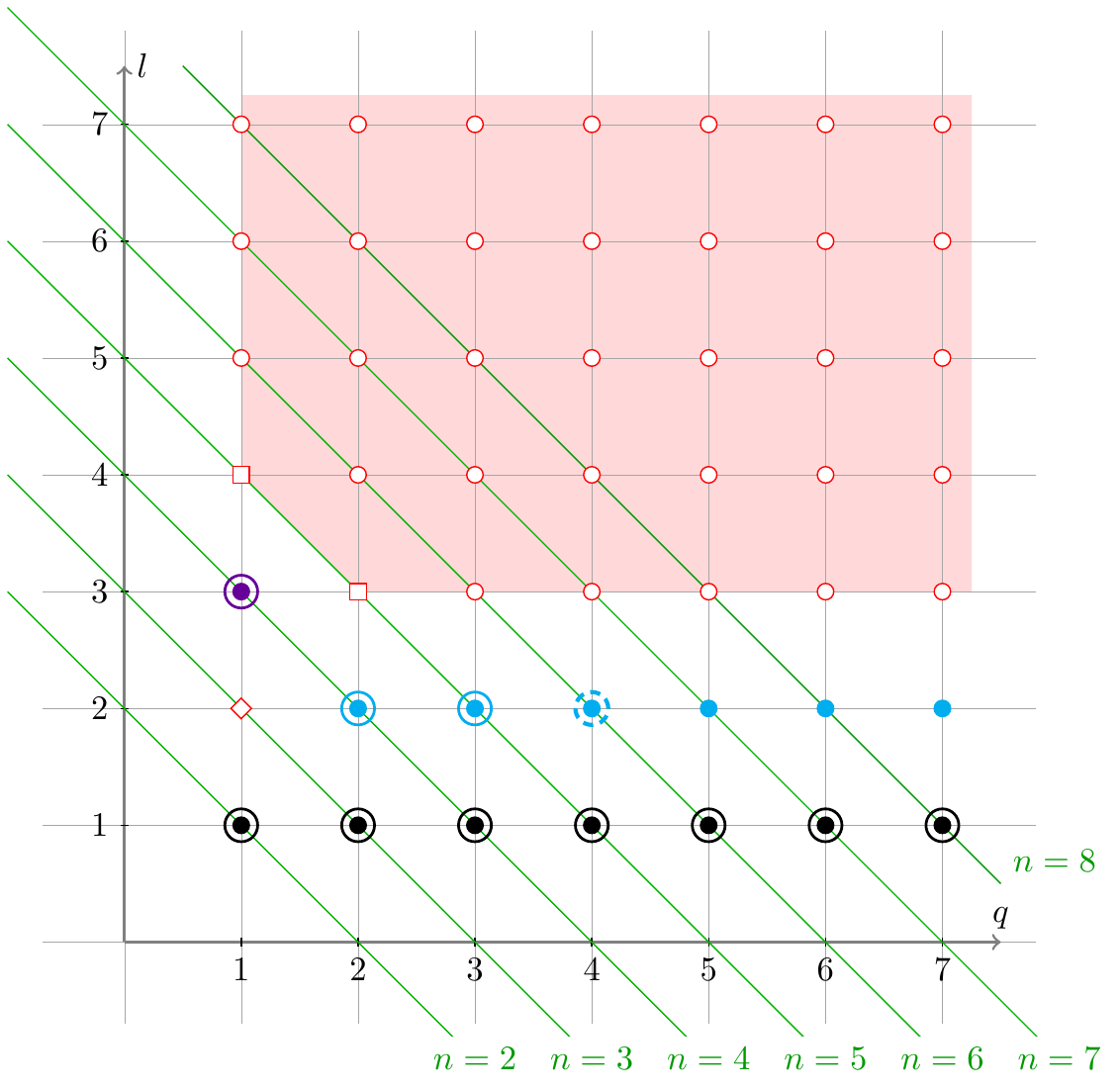}
\caption{\small Allowed values of $(q,l)$ in the double BPS limit that lead to a finite $E_{\perp}$ as $\nu\to\infty$ are represented by the solid circles. The notational conventions used in  figure~\ref{fig:allowed} are still valid in this figure.  The main difference between the two figures is that there is no double BPS solution if $(q,l)=(1,2)$ as denoted by the open red  diamond above.  The concentric circles indicate values of $(q,l)$ where we have found exact solutions to the double BPS equations of motion in terms of well known special functions: trigonometric, hyperbolic, Bessel, and hypergeometric. The dashed circle is an exact solution for the case $(4,2)$ in terms of the not well known confluent Heun functions; this explicit solution is not very illuminating.}%
\label{fig:allowed-dbps}
\end{figure}

In this  double BPS limit, the equations of motion become
	\begin{subequations}\label{eq:eom-bps}
	\begin{align}
	0&= - \frac{\dd^{2}\phi}{\dd\nu^{2}} - \left( q\,\tanh\nu + (l-1)\, \coth\nu\right) \frac{\dd\phi}{\dd\nu} 
	   + \frac{(l-1)}{\left[ \sinh\nu\right]^{2}}\;h(\nu)^{2}\, \phi(\nu)\,, 
	\label{eq:eom1-bps}\\
		0 &= -\frac{\dd^{2}h}{\dd\nu^{2}} - \left( q\,\tanh\nu + (l-3)\, \coth\nu\right) \frac{\dd h}{\dd \nu} 
	 + \frac{(l-2)}{\left[\sinh\nu\right]^{2}}\; \left(h(\nu)^{2}-1\right) h(\nu)\,.
	\label{eq:eom2-bps}
	\end{align}
\end{subequations}
Note that the $h$ equation of motion \eqref{eq:eom2-bps} has decoupled, yet it  is nonlinear. Thus we have a standalone second order ODE for $h$, whose solution can be inserted into eq.~\eqref{eq:eom1-bps} to obtain a standalone linear second order ODE for $\phi$.	If you are interested in pure Yang-Mills theory then you can just study  equation~\eqref{eq:eom2-bps} for $h$ and ignore the $\phi$ equation of motion. In solving the above we impose two point boundary conditions: $\phi(0)=0$, $\phi(\infty)=1$; $h(0)=1$, $h(\infty)=0$.
In this double BPS limit, the energy integrand asymptotics \eqref{eq:phi-energy} and \eqref{eq:h-energy-l} become
	\begin{align}
	J(\nu)\,u_{\phi}(\nu) &\xrightarrow{\nu\to +\infty\,} 
	 V_{l-1}(S^{l-1})\, \frac{(n-1)^{2}}{2^{n}}\; \rho^{l-3}\; \psi_{*}^{2}\; \exp\left[ -(n-1)\nu/\rho\right] 
	 \label{eq:u-dbps-phi}\\
	J(\nu)\,u_{h}(\nu) &\xrightarrow{\nu\to +\infty\,}  V_{l-1}(S^{l-1})  \; \frac{(l-1)(n-3)^{2}}{2^{n-2}\; g^{2}}\; \rho ^{l-5}\; h_{*}^{2}\; \exp\left[- (n-3) \nu/\rho\right]
	\label{eq:u-dbps-h}
	\end{align}
Since we always have $n \ge 2$, we see that exponential decay is always guaranteed in \eqref{eq:u-dbps-phi}. On the other hand, we see that if a gauge field is present then we get exponential decay in \eqref{eq:u-dbps-h} if and only if $n >3$. The asymptotic analysis for the case of $n=3$ with $q=1$ and $l=2$ is delicate because we lose the exponential decay factor in \eqref{eq:u-dbps-h} but the expression has a factor of $(n-3)^{2}$. To assess this case correctly requires more detailed analysis, see section~\ref{sec:vortex-1},  and we will conclude that there is no acceptable solution with $q=1$ and $l=2$. This means that in the double BPS limit we lose the vortex solution with $q=1$ and $l=2$ which is a $0$-defect. This discussion is summarized in figure~\ref{fig:allowed-dbps}.

In this article we consider the first step in a perturbative analysis of eqs. \eqref{eq:eomx}. By setting the left hand side of these equations to zero we obtain the double BPS equations \eqref{eq:eom-bps}. A solution to the double BPS equations is the starting point for finding a perturbative power series solution to \eqref{eq:eomx} in terms of the mass parameters $(m_{\phi}\rho)^{2}$ and $(m_{A}\rho)^{2}$. The left hand side of \eqref{eq:eomx} is viewed a perturbation. Once we have a formal power series solution we can insert it into the transverse energy integral \eqref{eq:defect-action-3} and obtain a power series expansion for $E_{\perp}$. The leading term goes like $1/m_{A}^{2}$ and the subleading term goes like $(m_{A}^{2})^{0}(m_{\phi}^{2})^{0}$. Both of these arise from the double BPS solutions to \eqref{eq:eom-bps}. The corrections to the double BPS solution lead to a sum of contributions to the transverse energy of the form $(m_{\phi}^{2})^{j}(m_{A}^{2})^{k}$ where $j +k \ge 1$, with $j$  and $k$ being non-negative integers.

\subsection{The double BPS kink-like defects ($l=1$)}
\label{sec:bps-kink}

In the $l=1$ case we only have one equation of motion \eqref{eq:eom1-bps} that is easily solved with the required boundary conditions. Since $q \ge 1$, the solution to the equation of motion is given by
\begin{subequations}
\begin{align} 
\phi(\nu) &=  \mathcal{N}_{(q,1)}^{-1} \int_{0}^{\nu}\frac{\dd \nu'}{(\cosh \nu')^{q}}\,,
\label{eq:gud-q}\\
\mathcal{N}_{(q,1)} &= \int_{0}^{\infty}\frac{\dd \nu}{(\cosh \nu)^{q}}\,.
\end{align}
\end{subequations}
The integrals above may be performed exactly for integer $q \ge 1$ and the answer is given in terms of the hypergeometric function:
\begin{subequations}
\label{eq:leq1}
\begin{align}
\phi(\nu) &= \mathcal{N}_{(q,1)}^{-1}\, (\sinh\nu)  \; {}_2F_1\!\left(\frac{1}{2},\frac{q+1}{2};\frac{3}{2};-\sinh ^2\nu \right)\\
\mathcal{N}_{(q,1)}&= \frac{2^q}{q} \, {}_2F_1\!\left(\frac{q}{2},q;\frac{q+2}{2};-1\right)
\end{align}
\end{subequations}

\begin{table}[tbp]
    \centering
    \begin{tabular}{@{}lcr@{}}
\toprule
$q$ & $\mathcal{N}_{(q,1)}\,\phi(\nu) = \sinh\nu  \; {}_2F_1\!\left(\frac{1}{2},\frac{q+1}{2};\frac{3}{2};-\sinh ^2\nu \right)$ & $\mathcal{N}_{(q,1)}$ \\
\midrule
$1$ & $\tan ^{-1}(\sinh \nu )$ & $\pi/2$ \\
$2$ & $\tanh \nu$ & $1$ \\
$3$ & $\frac{1}{2} \left(\tan ^{-1}(\sinh \nu )+\tanh \nu  \sech\nu )\right)$  & $\pi/4$ \\
$4$ & $\displaystyle\frac{2\sinh \nu \left(1+ \cosh^{2}\nu\right)}{3\cosh^{3}\nu}$  & $2/3$ \\
$10$ & $ \displaystyle \frac{(128 +130 \cosh 2 \nu +46 \cosh 4 \nu +10 \cosh 6 \nu +\cosh 8 \nu) \tanh \nu  }{315 \cosh^{8}\nu} $ & $128/315$ \\
\bottomrule
	\end{tabular}	
    \caption{\small Exact double BPS kink solutions ($l=1$) for small values of $q$. Remember that a $p$-defect has $q=p+1$. }
\label{tbl:kink-solutions}
\end{table}

\begin{figure}[tbp]
    \centering
    \includegraphics[width=0.8\textwidth]{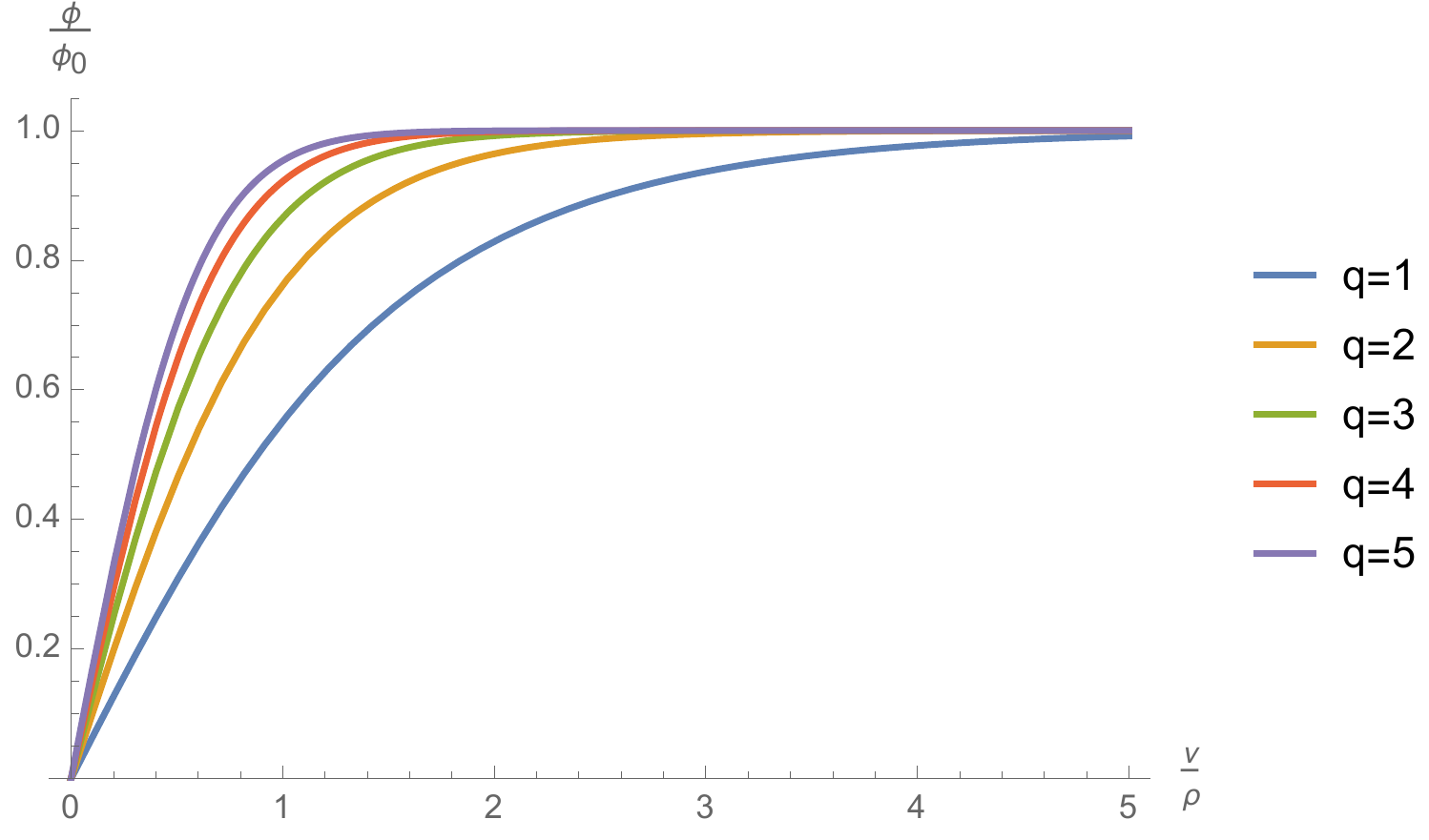}
    \caption{\small Graphs of the exact double BPS kink ($l=1$) for small values of $q$.}%
\label{fig:kink-graphs}
\end{figure}
For small values of $q$, the double BPS kink solutions $\phi$ are easily expressed in terms of elementary functions.  Some examples are tabulated in table~\ref{tbl:kink-solutions} and some are graphed in figure~\ref{fig:kink-graphs}.

\subsection{The double BPS vortex-like defects ($l=2$)}
\label{sec:bps-vortex}

The abelian vortex $l=2$ in the double BPS limit has various simplifications. First we remark that for a finite energy solution eq.~\eqref{eq:u-dbps-h} requires that the $p$-defect must have $p \ge 1$ or equivalently $q \ge 2$. The marginal case with $q=1$ is discussed in the next section. The $h$ equation of motion \eqref{eq:eom2-bps} is decoupled and linear since the abelian gauge field has no self-interactions:
\begin{equation}
		0 = \frac{\dd^{2}h}{\dd\nu^{2}}  +\left( q\,\tanh\nu - \coth\nu\right) \frac{\dd h}{\dd \nu} \,.
\end{equation}
The solution to this equation with the two point boundary conditions is
\begin{equation}
	h(\nu) = \frac{1}{(\cosh\nu)^{q-1}} \quad\text{for } q\ge 2\,.
	\label{eq:sol-h-l2}.
\end{equation}
From this we see that $f(0) = q-1$. 
Inserting this solution for $h$ in the $\phi$ equation of motion \eqref{eq:eom1-bps} leads to the linear ordinary differential equation
\begin{equation}
	0= - \frac{\dd^{2}\phi}{\dd\nu^{2}} - \left( q\,\tanh\nu + \coth\nu\right) \frac{\dd\phi}{\dd\nu} 
+ \frac{1}{\left( \sinh\nu\right)^{2} \left(\cosh\nu\right)^{2q-2}}\;\phi(\nu)\,. 
\label{eq:eom1-l2}
\end{equation}
The solution to this equation must satisfy the two sided boundary conditions $\phi(0)=0$ and $\phi(\infty)=1$. We point out that this ODE has a regular singular point at $\nu=0$. The behavior near the origin may be studied using the Frobenius power series method which leads to an indicial equation with indices $\pm 1$. 
We consider the regular solution that begins like $\nu$ and ignore the singular solution.

\subsubsection{Case of $q=1$ (no acceptable solution)}
\label{sec:vortex-1}

The analysis of $q=1$ requires the observation that the general solution to \eqref{eq:eom2-bps} is $h(\nu) = c_{1} \cdot 1 + c_{2} \ln\cosh\nu$. Note that $h(\nu) = O(\nu)$ as $\nu\to\infty$. This function cannot satisfy the two point boundary conditions on the $h$ field. We can require $h$ to have the correct behavior near $\nu=0$, and we write $h(\nu) = 1- f_{0} \ln\cosh\nu$ where $f_{0}$ is a constant. The asymptotic behavior is given by $h(\nu) \sim -f_{0}\nu$ as $\nu\to\infty$, and this is problematic because the contribution to $E_{\perp}$ in \eqref{eq:defect-action-2} from second summand diverges cubically while the contribution from the fourth summand diverges linearly. Thus there is no acceptable vortex solution for $q=1$ and $l=2$ in the double BPS limit.

\subsubsection{Case of $q=2$}
\label{sec:vortex-2}

\begin{figure}[tbp]
    \centering
    \includegraphics[width=0.8\textwidth]{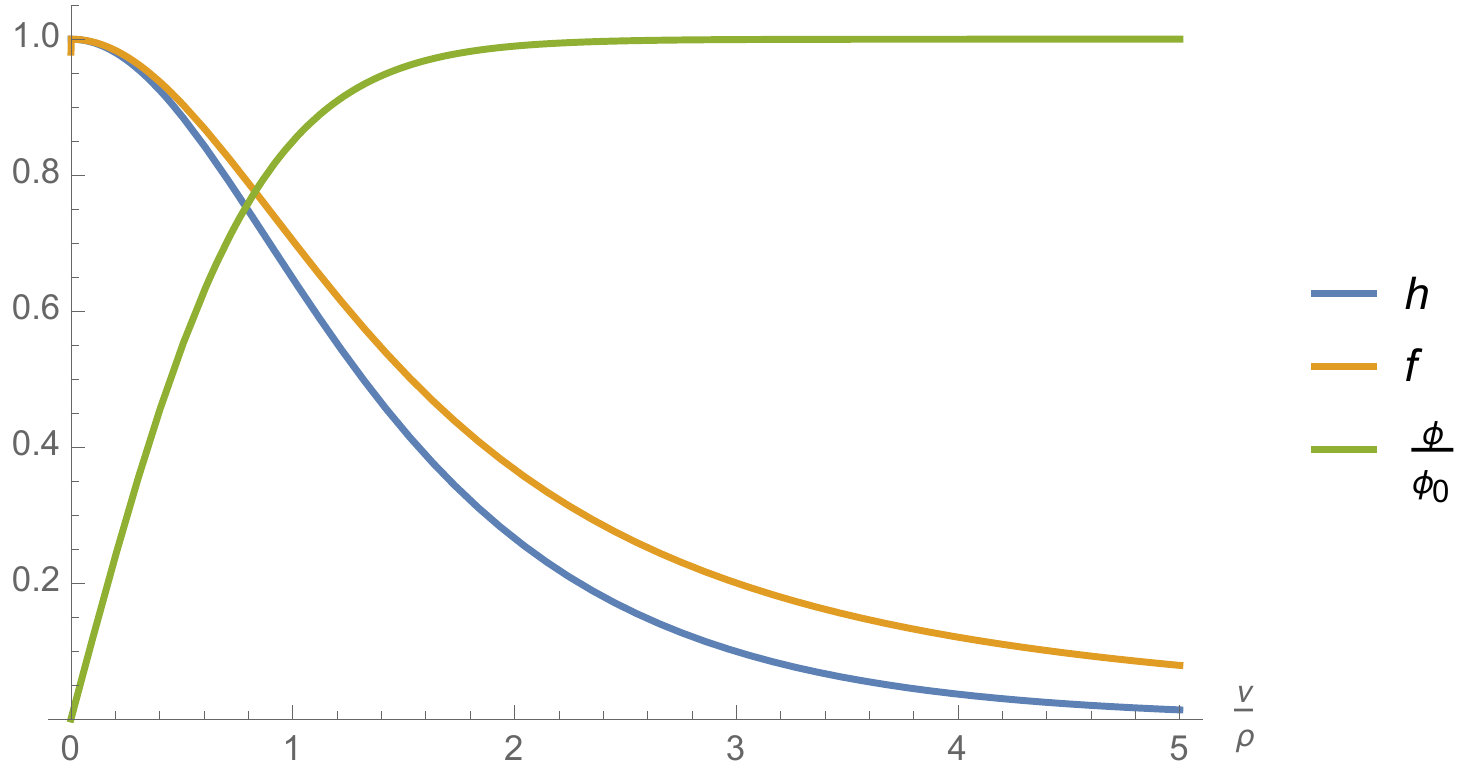}
    \caption{\small Graph of the exact double BPS Nielsen-Olesen vortex solution for $(q,l) = (2,2)$.}%
\label{fig:vortex-graphs-2}
\end{figure}

There is an exact analytic solution to \eqref{eq:eom1-l2} for the case $q=2$ that can be expressed by a hypergeometric function
\begin{subequations}\label{eq:exact-q2-l2}
\begin{align}
\phi(\nu) &= \mathcal{N}_{(2,2)}^{-1}\tanh\nu \;
   {}_2F_1\left(\frac{1}{4}-\frac{\sqrt{5}}{4},\frac{1}{4}+\frac{\sqrt{5}}{4};2;
   \tanh^2\nu\right)\,, 
   \label{eq:phi-q2} \\
   h(\nu) &= \frac{1}{\cosh\nu}\,,
\end{align}
\end{subequations}
where the normalization factor is given by
\begin{equation}
\mathcal{N}_{(2,2)} = \strut\ {}_2F_1\left(\frac{1}{4}-\frac{\sqrt{5}}{4},\frac{1}{4}+\frac{\sqrt{5}}{4};2; 1\right) =\frac{\sqrt{\pi }}{2\, \Gamma \left(\frac{7}{4}-\frac{\sqrt{5}}{4}\right) \Gamma \left(\frac{7}{4}+\frac{\sqrt{5}}{4}\right)} \approx
0.8206\,.
\end{equation}
A plot of the solution is given in figure~\ref{fig:vortex-graphs-2}.

\subsubsection{Case of $q=3$}
\label{sec:vortex-3}

\begin{figure}[tbp]
    \centering
    \includegraphics[width=0.8\textwidth]{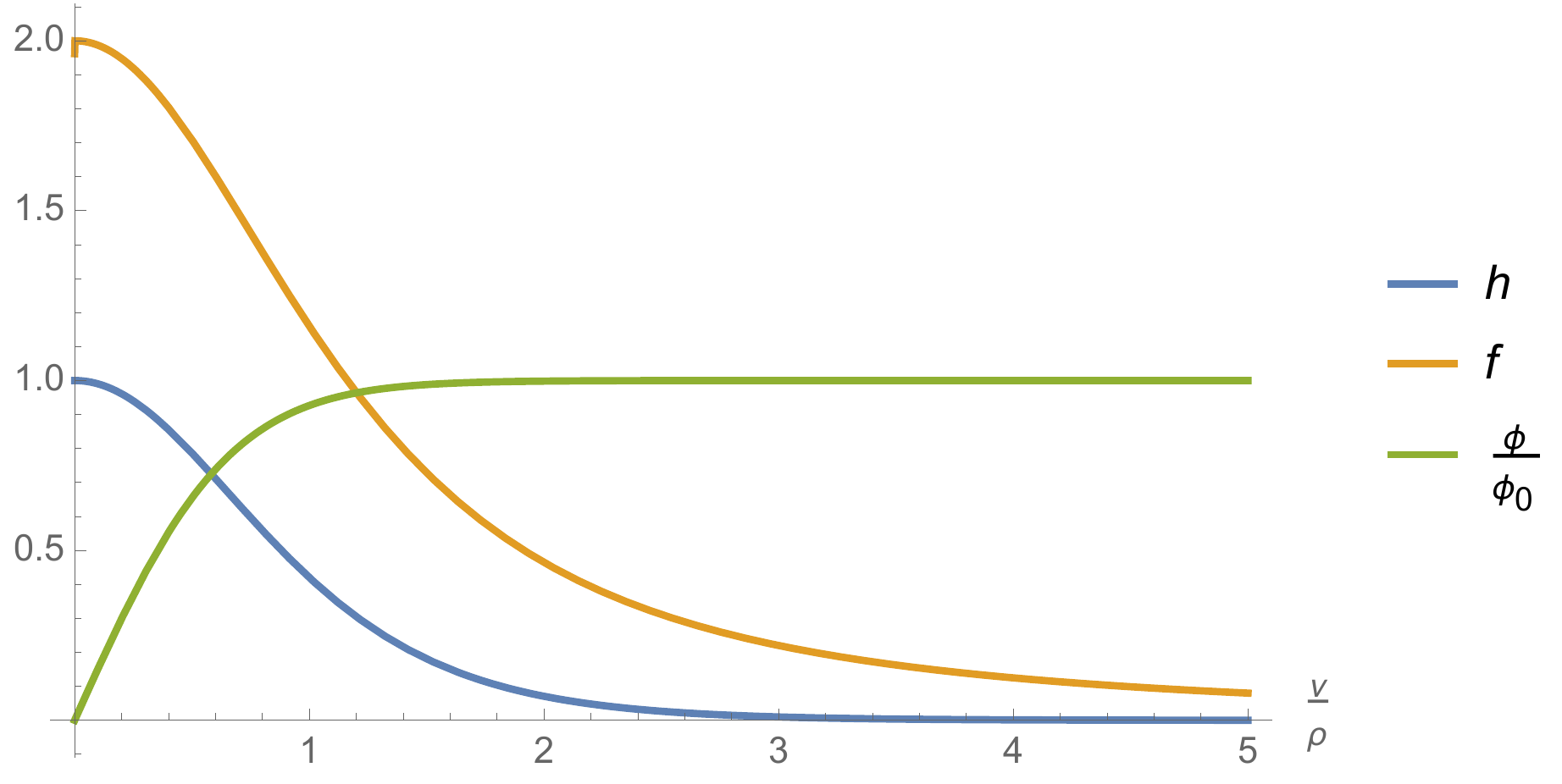}
    \caption{\small Graph of the exact double BPS vortex solutions for $(q,l) = (3,2)$.}%
\label{fig:vortex-graphs-3}
\end{figure}

There is an exact analytic solution to \eqref{eq:eom1-l2} for the case $q=3$ that can be expressed in terms of Bessel functions
\begin{subequations}
\begin{align}
\phi(\nu) &= \mathcal{N}_{(3,2)}^{-1}\; 2\bigl( \tanh\nu \; J_{0}(\tanh\nu) - J_{1}(\tanh\nu)
   \bigr)
= \mathcal{N}_{(3,2)}^{-1}\; 2 \tanh\nu\; J_{1}'(\tanh\nu)\,,
   \label{eq:phi-q3} \\
   h(\nu) &= \frac{1}{\cosh^{2}\nu}\,,
\end{align}
\end{subequations}
where the normalization factor is given by
\begin{equation}
\mathcal{N}_{(3,2)} = 2 \bigl(J_{0}(1)-J_{1}(1) \bigr) \approx
0.650294\,.
\end{equation}
A plot of the solution is given in figure~\ref{fig:vortex-graphs-3}.

\subsubsection{Case of $q \ge 4$}
\label{sec:vortex-4}

\begin{figure}[tbp]
    \centering
    \includegraphics[width=0.7\textwidth]{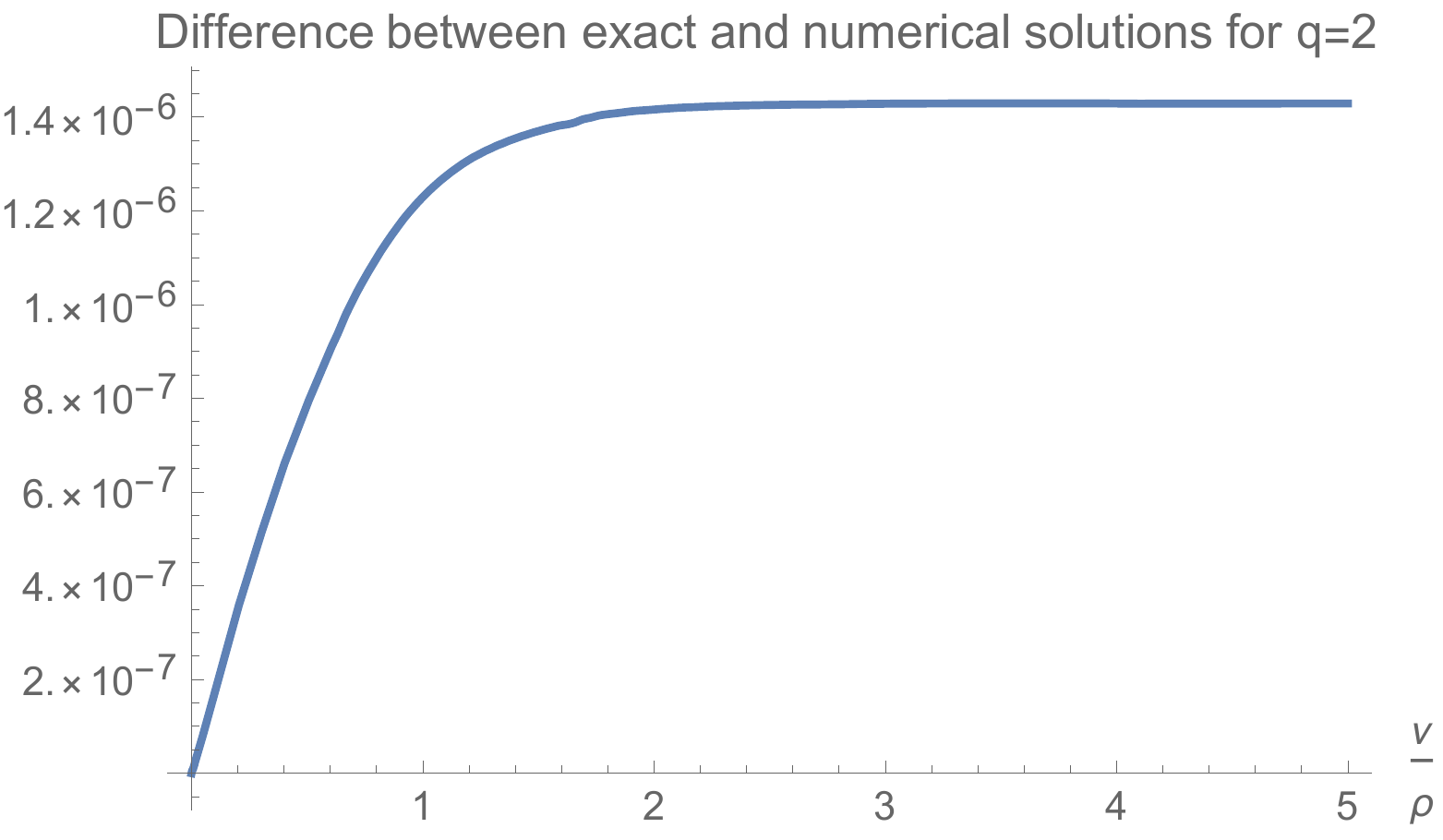}
    \caption{\small Case $q=2$ and $l=2$: plot of the difference between the exact solution \eqref{eq:phi-q2}, and the normalized numerical solution for the scalar field $\phi$ with initial data specified at $\nu_{*}= 1 \times 10^{-6}$ and determined by the truncated power series solution. }
    \label{fig:difference-solutions}
\end{figure}

\begin{figure}[tbp]
    \centering
    \includegraphics[width=0.8\textwidth]{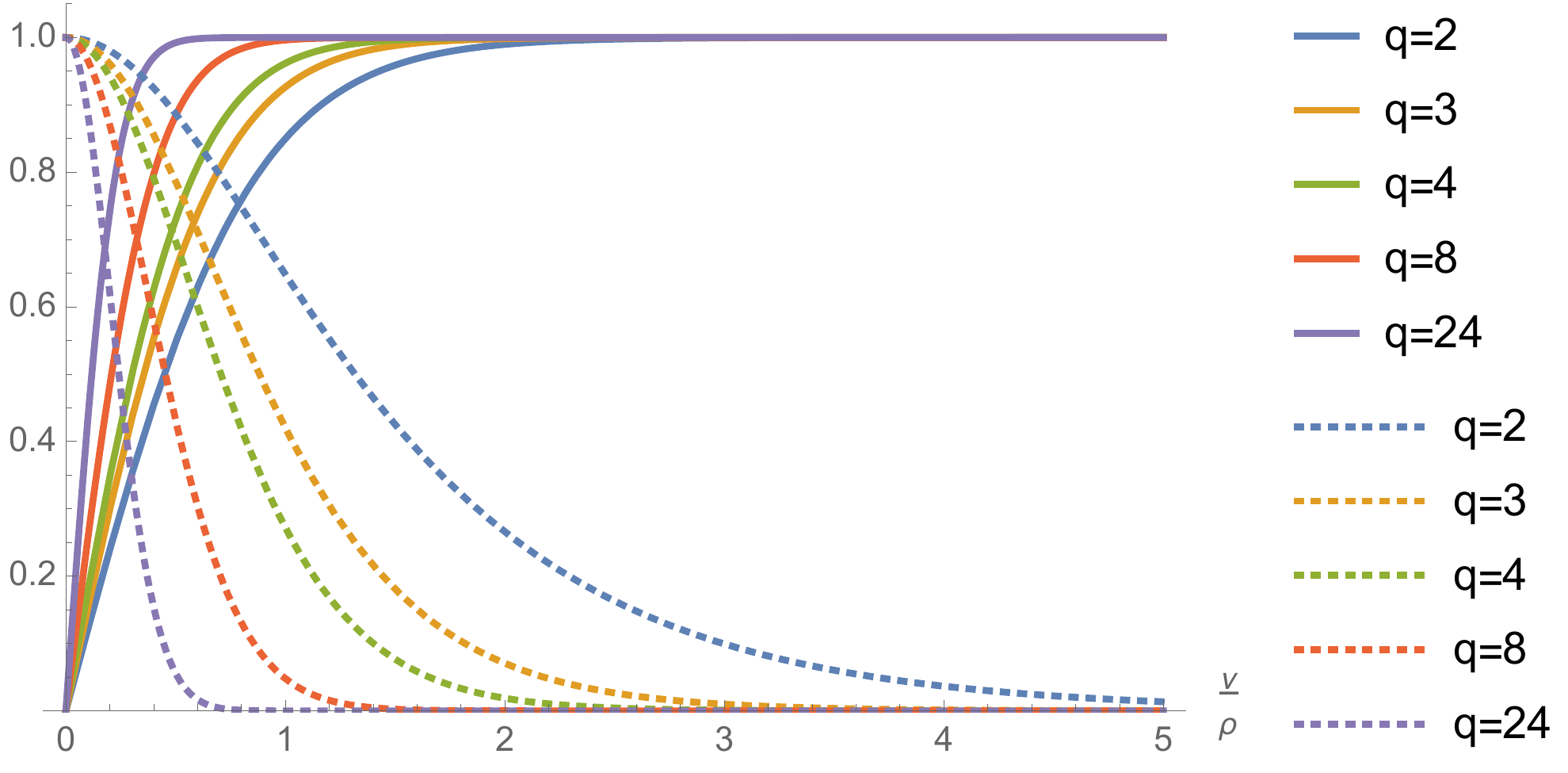}
    \caption{\small Plots of the $\phi$ and $h$ fields in the vortex case $l=2$ for various values of $q$. The $\phi$ fields are the solid curves, and the $h$ fields are the dotted curves. Remember that the $h$ solution is exact, and $\phi$ is obtained numerically. The $\phi$ numerical solutions are normalized to the correct behavior at $\nu=\infty$.}
    \label{fig:vortex-graphs-1}
\end{figure}

In the case of $q=4$, we found an exact solution in terms of confluent Heun functions,  see the  NIST Digital Library of Mathematical Functions \cite[\href{http://dlmf.nist.gov/31.12}{\S~31.12}]{NIST:DLMF}. This $(q,l)=(4,2)$ solution is not very illuminating because of the unfamiliarity of the confluent Heun functions. For the case $q \ge 5$ we were unable to find exact solutions to the linear differential equation \eqref{eq:eom1-l2}, but we constructed numerical solutions for many $q \ge 2$. There are well known subtleties in trying to construct numerical solutions because the natural initial conditions $\phi(0)=0$ and $\phi'(0)=1$ are numerically unstable due to the regular singular point at $\nu=0$. To get around this, we compute the terms of the regular power series solution $\phi_{\text{ps}}$ to \eqref{eq:eom1-l2} to $O(\nu^{7})$ with initial conditions $\phi(0)=0$ and $\phi'(0)=1$:\begin{equation}
\phi_{\text{ps}}(\nu) = \nu +\frac{1-6 q}{24}\; \nu ^3 +\frac{60 q^2-10 q+3}{960}\; \nu ^5 +\frac{-4480
   q^3+420 q^2-1036 q-69}{322560}\; \nu ^7 \,. 
\end{equation}
You can verify that the above agrees with the power series expansion of the exact solution \eqref{eq:phi-q2} for the case $q=2$, and \eqref{eq:phi-q3} for the case $q=3$. The idea is to replace the numerically unstable initial conditions at $\nu=0$ with nearby initial conditions at $\nu_{*}$ where  $0 < \nu_{*} \lll 1$. The initial conditions are set by the truncated power series expansion: $\phi(\nu_{*}) =  \phi_{\text{ps}}(\nu_{*})$ and $\phi'(\nu_{*}) =  \phi'_{\text{ps}}(\nu_{*})$. Typically we chose $\nu_{*}$ from $10^{-8}$ to $10^{-6}$. The numerical solution $\phi_{\text{num}}$ thus obtained will not have the correct normalization at $\nu=\infty$ but since we have a linear ODE we know that the correct normalized solution will be $\phi_{\text{norm}}(\nu) = \phi_{\text{num}}(\nu)/\phi_{\text{num}}(\infty)$. The difference between the exact solution in $q=2$ and the numerical solution is plotted in figure~\ref{fig:difference-solutions}. The normalized numerical solutions for $q=2,3,4,8,24$ are shown in figure~\ref{fig:vortex-graphs-1}.

\subsection{Double BPS hedgehog-like defects ($l=3$)}
\label{sec:bps-hedgehog}

The discussion about the finiteness of the energy in section~\ref{sec:asymp-l3} tells us that we only have to consider the case of a soliton ($p=0$ or equivalently $q=1$) embedded in $\AdS_{4}$. The equations of motion become
\begin{align}
	0&= - \frac{\dd^{2}\phi}{\dd\nu^{2}} - \left( \tanh\nu + 2\, \coth\nu\right) \frac{\dd\phi}{\dd\nu} 
	   + \frac{2}{\left[ \sinh\nu\right]^{2}}\;h(\nu)^{2}\, \phi(\nu)\,, 
	\label{eq:eom1-bps-l3-q1}\\
		0 &= -\frac{\dd^{2}h}{\dd\nu^{2}} -  \tanh\nu\, \frac{\dd h}{\dd \nu} 
	 + \frac{1}{\left[\sinh\nu\right]^{2}}\; \left(h(\nu)^{2}-1\right) h(\nu)\,.
	\label{eq:eom2-bps-l3-q1}
\end{align}
You can check that the solution to \eqref{eq:eom2-bps-l3-q1} is
\begin{equation}
h(\nu) = \frac{1}{\cosh\nu} \,,
\label{eq:sol-q1-l3-h}
\end{equation}
and we are left with an uncoupled linear ODE for $\phi$:
\begin{equation}
	0= - \frac{\dd^{2}\phi}{\dd\nu^{2}} - \left( \tanh\nu + 2\, \coth\nu\right) \frac{\dd\phi}{\dd\nu} + \frac{2}{\left[ \sinh\nu\cosh\nu\right]^{2}} \phi(\nu)\,, 
	\label{eq:eom1-bps-l3-q1-ode}
\end{equation}
This ODE admits even and odd solutions, and it has a regular singular point at $\nu=0$ with Frobenius indices $\alpha_{+}= 1$ and $\alpha_{-}=-2$. The $\alpha=1$ solution may be taken to be an odd function, and we ignore the $\alpha_{-}=-2$ solution that may be taken to be an even function. With our initial conditions $\phi(0)=0$ and $\phi'(0)=1$, we find the power series solution
\begin{equation}
\phi_{\text{ps}}(\nu) =
\nu -\frac{13 \nu^3}{30}+\frac{5 \nu^5}{24}-\frac{7279 \nu^7}{75600}+\frac{172219 \nu^9}{3991680} -\frac{16386323
   \nu^{11}}{864864000}+O\left(\nu^{13}\right).
   \label{eq:phi-q1-l3-ps}
\end{equation}
The exact normalized solution is given by the odd function
\begin{equation}
\phi_{\text{norm}}(\nu) = \mathcal{N}_{(1,3)}^{-1}\;\sinh\nu\, \sech^{4}\nu\; {}_{2}F_{1}\!\left(2+\half\sqrt{2},2-\half\sqrt{2};\; \frac{5}{2};\; \tanh^{2}\nu \right)
\label{eq:phi-q1-l3-exact}
\end{equation}
where
\begin{equation}
\mathcal{N}_{(1,3)}=\frac{3 \sin \left(\pi/\sqrt{2}\right)}{2 \sqrt{2}} \approx 0.84396\,.
\end{equation}
It is easy to verify that the power series  expansion \eqref{eq:phi-q1-l3-ps}, derived directly from ODE \eqref{eq:eom1-bps-l3-q1-ode},  agrees with the power series  expansion of \eqref{eq:phi-q1-l3-exact} if you remove the normalization factor $\mathcal{N}_{(1,3)}$ in order that both functions have derivative $1$ at $\nu=0$.
The magnetic charge for the spherically symmetric ansatz is discussed in appendix~\ref{sec:bogomolny}.

\begin{figure}[tbp]
    \centering
    \includegraphics[width=0.8\textwidth]{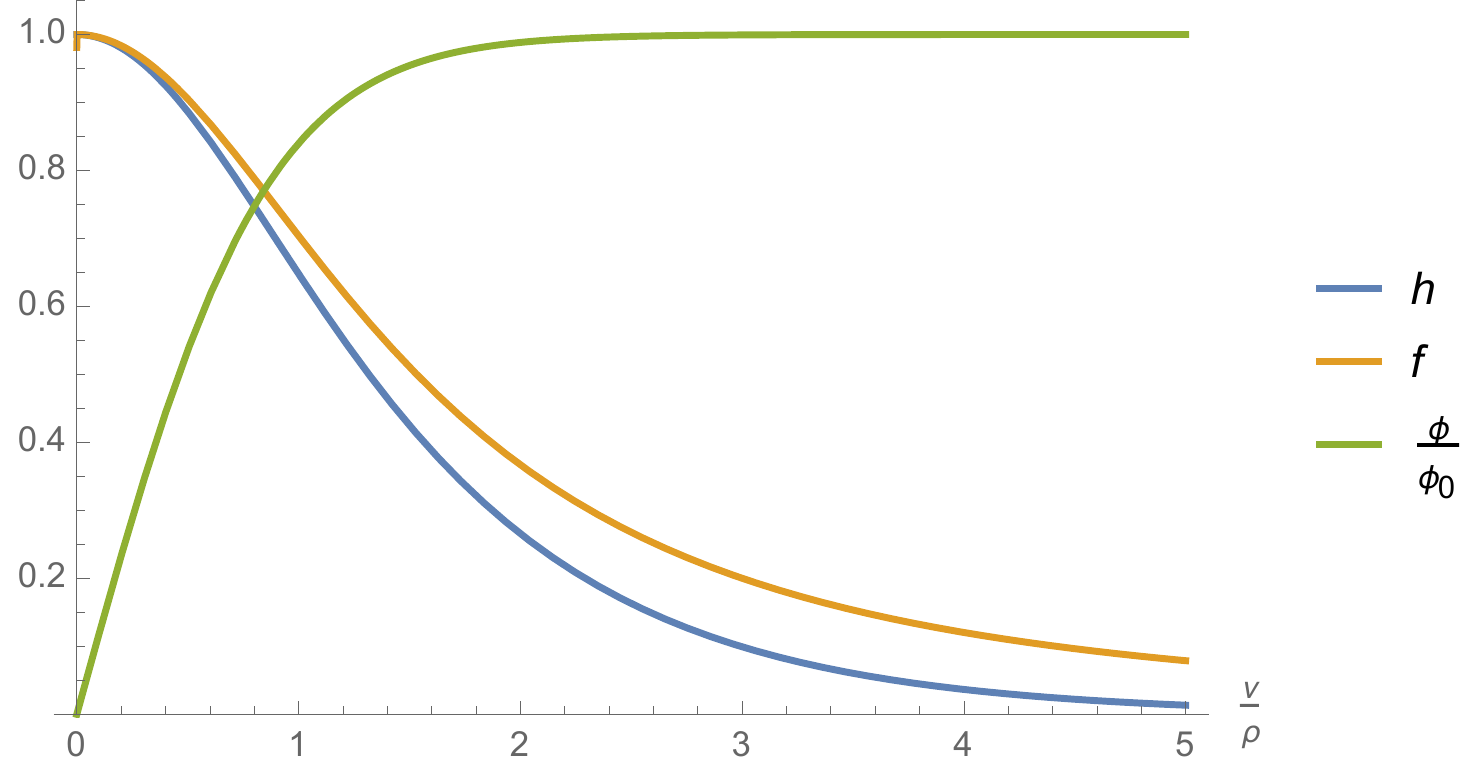}
    \caption{\small Graph of the exact double BPS hedgehog solutions ($l=3$) for $q=1$.}
    \label{fig:hedgehog-graphs}
\end{figure}

\begin{figure}[tbp]
    \centering
    \includegraphics[width=0.65\textwidth]{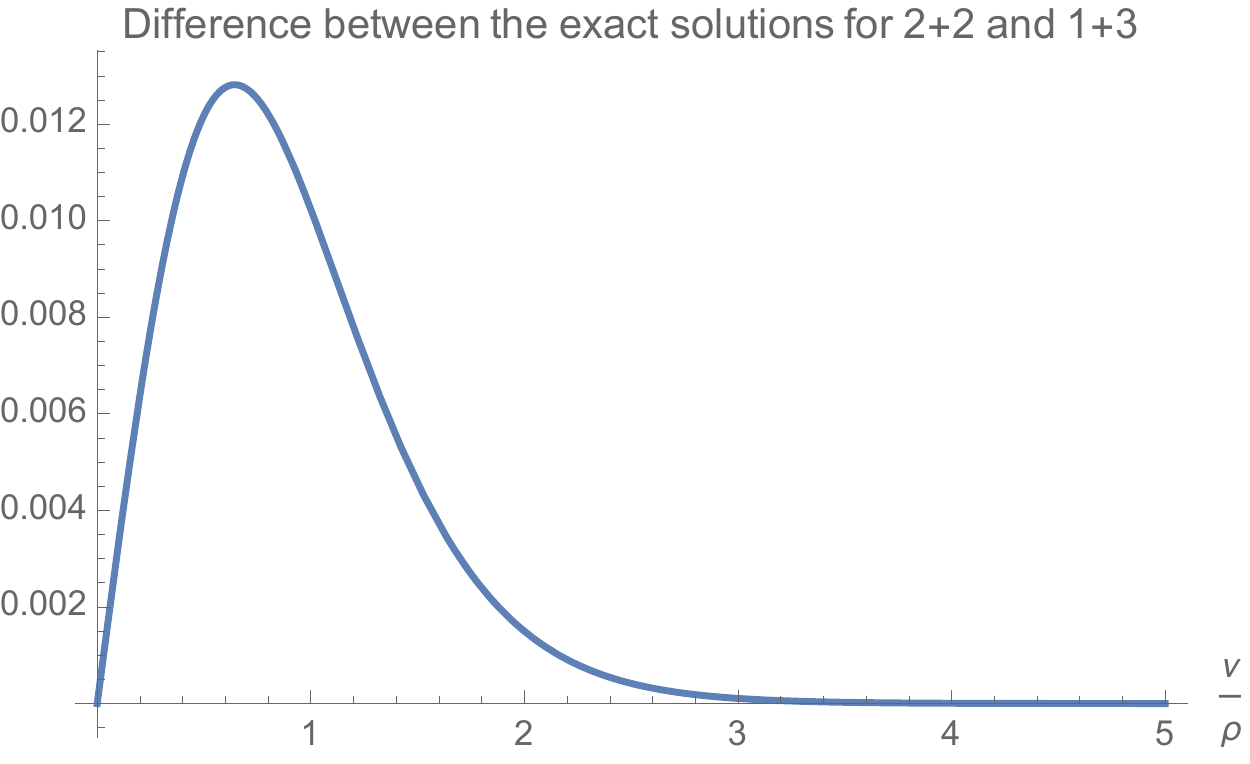}
    \caption{\small Absolute value difference between the two exact solutions for $\phi$ in the case where $n=4$ with $(q,l)=(2,2)$ and $(q,l)=(1,3)$. In both cases, the field $\phi$ satisfies the same two point boundary conditions, also eq.~\eqref{eq:def-m-star} tells us that both solutions decay exponentially at the same rate $e^{-3\nu/\rho}$. It is not surprising that both solutions will be similar but the 1\% discrepancy is a bit surprising.}
    \label{fig:differences-exact}
\end{figure}

Surprisingly, the exact scalar field profiles for $(q,l)=(2,2)$ and $(q,l)=(1,3)$ are very similar, see figure~\ref{fig:differences-exact}.

\subsection{Static spherically symmetric classical glueballs } 
\label{sec:post-arxiv}

\emph{Note added after submission to arXiv:} On the day our manuscript appeared on the arXiv, there was also a manuscript by Ivanova, Lechtenfeld and Popov~\cite{Ivanova:2017wun} where they study finite action and finite energy solutions of Yang-Mills theory in $\AdS_{4}$. This work is based on an earlier paper~\cite{Ivanova:2017xjp}. We did not realize at the time of submission that our double BPS limit monopole solution automatically gives a static spherically symmetric solution of the $\SO(3)$ Yang-Mills equations in $\AdS_{4}$ with finite energy. It should have been obvious.  The reason is that in the double BPS limit, eq.~\eqref{eq:eom2-bps-l3-q1} does not depend on the scalar field $\phi$. This equation is the Yang-Mills equation. All we have to do is ignore the scalar field $\phi$ and its equation of motion~\eqref{eq:eom1-bps-l3-q1}. Equivalently, you can look at the transverse energy \eqref{eq:defect-action-3} and ignore the scalar field $\phi$. We know the solution is $h(\nu)=\sech\nu$, see~\eqref{eq:sol-q1-l3-h}. The mass of this static spherically symmetric Yang-Mills solution is
\begin{equation}
E_{\perp} = \frac{3\pi^{2}}{2\,g^{2}\rho}\,.
\end{equation}
This solution is a classical glueball in $\AdS_{4}$. We do not know if this solution is stable. Also, this solution goes away in the flat space limit $\rho\to\infty$ in accordance with known theorems.  We have not explored whether there is a generalization to $\AdS_{4}$ of the Minkowski space theorems of Coleman~\cite{coleman:1977hd}, and Coleman and Smarr~\cite{coleman:1977yb}   that constrain the existence and the properties of glueballs. It is not immediately apparent how to relate the static Yang-Mills solution in \cite[section~6]{Ivanova:2017wun} to the static solution presented here.

Another observation is that the discussion of the double BPS limit in the $\SO(2)$ case in section~\ref{sec:bps-vortex} provides exact maximally symmetric purely electromagnetic $p$-defects associated with an embedding $\AdS_{q} \hookrightarrow \AdS_{n}$ given by eq.~\eqref{eq:sol-h-l2}. The transverse energy density of this solution is given by
\begin{equation}
E_{\perp}(q) = (q-1) \frac{\pi}{g^{2}\rho^{2}}= p\; \frac{\pi}{g^{2}\rho^{2}}\,,\quad\text{where $q \ge 2$}.
\end{equation}
It is interesting that the transverse energy density is proportional to the dimensionality $p$ of the defect.

\section{$\rho^{2} \neq 0$ and $m_{A}^{2} \downarrow 0$}
\label{sec:pink-plane}

Here we briefly discuss solutions to the spherically symmetric ansatz in the parameter plane defined by $m_{A} \downarrow 0$ with $1/\rho^{2} \neq 0$, see the pinkish plane in figure~\ref{fig:bps-parameter}. The equations of motion in this limit become
\begin{subequations}\label{eq:eomz}
	\begin{align}
	- \half (m_{\phi}\rho)^{2}(\phi^{2}-1)\phi&= - \frac{\dd^{2}\phi}{\dd\nu^{2}} - \left( q\,\tanh\nu + (l-1)\, \coth\nu\right) \frac{\dd\phi}{\dd\nu} 
	\nonumber\\
	&\quad   + \frac{(l-1)}{\left[ \sinh\nu\right]^{2}}\;h(\nu)^{2}\, \phi(\nu)\,, 
	\label{eq:eom1z}\\
		0 &= -\frac{\dd^{2}h}{\dd\nu^{2}} - \left( q\,\tanh\nu + (l-3)\, \coth\nu\right) \frac{\dd h}{\dd \nu} 
	\nonumber\\
	&\quad  + \frac{(l-2)}{\left[\sinh\nu\right]^{2}}\; \left(h(\nu)^{2}-1\right) h(\nu)\,.
	\label{eq:eom2z}
	\end{align}
\end{subequations}
Equation~\eqref{eq:eom2z} already appeared  as eq.~\eqref{eq:eom2-bps} in section~\ref{sec:double} where we  analyzed its solutions in detail. There we learned that the solution is given by $h(\nu) = 1/(\cosh\nu)^{q-1}$ if $l=2$ and $q \ge 2$, see~\eqref{eq:sol-h-l2}. For $l=3$, there is only a $q=1$ solution $h(\nu)=1/\cosh\nu$, see eq.~\eqref{eq:sol-q1-l3-h}. The strategy is to insert the known $h$ solution into ODE \eqref{eq:eom1z}, and look for a solution that satisfies the two point boundary conditions  $\phi(0)=0$ and $\phi(\infty)=1$. The analysis is more complicated than in the double BPS limit because the field $\phi$ now satisfies a non-linear ODE; the solution can be determined numerically. We use a modification of the power series technique discussed in section~\ref{sec:vortex-4} to get around the numerically unstable initial conditions at $\nu=0$. 

\subsection{Vortex-like defects ($l=2$)}

\begin{figure}
\centering
\includegraphics[width=0.975\textwidth]{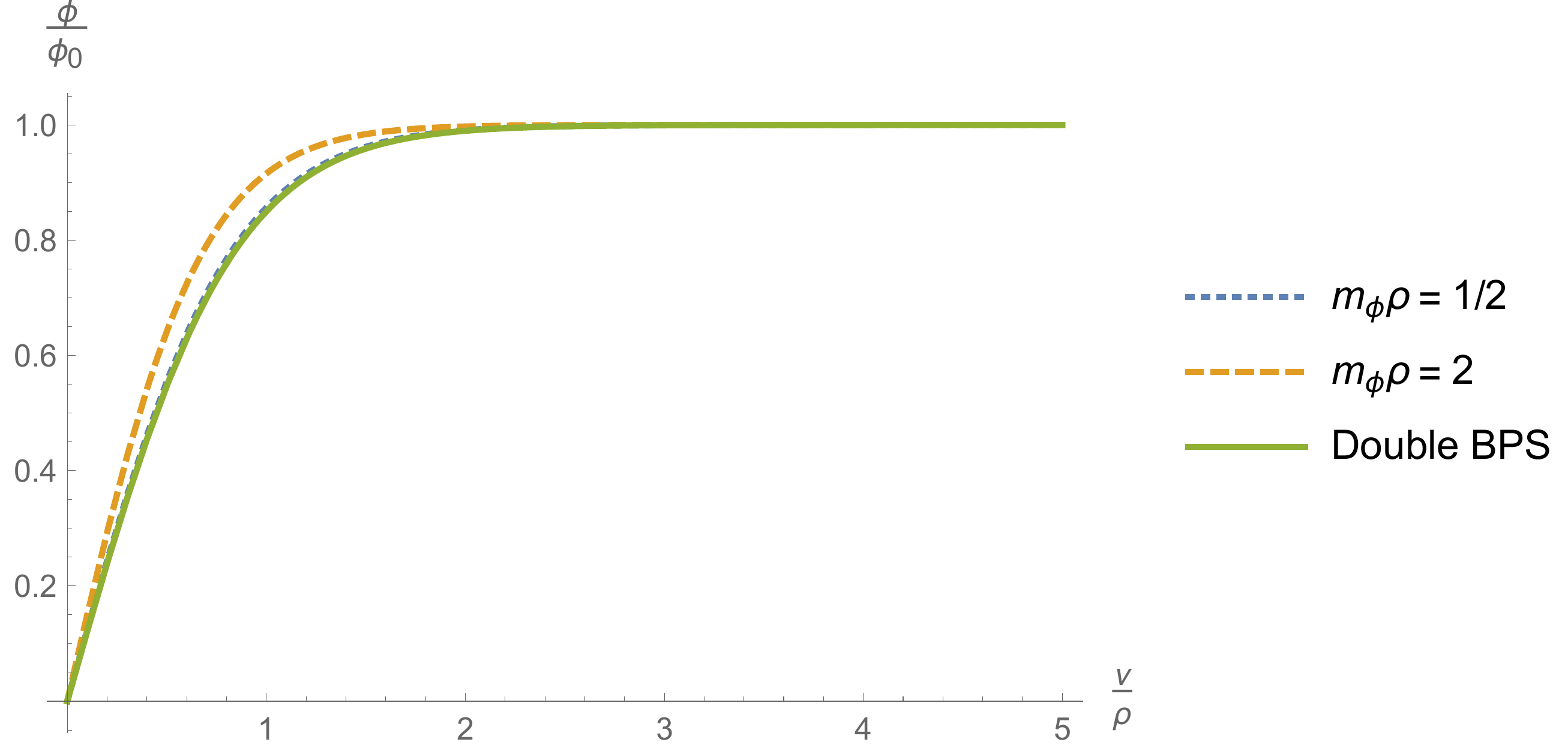}
\caption{\small Numerical solution for the Nielsen-Olesen vortex $(q,l)=(2,2)$ in anti de~Sitter space in the $m_{A}\downarrow 0$ limit with the Higgs field Compton wavelength $1/m_{\phi}=2\rho$, and $1/m_{\phi}=\rho/2$. The solutions are compared to the exact double BPS solution.}
\label{fig:x1}
\end{figure}

\begin{figure}
\centering
\includegraphics[width=0.975\textwidth]{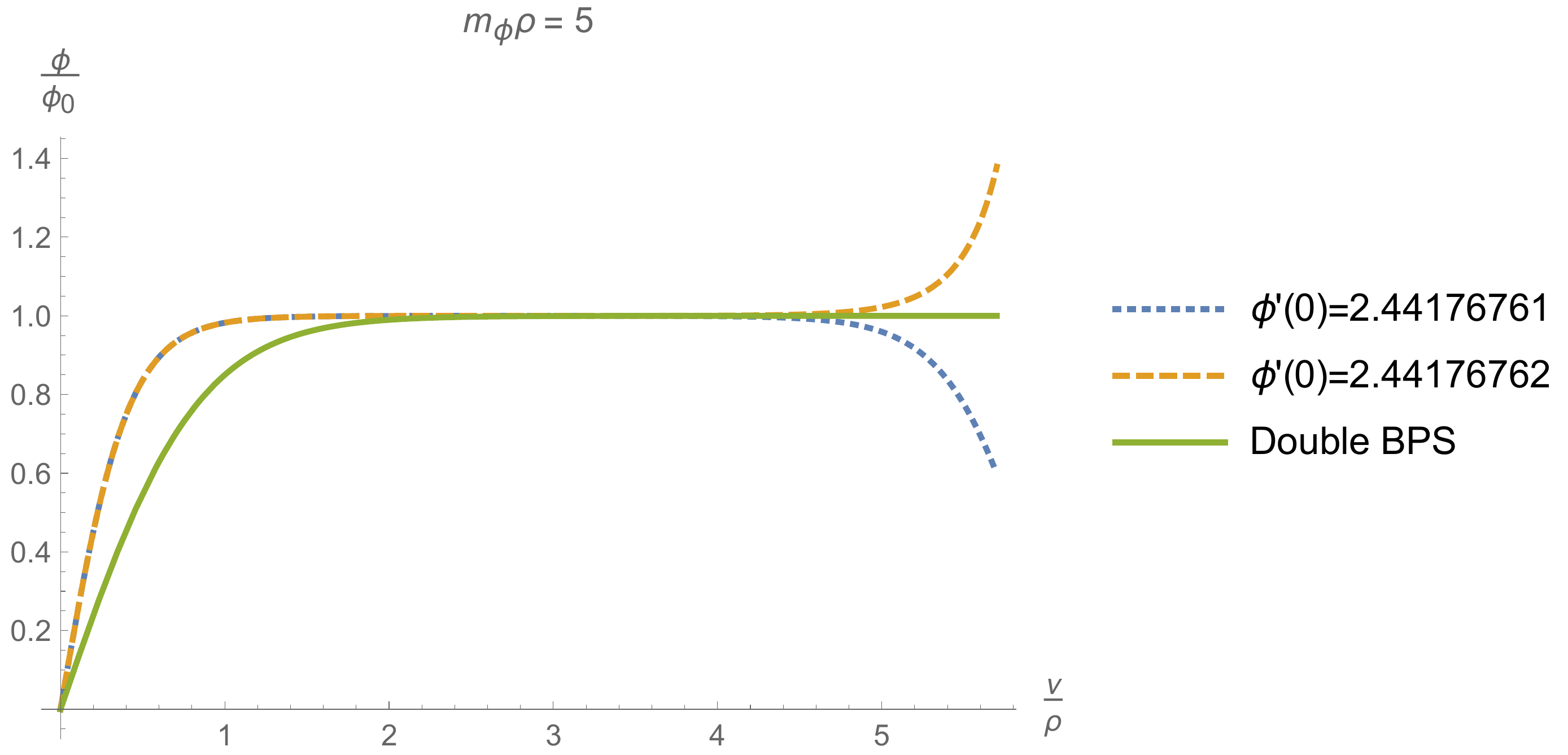}
\caption{\small Numerical solution for the Nielsen-Olesen vortex $(q,l)=(2,2)$ in anti de~Sitter space in the $m_{A}\downarrow 0$ limit with the Higgs field Compton wavelength $1/m_{\phi}=\rho/5$. As expected, the decay behavior of the $\phi$ field is dominated by the Compton wavelength. As a comparison, the curve in green is the exact double BPS solution where the the Higgs field has infinite Compton wavelength and the length scale is set by the radius of curvature $\rho$. The shooting method is very sensitive to the initial condition $\phi'(0)$ due to the potential presence of exponential growth terms; compare the dotted blue curve with the dashed orange curve. Numerical methods become more finicky as $m_{\phi}\rho$ increases.}
\label{fig:stability2}
\end{figure}

\begin{figure}
\centering
\begin{subfigure}[b]{0.48\textwidth}
\centering
\includegraphics[width=\textwidth]{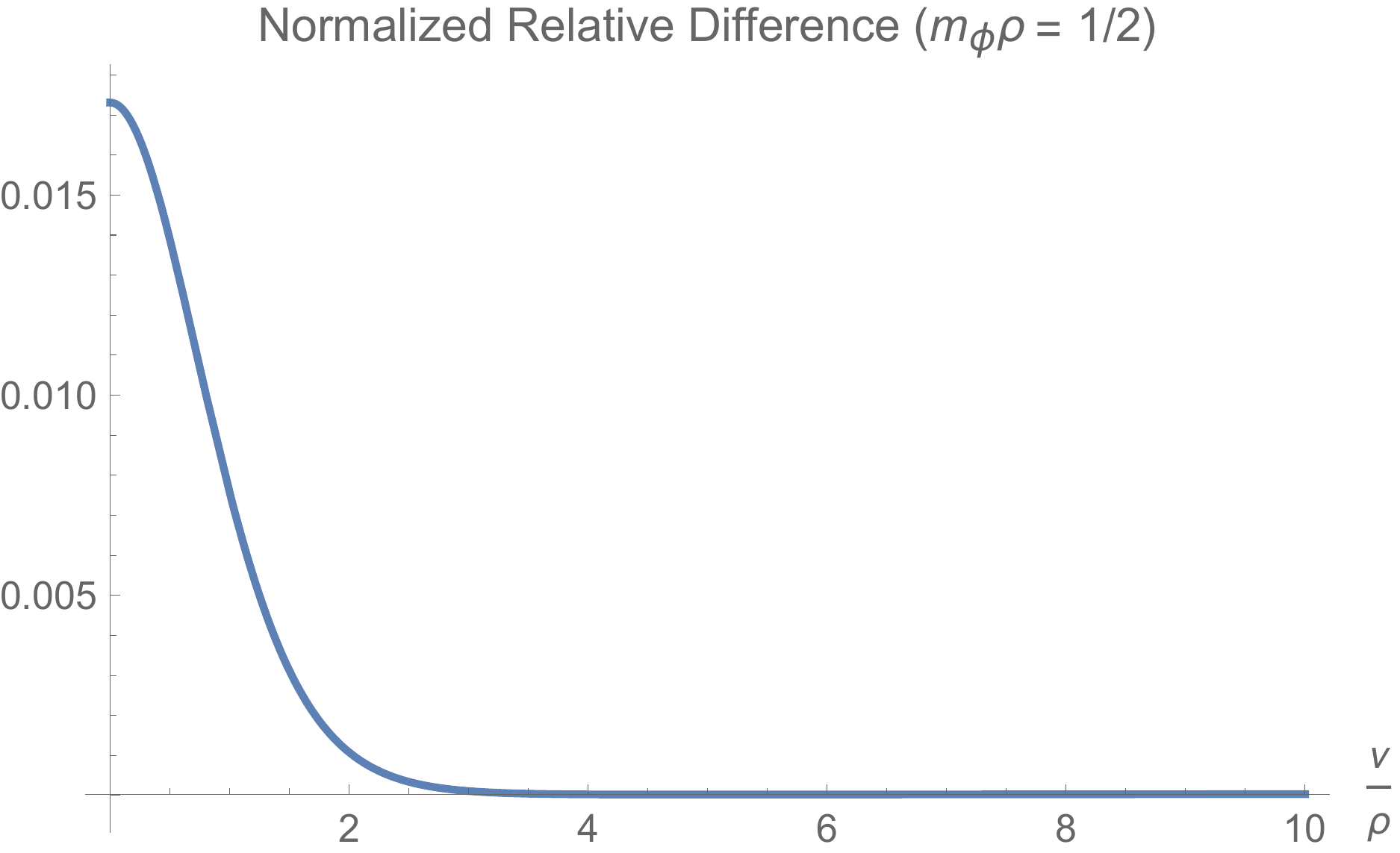}
\caption{\small Normalized relative difference for $m_{\phi}\rho=1/2$. The two solutions differ by up to 1\%.}
\label{fig:x2}
\end{subfigure}%
\hfil
\begin{subfigure}[b]{0.48\textwidth}
\centering
\includegraphics[width=\textwidth]{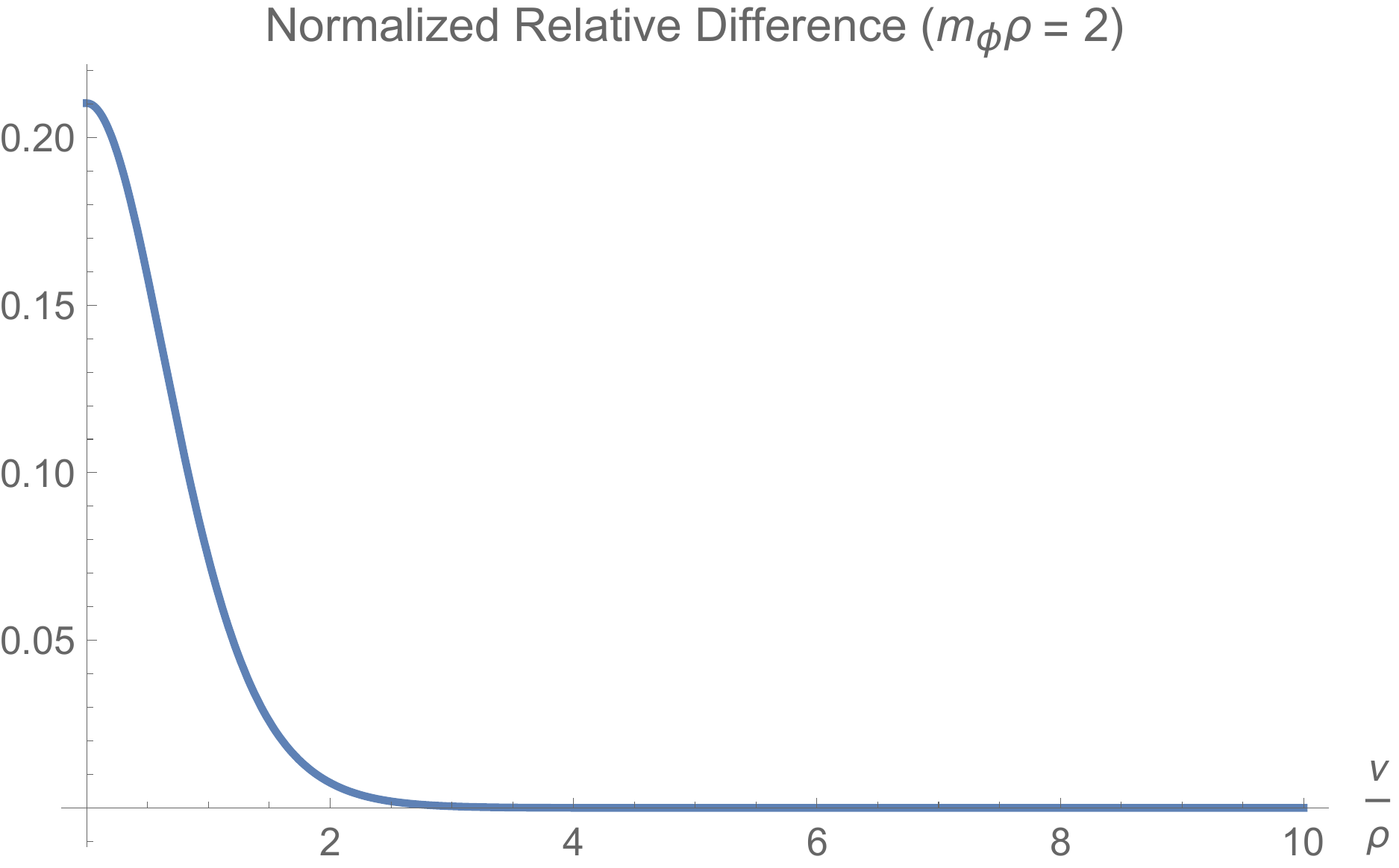}
\caption{\small Normalized relative difference for $m_{\phi}\rho=2$. The two solutions differ by up to 20\%.}
\label{fig:x3}
\end{subfigure}
\caption{\small The normalized relative difference is defined by $ \frac{\left\lvert \phi(\nu)-\phi_{\text{dBPS}}(\nu)\right\rvert}{\frac{1}{2}\left(\phi(\nu)+\phi_{\text{dBPS}}(\nu)\right)}$, where $\phi_{\text{dBPS}}$ is the exact double BPS solution. Numerical experiments in the $m_{A} \downarrow 0$ limit indicate that if $m_{\phi}\rho \lesssim 1$ then the numerical solution is well approximated by the exact double BPS solution. If $m_{\phi}\rho>1$ then the numerical solution begins to differ from the exact double BPS one, \emph{e.g.}, see figure~\ref{fig:stability2}.}
\label{fig:xyz}
\end{figure}

The power series solution to the $\phi$ equation of motion with initial condition $\phi(0)=0$ and $\phi'(0)$ arbitrary is given to $O(\nu^{7})$ by
\begin{multline*}
\phi_{\text{ps}}(\nu) = \nu  \phi '(0) 
    +\frac{1}{48} \nu ^3 \left(-12 q \phi '(0)+2 \phi
   '(0)  -3 \epsilon ^2 \phi '(0)\right) \\
 +\frac{1}{3840}\nu ^5 \left(240 q^2 \phi '(0)-40 q \phi '(0)+60 q
   \epsilon ^2 \phi '(0)+12 \phi '(0)+5 \epsilon ^4 \phi '(0)+80
   \epsilon ^2 \phi '(0)^3\right)\\
    + \frac{1}{2580480} \nu ^7 \left(-35840 q^3
   \phi '(0)+3360 q^2 \phi '(0)-8400 q^2 \epsilon ^2 \phi '(0)-8288 q
   \phi '(0) \right. \\
     -840 q \epsilon ^4 \phi '(0)-26880 q \epsilon ^2 \phi
   '(0)^3-2240 q \epsilon ^2 \phi '(0)-552 \phi '(0)-35 \epsilon ^6
   \phi '(0) \\
   \left. -5600 \epsilon ^4 \phi '(0)^3-70 \epsilon ^4 \phi '(0)+2240
   \epsilon ^2 \phi '(0)^3-532 \epsilon ^2 \phi
   '(0)\right),
\end{multline*}
where $\epsilon = m_{\phi}\rho$. Because the ODE is nonlinear, we see  nonlinear behavior on the initial condition $\phi'(0)$ beginning at $O(\nu^{5})$. We replace the initial conditions at $\nu=0$ with those given by the power series at a nearby point $\nu_{*} = 1\times 10^{-8}$. We use~\cite{press:1992} the ``shooting method'' where we vary $\phi'(0)$ numerically until we find a solution with approximately the correct asymptotic behavior $\phi(\infty)=1$, see figure~\ref{fig:x1} and figure~\ref{fig:stability2}. A comparison between the numerical solutions and the exact double BPS solution is given in figure~\ref{fig:xyz}.

\subsection{Monopole defects $(q,l)=(1,3)$}
\begin{figure}
\centering
\includegraphics[width=0.975\textwidth]{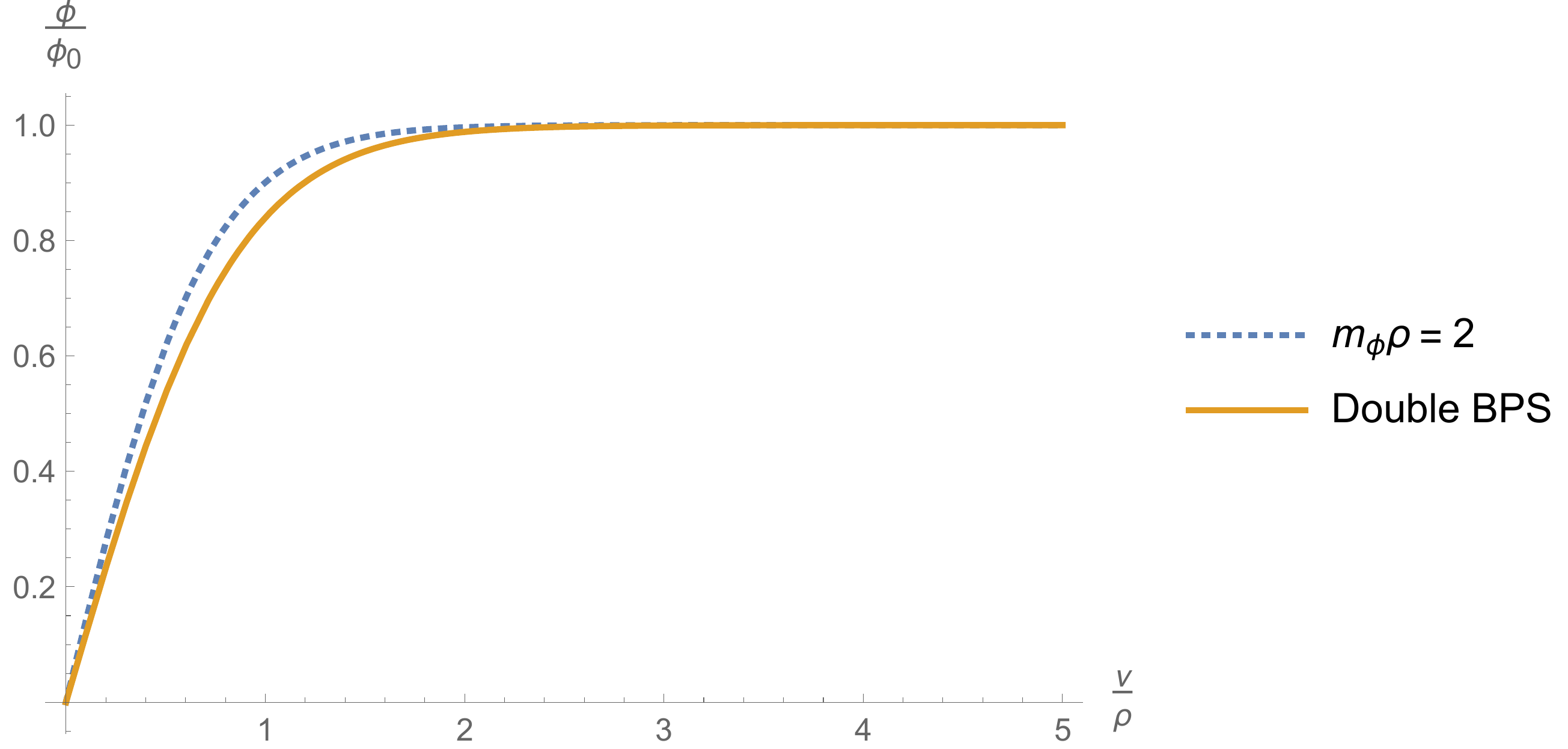}
\caption{\small Numerical solution for the 'tHooft-Polyakov monopole $(q,l)=(1,3)$ in anti de~Sitter space in the $m_{A}\downarrow 0$ limit with the Higgs field Compton wavelength $1/m_{\phi}=\rho/2$. The solution is compared to the exact double BPS monopole solution. The initial derivative of the numerical solution is $\phi'(0)= 1.424467087$.}
\label{fig:y1}
\end{figure}

\begin{figure}
\centering
\includegraphics[width=0.65\textwidth]{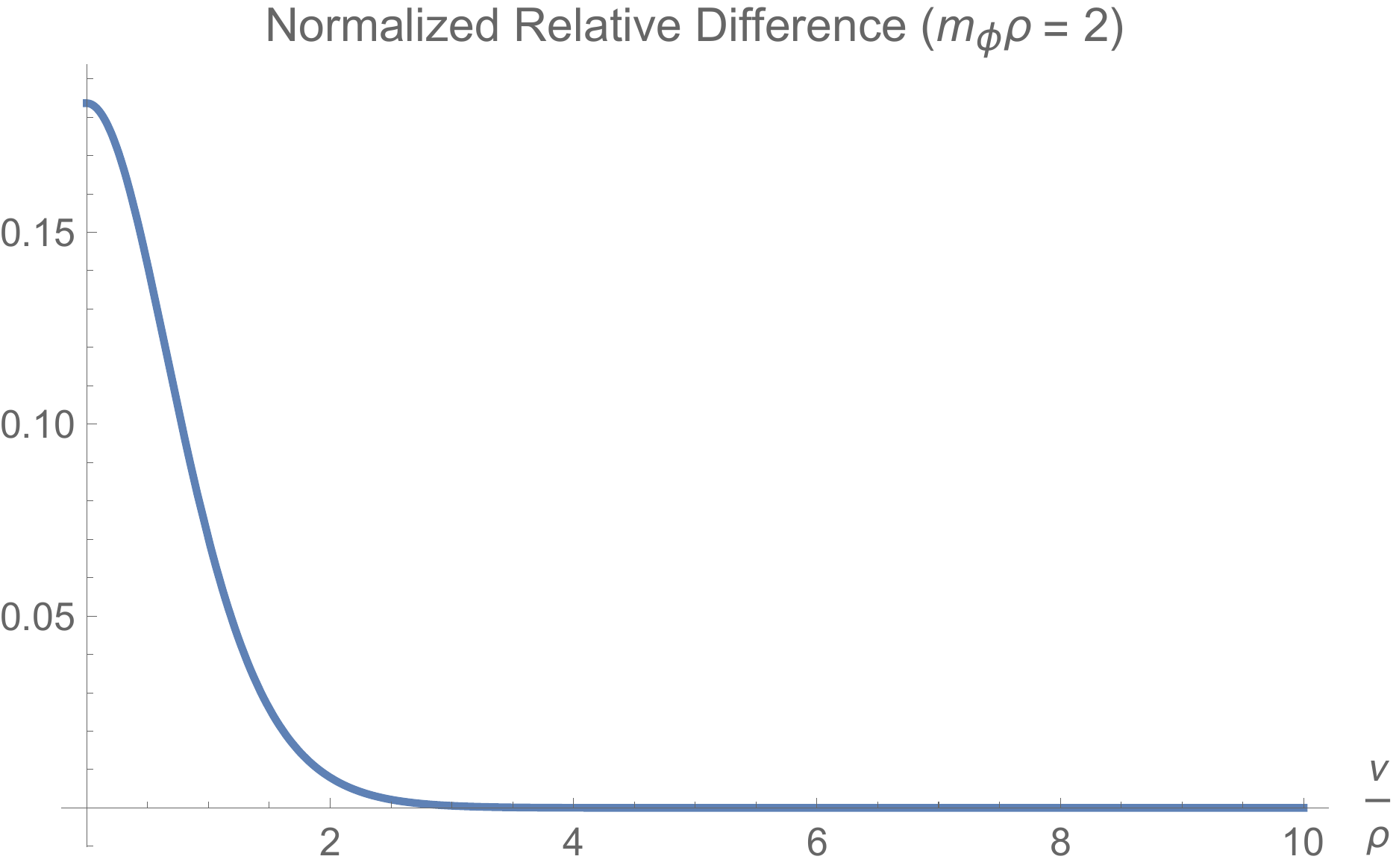}
\caption{\small Normalized relative difference between the numerical monopole solution and the exact double BPS solution. The normalized relative difference is defined by $ \frac{\left\lvert \phi(\nu)-\phi_{\text{dBPS}}(\nu)\right\rvert}{\frac{1}{2}\left(\phi(\nu)+\phi_{\text{dBPS}}(\nu)\right)}$, where $\phi_{\text{dBPS}}$ is the exact double BPS solution. With $m_{\phi}\rho=2$, the normalized relative difference is less than 20\%.}
\label{fig:y2}
\end{figure}

In this case, the power series solution to the $\phi$ equation of motion with initial condition $\phi(0)=0$ and $\phi'(0)$ arbitrary is given to $O(\nu^{7})$ by
\begin{multline*}
\phi_{\text{ps}}(\nu) = \nu  \phi '(0)+\frac{1}{60} \nu ^3 \left(-26 \phi '(0)-3
   \epsilon ^2 \phi '(0)\right) \\
   +\frac{1}{3360}\nu ^5 \left(700
   \phi '(0)+3 \epsilon ^4 \phi '(0)+60 \epsilon ^2
   \phi '(0)^3+72 \epsilon ^2 \phi
   '(0)\right) \\
   +\frac{1}{604800}\nu ^7 \left(-58232 \phi '(0)-5
   \epsilon ^6 \phi '(0)-940 \epsilon ^4 \phi '(0)^3 \right.
   \\ \left. -230
   \epsilon ^4 \phi '(0)-9480 \epsilon ^2 \phi
   '(0)^3-5636 \epsilon ^2 \phi '(0)\right)
\end{multline*}
where, once again, $\epsilon = m_{\phi}\rho$. We see a similar nonlinear behavior to that of the previous case. We use the initial conditions provided by our power series at a sufficiently nearby point. Here we use $\nu_{*} = 1\times 10^{-8}$. See figure~\ref{fig:y1} for a comparison between our numerical solution and the double BPS solution when $1/m_{\phi} = \rho/2$, and figure~\ref{fig:y2} for the relative difference between these two solutions.


\section{Flat space equations of motion}

For completeness, the flat space equations of motion for spherically symmetric defects are given by taking the $\rho\uparrow \infty $ limit of eqs.~\eqref{eq:eomx}:
\begin{subequations}\label{eq:eomy}
	\begin{align}
	- \half m_{\phi}^2\, (\phi^{2}-1)\phi&= - \frac{\dd^{2}\phi}{\dd\nu^{2}} - \frac{(l-1)}{\nu}\,  \frac{\dd\phi}{\dd\nu} 
   + \frac{(l-1)}{\nu^2}\;h(\nu)^{2}\, \phi(\nu)\,, 
	\label{eq:eom1y}\\
		- m_{A}^{2}\, \phi(\nu)^{2}\, h(\nu) &= -\frac{\dd^{2}h}{\dd\nu^{2}} - \frac{(l-3)}{\nu}\,  \frac{\dd h}{\dd \nu} 
	 + \frac{(l-2)}{\nu^2}\; \left(h(\nu)^{2}-1\right) h(\nu)\,.
	\label{eq:eom2y}
	\end{align}
\end{subequations}
These equations are well known and have been thoroughly studied over the past 40 years.

The Prasad-Sommerfield monopole solution ($l=3$) is given by
\begin{align}
\phi(\nu)&= \coth m_{A}\nu - \frac{1}{m_{A}\nu}\,, & h(\nu)&= \frac{m_{A}\nu}{\sinh m_{A}\nu}\,, 
\label{eq:bps-soln}\\
&= 1 - \frac{1}{m_{A}\nu} +O\left( e^{-2m_{A}\nu}\right)\,. 
& & = 2m_{A}\nu \, e^{-m_{A}\nu} +O\left( \nu\, e^{-3m_{A}\nu} \right)\nonumber
\end{align}
The scalar field $\phi$ has a $1/\nu$ Coulomb tail because $m_{\phi}=0$.

\section{Conclusions}
\label{sec:conclusions}

In this article we studied the equations of motion for maximally symmetric $p$-defects in $\AdS_{n}$. In the double BPS limit the radius of curvature $\rho$ is the only length scale that appears in the equations of motion, and we were able to find exact analytic solutions in many cases. We also saw that the radial exponential increase in volume in $\AdS_{n}$ plays a crucial role, and requires a case by case study for admissible values of $(q,l)$.

The method we advocate in this paper is part of a broader strategy to study solutions of the Yang-Mills Higgs system in $\AdS_{n}$. The solutions found in this double BPS limit are the first step in a perturbative expansion in small parameters $m_{\phi}\rho$ and $m_{A}\rho$ for the full equations of motion.

\appendix

	\section{Maximally symmetric submanifolds of maximally symmetric spaces}
	\label{sec:max-sym}
	
	In this section we derive necessary conditions satisfied by a maximally symmetric submanifold of a maximally symmetric manifold.
		
		Any local orthonormal coframe $\theta^{\mu}$ on a constant curvature manifold $M^{n}$ with associated Levi-Civita connection $\omega_{\mu\nu}= -\omega_{\nu\mu}$ will satisfy the Cartan structural equations for a manifold of constant sectional curvature $k$:
	\begin{subequations}\label{eq:cartan-structure}
		\begin{align}
		\dd{\theta^\mu} &= -\omega_{\mu\nu}\wedge\theta^{\nu} \label{eq:cartan-torsion}\\
		\dd{\omega_{\mu\nu}} &= -\omega_{\mu\lambda}\wedge\omega_{\lambda\nu} + k\, \theta^{\mu}\wedge\theta^{\nu} \label{eq:cartan-curvature}
		\end{align}
	\end{subequations}
These constant curvature manifolds are maximally symmetric spaces with $\dim\Iso(M^{n}) = \frac{1}{2}n(n+1)$, where $\Iso(M^{n})$ is the component of the isometry group of $M^{n}$ that is connected to the identity. In fact, eqs.~\eqref{eq:cartan-structure} are the Maurer-Cartan equations for a Lie group.

Next we use the Cartan structural equations and a bit of the theory of exterior differential systems~\cite{BCG3}, mostly the Frobenius theorem for integrability of a Pfaffian system of equations. Assume $\Sigma^{q}$ is an isometrically embedded $q$-submanifold of the constant curvature $M^{n}$. If we use the  index conventions that latin indices from the beginning of the alphabet $a,b,c,d$ run from $1,2,\dotsc,q$ and latin indices from the middle of the alphabet $i,j,k,\dotsc$ take $l=n-q$ values from $q+1,\dotsc,n$ then the structural equations in an  orthonormal coframe adapted to the tangent bundle of the submanifold may be written as
	\begin{subequations}\label{eq:cartan-submanifold}
		\begin{align}
		\dd \theta^{a} &= -\omega_{ab} \wedge \theta^{b} -\omega_{ai} \wedge \theta^{i} \,, \\
		\dd \theta^{i} &= -\omega_{ij}\wedge\theta^{j} +\omega_{ai}\wedge\theta^{a}\,, \label{eq:cs-i}\\
		\dd\omega_{ab} &= - \omega_{ac}\wedge \omega_{cb} + \omega_{ai}\wedge \omega_{bi} + k\, \theta^{a} \wedge \theta^{b}\,, \\
		\dd\omega_{ai} &= -\omega_{ab}\wedge\omega_{bi} - \omega_{aj}\wedge\omega_{ji} + k\, \theta^{a} \wedge \theta^{i}\,, \\
		\dd \omega_{ij} &= -\omega_{ik}\wedge\omega_{kj} +\omega_{ai}\wedge\omega_{aj} + k\, \theta^{i}\wedge\theta^{j}\,.
		\end{align} 
	\end{subequations}	
The submanifold $\Sigma^{q}$ in this adapted coframe is given by the exterior differential system $\theta^{i}=0$ and consequently $\dd\theta^{i}=0$. Using these conditions in \eqref{eq:cartan-submanifold} we find
	\begin{subequations}\label{eq:cartan-submanifold-x}
		\begin{align}
		\dd \theta^{a} &= -\omega_{ab} \wedge \theta^{b}  \,, \\
		0 &= \omega_{ai}\wedge\theta^{a}\,, \label{eq:cs-i-x}\\
		\dd\omega_{ab} &= - \omega_{ac}\wedge \omega_{cb} + \omega_{ai}\wedge \omega_{bi} + k\, \theta^{a} \wedge \theta^{b}\,, \\
		\dd\omega_{ai} &= -\omega_{ab}\wedge\omega_{bi} - \omega_{aj}\wedge\omega_{ji} \,, \\
		\dd \omega_{ij} &= -\omega_{ik}\wedge\omega_{kj} +\omega_{ai}\wedge\omega_{aj} \,.
		\end{align} 
	\end{subequations}	
Applying Cartan's Lemma to \eqref{eq:cs-i-x}, we conclude that on $\Sigma^{q}$ we have 
\begin{equation}
\omega_{ai} = K_{ab}{}^{i}\,\theta^{b}\,,
 \label{eq:def-K}
\end{equation}
where $K_{ab}{}^{i}= K_{ba}{}^{i}$ are tensors on $\Sigma^{q}$ called the extrinsic curvatures or the second fundamental form. When restricted to $\Sigma^{q}$, the Cartan structural equation for the intrinsic curvature is $\dd \omega_{ab} + \omega_{ac}\wedge \omega_{cb} = \half R^{\Sigma}_{abcd}\, \theta^{c}\wedge\theta^{d}$. Thus we obtain the Gauss equation for embedding in a space of constant curvature:
\begin{equation}
    R^{\Sigma}_{abcd}= K_{ac}{}^{i}K_{bd}{}^{i}-K_{ad}{}^{i}K_{bc}{}^{i} + k\left( \gsig_{ac}\,\gsig_{bd} - \gsig_{ad}\,\gsig_{bc}\right)\,,
    \label{eq:Gauss-eq-x}
\end{equation}
where $\gsig_{ab}$ is the induced metric on $\Sigma^{q}$ due to the isometric embedding.

	\subsection{Totally geodesic submanifolds}
		\label{sec:geodesic}

	A submanifold $\Sigma^{q}$ of a general manifold $M^{n}$ is said to be totally geodesic if every geodesic (with respect to the induced metric) on $\Sigma^{q}$ is also a geodesic on $M^{n}$. If $D^{\Sigma}$ and $D^{M}$ are respectively the Levi-Civita connections on the respective manifolds then the definition of the second fundamental form (extrinsic curvatures) tells us that for the tangent vector field $X$ to a curve on $\Sigma$ we have $D^{M}_{X} X = D^{\Sigma}_{X} X - K^{i}(X,X)\, \hat{\vb{n}}_{i}$, where $\{ \hat{\vb{n}}_{i}\}$ is a local orthonormal frame for the normal bundle $(T\Sigma)^{\perp}$. The definition of totally geodesic implies that $K^{i}(X,X)=0$ for all $X(\sigma) \in T_{\sigma}\Sigma$ and so we conclude that $K_{ab}{}^{i}=0$. It is easy to show that totally geodesics submanifolds exist by  constructing an example directly. Fix a point $x\in M$ and a $q$-dimensional vector subspace $V_{x} \subset T_{x}M$. Consider all the geodesics in $M^{n}$ that begin at $x$ with initial velocity in $V_{x}$. The locus of  all these geodesics\footnote{The geodesics should be short in an appropriate sense.}  is a $q$-dimensional totally geodesic submanifold $\Sigma^{q} \subset M^{n}$.
From the viewpoint of standard General Relativity, totally geodesic submanifolds are very desirable because if a test mass in $\Sigma$ is given an initial velocity tangential to $\Sigma$ then its motion will be restricted to $\Sigma$.

	Next we ask what are the totally geodesic submanifolds of a constant curvature manifold $M^{n}$. 
The totally geodesic condition $\omega_{ai}=0$ means that  equations \eqref{eq:cartan-submanifold-x} restricted to $\Sigma^{q}$ become 
	\begin{subequations}\label{eq:cartan-submanifold-1}
		\begin{align}
		\dd \theta^{a} &= -\omega_{ab} \wedge \theta^{b} \, \\
		\dd\omega_{ab} &= - \omega_{ac}\wedge \omega_{cb}  + k\, \theta^{a} \wedge \theta^{b}\,, \\
		\dd \omega_{ij} &= -\omega_{ik}\wedge\omega_{kj} \,.
		\label{eq:cs1-ij}
		\end{align} 
	\end{subequations}	
	Thus we have derived necessary conditions for submanifold $\Sigma^{q}$ to be a totally geodesic submanifold of the constant curvature manifold $M^{n}$. The first two equations in \eqref{eq:cartan-submanifold-1} are the Cartan structural equations for  a $q$-dimensional constant curvature manifold $\Sigma^{q}$ with the same constant sectional curvature $k$ as $M^{n}$. These totally geodesic submanifolds are maximally symmetric spaces with $\dim\Iso(\Sigma^{q}) = \frac{1}{2}q(q+1)$. Equation~\eqref{eq:cs1-ij} is the statement that the connection on the normal bundle $(T\Sigma)^{\perp}$ is flat.

	\subsection{Examples of totally geodesic submanifolds}
	\label{sec:ex-tot-geo}
	
	First we consider the example of the sphere of radius $a$, $M^{n} = S^{n}_{a}$. The main observation is that the equatorial sphere $S^{n-1}_{a}$ also has radius $a$. A geodesic in $S^{n-1}_{a}$ is a great circle in $S^{n-1}_{a}$, and this great circle is also a great circle in $S^{n}_{a}$. Thus the equatorial $S^{n-1}_{a}$ is a totally geodesic submanifold of $S^{n}_{a}$. Repeating the argument we see that the equatorial sphere $S^{n-2}_{a}$ of $S^{n-1}_{a}$ is a totally geodesic submanifold of $S^{n-1}_{a}$ and consequently a totally geodesic submanifold of $S^{n}_{a}$. We can repeat this argument until get get down to the equatorial $S^{1}_{a}$. Thus we have shown that there exists an ``equatorial'' $S^{q}_{a}$ that is a totally geodesic submanifold of $S^{n}_{a}$.
	
	To extend the result above to arbitrary $k$ and Euclidean or Minkowski signature for the metric, we repeat the discussion in the previous paragraph in terms of equations. The $n$-sphere of radius $a$ is the set of points in $\mathbb{E}^{n+1}$ that satisfies the equation $(x^{1})^{2} + (x^{2})^{2}+ \dotsb + (x^{n})^{2} + (x^{n+1})^{2}=a^{2}$. The isometry group of $S^{n}$ is $\SO(n+1)$, the stability group of the North Pole $(0,0,\dotsc,a)$ is $\SO(n)$, and $S^{n} \approx \SO(n+1)/\SO(n)$. A polar $(n-1)$-sphere\footnote{We replace equatorial great spheres by polar great spheres to simplify the equations.} has radius $a$ and contains the point $(0,\dotsc,0,a)$, the North Pole. It is obtained by adjoining to the spherical constraint equation an additional equation $x^{n}=0$. To see that this ``polar'' $S^{n-1}$ is totally geodesic, we observe that the the vector field $\partial/\partial x^{n}$ along this $S^{n-1}$ is  tangent to $S^{n}$ and is the unit normal vector field to the polar $S^{n-1}$.  This vector is parallel in $\mathbb{E}^{n+1}$ and thus $D^{S^{n}}_{X}(\partial/\partial x^{n}) =0$ where the vector field $X$ is tangent to $S^{n-1}$, and $D^{S^{n}}$ is the connection  induced on $S^{n}$ from the Euclidean structure of $\mathbb{E}^{n+1}$. Therefore, the extrinsic curvature vanishes. You can inductively repeat this argument. A polar $(n-2)$ sphere containing $(0,\dotsc,0,a)$ is obtained by adjoining an additional constraint $x^{n-1}=0$. You can either do an inductive argument or note that the extrinsic curvature vanishes again since the normal bundle is spanned by $\partial/\partial x^{n}$ and $\partial/\partial x^{n-1}$. In this way we can go all the way down to a polar $S^{1}$. This sequence of ``polar spheres'' is a collection of totally geodesic submanifolds of $S^{n}$. You can move these totally geodesic spheres to other locations by using a transformation in the isometry group $\Iso(S^{n})= \SO(n+1)$.
	
	The argument for the vanishing of the extrinsic curvatures is identical in the three cases below and we skip it. The geodesic submanifolds that we construct can be moved by using the appropriate isometry group. 
	
	For hyperbolic space $H^{n}$ (Euclidean signature and $k<0$), we consider the connected set of points that satisfies $(x^{1})^{2} + (x^{2})^{2}+ \dotsb + (x^{n})^{2} - (x^{n+1})^{2}=-a^{2}$ in $\mathbb{R}^{n+1}$ with the signature of the $\mathbb{R}^{n+1}$ metric being $(+,+,\dotsc,+,-)$ and contains the ``North Pole'' $(0,\dotsc,0,a)$. The isometry group of $H^{n}$ is $\SO(n,1)$, the stability group of the North Pole is $\SO(n)$, and thus $H^{n} \approx \SO(n,1)/\SO(n)$. The ``polar'' $H^{n-1}$ is obtained by adjoining the constraint $x^{n}=0$. The ``polar'' $H^{n-2}$ is obtained by adjoining the additional constraint $x^{n-1}=0$. This argument can be repeated until you get down to a ``polar'' $H^{1} \approx \mathbb{R}$. The $H^{q}$ constructed this way are all totally geodesic submanifolds of $H^{n}$.
	
	If we are looking for manifolds of Lorentzian signature then we get de~Sitter space $\dS_{n}$ $(k>0)$ or anti de~Sitter space $\AdS_{n}$ $(k<0)$.
	
	For de~Sitter space $\dS_{n}$, we consider the connected set of points that satisfies $-(x^{1})^{2} + (x^{2})^{2}+ \dotsb + (x^{n})^{2} + (x^{n+1})^{2}=a^{2}$ in $\mathbb{R}^{n+1}$ with signature of the $\mathbb{R}^{n+1}$ metric being $(-,+,\dotsc,+,+)$ and contains the point $(0,\dotsc,0,a)$. The isometry group of $\dS_{n}$ is $\SO(n,1)$, the stability group of the North Pole is $\SO(n-1,1)$, and $\dS_{n} \approx \SO(n,1)/\SO(n-1,1)$. The ``polar'' $\dS_{n-1}$ is obtained by adjoining the constraint $x^{n}=0$. The ``polar'' $\dS_{n-2}$ is obtained by adjoining the additional constraint $x^{n-1}=0$. This argument can be repeated until you get down to a ``polar'' $\dS_{1}$. The $\dS_{q}$ constructed this way are all totally geodesic submanifolds of $\dS_{n}$.

	For anti de~Sitter space $\AdS_{n}$, we consider the connected set of points that satisfies $-(x^{1})^{2} + (x^{2})^{2}+ \dotsb + (x^{n})^{2} - (x^{n+1})^{2}= -a^{2}$ in $\mathbb{R}^{n+1}$ with signature of the $\mathbb{R}^{n+1}$ metric being $(-,+,\dotsc,+,-)$ and contains the point $(0,\dotsc,0,a)$\footnote{In this model for anti de~Sitter space, $\AdS_{1}$ is a  closed timelike  curve. In fact for any $n \ge 1$, this model has a closed timelike curve, and for this reason we always implicitly assume we are working in the simply connected universal covering space. The universal cover of $\AdS_{1}$ is a timelike line.}. The isometry group of $\AdS_{n}$ is $\SO(n-1,2)$, the stability group of the North Pole is $\SO(n-1,1)$, and $\AdS_{n} \approx \SO(n-1,2)/\SO(n-1,1)$.  The ``polar'' $\AdS_{n-1}$ is obtained by adjoining the constraint $x^{n}=0$. The ``polar'' $\AdS_{n-2}$ is obtained by adjoining the additional constraint $x^{n-1}=0$. This argument can be repeated until you get down to a ``polar'' $\AdS_{1}$. $\AdS_{q}$ constructed this way are all totally geodesic submanifolds of $\AdS_{n}$. We remind the reader that we always have in mind the simply connected universal cover of the quadric hypersurface in question.

	\subsection{Intrinsically flat submanifolds}
	\label{sec:flat-submanifolds}

	Next we show that there are intrinsically flat submanifolds that can be isometrically embedded in negative constant curvature spaces. These manifolds are not totally geodesic submanifolds. We impose the intrinsic flatness condition $R^{\Sigma}_{abcd}=0$. We also require a special form for the extrinsic curvature tensor $K_{ab}{}^{i} = h_{ab}\, C^{i}$, where $h_{ab}$ is the flat metric on $\Sigma$. Note that $C^{i}$ is the mean curvature vector. The reason for this form is that we require the extrinsic curvature to be compatible with the isometries of the vector space $\Sigma^{q}$. Inserting these conditions into \eqref{eq:cartan-submanifold-x}, we obtain
\begin{subequations}\label{eq:cartan-submanifold-y}
	\begin{align}
		\dd \theta^{a} &= -\omega_{ab} \wedge \theta^{b}  \,, \\
		0 &= \left( \lVert C \rVert^{2} + k \right) \theta^{a} \wedge \theta^{b}\,, 
		\label{eq:curv-cond}\\
		\left(DC^{i}\right)\wedge \theta^{a} &= 0 \,, 
		\label{eq:cov-DC}\\
		\dd \omega_{ij} &= -\omega_{ik}\wedge\omega_{kj} \,.
		\label{eq:norm-triv}
	\end{align} 
\end{subequations}	
To satisfy the structural equations above, we require from \eqref{eq:curv-cond} that $k = - \lVert C \rVert^{2} \le 0$. Thus, we have our first result that the space $M^{n}$ must have constant negative curvature\footnote{If $M^{n}$ has Lorentzian signature and $\Sigma^{q}$ is a timelike submanifold then $(T_{\sigma}M)^{\perp}$ has Euclidean signature and $\lVert C \rVert^{2} \ge 0$.} since we are restricting our analysis in this paper to $k\neq 0$. Next, we observe that since the $\{ \theta^{a} \}$ are linearly independent, eq.~\eqref{eq:cov-DC} implies that $0 = DC^{i} = \dd C^{i} + \omega_{ij}C^{j}$. Thus the mean curvature vector $\bC = C^{i}\, \hat{\vb{n}}_{i}$ is covariantly constant and we have a preferred direction in the flat normal bundle $(T\Sigma)^{\perp}$, see condition~\eqref{eq:norm-triv}.  If $M^{n}$ has Lorentzian signature then the maximal  possible symmetry group of a defect is $\mathcal{P}(q) \times \SO(l-1)$ as opposed to the $\AdS_{q}$ case where  you can have $\SO(2,q-1)\times \SO(l)$. There is no spherically symmetric defect in the case of $\Sigma \approx \mathbb{M}^{q}$. Said differently, the ``transverse part of the Lagrangian'' is at most $\SO(l-1)$ invariant.

\begin{rem}\label{rem:maximal}
If $\AdS_{q}$ is not a totally geodesic submanifold then there is a non-zero mean curvature vector that gives  a preferred normal direction. The Gauss equation in this case becomes $k_{\Sigma^{q}}= \lVert C\rVert^{2} + k_{M^{n}}$. The symmetry group of the $p$-defect is maximally $\Iso(\Sigma^{q}) \times \SO(l-1)$. There are no maximally symmetric solutions in this case.
\end{rem}

You can verify that the world branes in references \cite{Dehghani:2001ft,Lugo:1999fm,Lugo:1999ai} are totally geodesic submanifolds.

It is easy to discuss a basic example that is commonly used in physics. Consider $\AdS_{n}$ with embedded submanifold $\Sigma^{q} \approx \Mink^{n-1}$, \emph{i.e.}, $q=n-1$. The metric on $\AdS_{n}$ is written in the ``upper half space'' form
\begin{equation}
ds^{2}_{\AdS_{n}} = \frac{\mink_{ab}\, \dd\sigma^{a}\otimes \dd\sigma^{b} + (dy)^{2}}{y^{2}} \quad\text{with } y>0\,.
\label{eq:uhp-metric}
\end{equation}
If we restrict to the $q=n-1$ dimensional submanifold $\Sigma^{q}$ defined by $y = y_{0}$, $y_{0}$ constant, then the induced metric on it, 
\begin{equation}
ds^{2}_{\Sigma} = \frac{1}{y_{0}^{2}}\; \mink_{ab}\, \dd\sigma^{a}\otimes \dd\sigma^{b}\,,
\end{equation}
is a rescaled Minkowski metric. The mean curvature vector is parallel to the unit normal to $\Sigma$. Any vector subspace of this $\Sigma^{n-1}$ is automatically a flat submanifold of $\AdS_{n}$.
\begin{figure}
\centering
\includegraphics[width=0.7\textwidth]{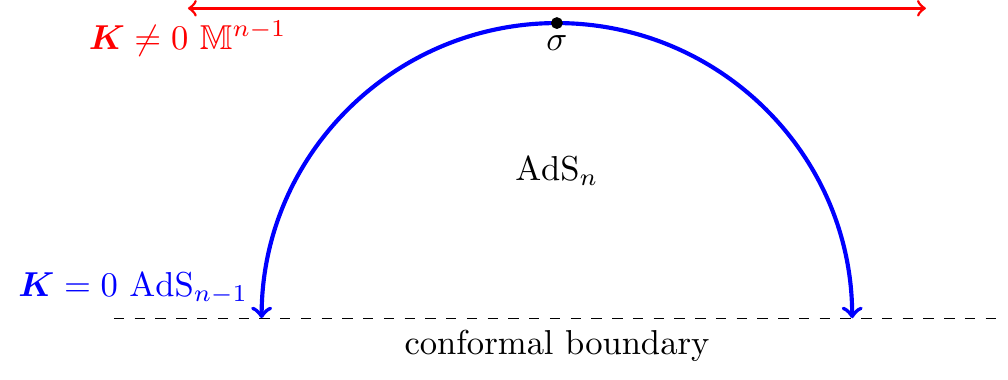}
\caption{\small Cartoonish representation of a totally geodesic $\AdS_{n-1}$ and a flat $\mathbb{M}^{n-1}$ embedded in $\AdS_{n}$ in the upper half space representation with metric \eqref{eq:uhp-metric}. Choose a point $\sigma \in \AdS_{n-1}$ and consider all the geodesics beginning at $\sigma$ with \emph{spacelike} initial  velocity tangential to $\AdS_{n-1}$. The locus of all these geodesics is a submanifold of the totally geodesic $\AdS_{n-1}$ represented by the blue curve. A physical deformation (motion slower than the speed of light) of $\AdS_{n-1}$ into $\mathbb{M}^{n-1}$ would take an infinite amount of time because the asymptotic parts of both manifolds are infinitely far apart. The distances between the corresponding arrowheads in $\mathbb{M}^{n-1}$ and in $\AdS_{n-1}$ are infinite.}%
\label{fig:submanifolds}
\end{figure}
In figure~\ref{fig:submanifolds} we compare a totally geodesic constant curvature submanifold and a  zero curvature submanifold of $\AdS_{n}$ in the ``upper half space'' representation.

\section{Relation to  the work of Lugo, Moreno and Schaposnik}
\label{sec:translation}

\begin{table}[tbp]
    \centering
    \begin{tabular}{@{}cc@{}}
\toprule
LMS & This article \\
\midrule
$\lambda_{\text{LMS}}$ & $\half \lambda$\\
$e$ & $g$ \\
$\vec{H} = h_{0}H(x)$ & $\Phi^{I}$ \\
$h_{0}$ & $\phi_{0}$ \\
$r_{0}$ & $\rho$ \\
$\gamma_{0}=1/(g\phi_{0}\rho)^{2}$ & $\gamma_{0}= 1/(m_{A}\rho)^{2}$\\
$r$ & $r/\rho = \sinh(\nu/\rho) \xrightarrow{\nu\to\infty\;} \half e^{\nu/\rho}$ \\
$x=m_{A}r$, $[x]=M^{0}$ &  $x= (m_{A}\rho) \sinh(\nu/\rho) \xrightarrow{\nu\to\infty\;} \half (m_{A}\rho) e^{\nu/\rho}$\\
$\gamma_{0}x^{2}$ & $r^{2}/\rho^{2}$\\
their $\nu$  & $\nua $ \\
$K(x)$ & $-h(\nu)$\\
\bottomrule
	\end{tabular}	
    \caption{\small Translation dictionary  of notation between the LMS papers and us.  Both of us have  $k = \Lambda/3<0$ and $\lvert k \rvert = 1/\rho^{2}=1/r_{0}^{2}$. Since we use $\nu$ for the radial distance from the defect world brane, we will use $\nua$ for the LMS asymptotic decay index to avoid confusion.}
\label{tbl:translation}
\end{table}
\begin{figure}[tbp]
\centering
\includegraphics[width=0.65\textwidth]{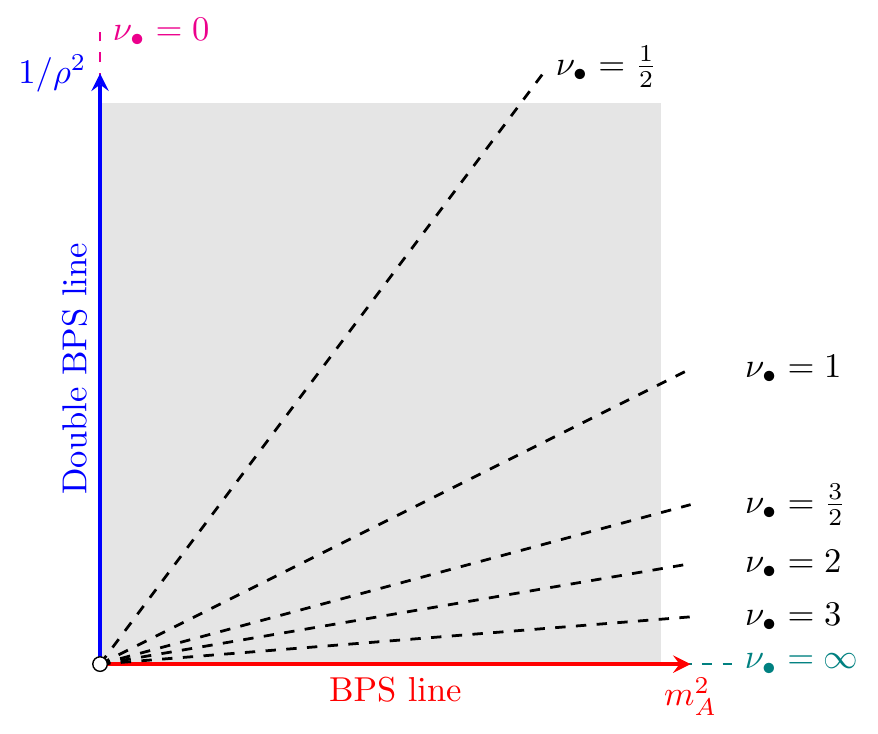}
\caption{\small The gray plane above, taken from figure~\ref{fig:bps-parameter}, corresponds to the generalized BPS limit where $m_{\phi}^{2}\downarrow 0$. We have reinterpreted the results of LMS as the dashed straight lines $m_{A}^{2} = \nua(\nua+1)/\rho^{2}$ 
in this plane, where LMS discovered the existence of a power series solution if $\nua$ is a positive half integer. LMS also studied other values of $\nua$ numerically in \cite{Lugo:1999ai}. The BPS line corresponds to formally setting $\nua=\infty$, which is the same as sending $\rho\uparrow \infty$. The double BPS line corresponds to formally setting $\nua=0$, which is the same as sending $m_{A}\downarrow 0$. The work of LMS seems to indicate that solutions along the dashed lines are special but we have not explored this. Note that those dashed lines interpolate between the two lines where we have exact analytic solutions. LMS found reasonable numerical solutions using a relaxation method with starting point the BPS solution; this is not surprising in light of this figure. On the other hand, this figure suggests that in some sense, the double BPS solutions are far away from the Prasad-Sommerfield solutions. }
\label{fig:LMS-parameter}
\end{figure}

This is an  attempt at relating the work of Lugo, Moreno and Schaposnik (LMS) with ours. They choose a different coordinate system for $\AdS_{4}$ where the metric is
\begin{equation}
\dd s_{\AdS_{4}}^{2} = -\left(1+r^{2}/\rho^{2}\right) \dd t^{2} + 
\left(1+r^{2}/\rho^{2}\right)^{-1} \dd r^{2} + r^{2} \left( \dd \theta^{2} + \sin^{2}\theta\; \dd\varphi^{2}\right).
\end{equation}
It is important to notice that $r$ is not the radial distance. You can verify that $\Sigma^{1}$, the ``$t$-axis'', is a totally geodesic submanifold.
A notational dictionary is in table~\ref{tbl:translation}. 

A first observation is that we have a different viewpoint on what is BPS. Their philosophy is described in the paragraph following eq.~(4.32) in reference~\cite{Lugo:1999fm}. They assume that setting $\lambda_{\text{LMS}} = 0$ means that there is no potential and the asymptotic value of the Higgs field can float to some non-zero value $\bigl\lvert\vec{H}(\infty)\bigr\rvert$ as you solve the equations of motion. This is natural because in their numerical work  they are using a relaxation method~\cite{press:1992} taking the exact BPS solution as a starting point. They also postulated a power series expansion in $1/x$ about $x=\infty$, see eqs.~\eqref{eq:LMS}, and discovered that the expansion is well behaved if the asymptotic value of the Higgs field takes a very specific form
\begin{equation}
\left\lvert\vec{H}(\infty)\right\rvert^{2} = \nua\left(\nua +1\right) \frac{1}{(g\rho)^{2}} \quad\text{where}\quad \nua=\frac{1}{2},1,\frac{3}{2},2,\frac{5}{2},3,\dotsc,
\end{equation}
see~\cite[eq.~(29)]{Lugo:1999ai}. Here $\nua$ specifies the decay behavior of the $K$ field~\eqref{eq:K-decay}. For non half integer values of $\nua$, the power series may have to be generalized to include logarithms.

We interpret the BPS solution as a limiting solution in the limit $\lambda \downarrow 0$. This means that the potential \eqref{eq:phi-4} governs the asymptotic behavior and requires that $\lVert\Phi\rVert \to \phi_{0}$ as $\nu \to\infty$. Since the asymptotic value $\phi_{0}$ is arbitrary in our formalism, we can try to compare results by setting $\bigl\lvert\vec{H}(\infty)\bigr\rvert=\phi_{0}$, and we find
$(g\phi_{0})^{2} = m_{A}^{2} = \nua(\nua+1)/\rho^{2}$. Next we corroborate this identification by analyzing the asymptotic behavior of the fields. 
\begin{subequations}\label{eq:LMS}
\begin{align}
  \left\lvert \vec{H}(x)\right\rvert &= \left\lvert \vec{H}(\infty)\right\rvert + \frac{h_{0}H_{3}}{x^{3}} + \dotsb &   \phi(\nu) &\to \phi_{0}-\psi_{*} e^{-3\nu/\rho} 
  \label{eq:H-decay} \\
 K(x) &= \frac{K_{\nua+1}}{x^{\nua+1}}\left( 1 + O\left(1/x^{2}\right)\right) & h(\nu) &\to h_{*}e^{-\mu_{*}\nu} + \dotsb 
 \label{eq:K-decay}
\end{align}
\end{subequations}
Notice that the asymptotic behavior of the Higgs field agrees in both our computations since $x \sim e^{\nu/\rho}$. To make a connection with the second equation we note that \eqref{eq:def-mu-star} says that $\mu_{*}\rho = \half + \sqrt{\frac{1}{4} + (m_{A}\rho)^{2}}$, and thus we conclude that $\nua+1 = \mu_{*}\rho$. This relates the decay exponent $\nua$ to the flat space mass $m_{A}$ and the radius of curvature $\rho$ via the equation $\nua(\nua+1) = (m_{A}\rho)^{2}$. These ideas are summarized in figure~\ref{fig:LMS-parameter}. 

\section{The lack of Bogomolny equations and a partial bound}
\label{sec:bogomolny}

\subsection{The magnetic flux}
\label{sec:magnetic}

In this appendix we make an unsuccessful attempt to determine the stability of our double BPS monopole solution. We analyze whether ours is a minimum  energy static solution. We discuss various topics related to   static magnetic monopoles  in $\AdS_{4}$. We address whether in the $\lambda \downarrow 0$ Prasad-Sommerfield  limit~\cite{Prasad:1975kr} there is an analog of the first order Bogomolny equations~\cite{Bogomolny:1975de} that gives solutions of the equations of motion.  More precisely, we show that the arguments~\cite{Bogomolny:1975de,Coleman:1976uk} that lead to the Bogomolny bound in the $\mathbb{M}^{4}$ theory do not generalize to  the $\AdS_{4}$ case. In this section we ignore the potential term because we are always implicitly taking the BPS limit. The value of $g$ is generic.

First, we compute the magnetic flux. The metric on $M=\AdS_{4}$ may be written in the form
	\begin{equation}
	\dd s^{2}_{\AdS_{4}} = -\cosh^{2}\left( \lvert k \rvert^{1/2} \lVert\vb*{\nu}\rVert\right) \; \dd\tau^{2} + \dd\nu^{2} + \left(\frac{\sinh\left( \lvert k \rvert^{1/2} \lVert\vb*{\nu}\rVert\right)}{\lvert k \rvert^{1/2} \lVert\vb*{\nu}\rVert}\right)^{2}\;
	\nu^{2}\; \dd s^{2}_{S^{2}}\,,
	\label{eq:met-geod-four}
	\end{equation}
where the coordinate $\tau$ is the proper time along the $1$-dimensional defect world line $\Sigma^{1}=\AdS_{1}$.
The restriction of the $\AdS_{4}$ metric to the normal tangent space $(T_{\sigma}\Sigma)^{\perp}$ is given by
	\begin{equation}
	\dd s^{2}_{\AdS_{4}} \biggr\rvert_{(T_{\sigma}M)^{\perp}} =  \dd\nu^{2} + \left(\frac{\sinh\left( \lvert k \rvert^{1/2} \lVert\vb*{\nu}\rVert\right)}{\lvert k \rvert^{1/2} \lVert\vb*{\nu}\rVert}\right)^{2}\;
	\nu^{2}\; \dd s^{2}_{S^{2}}\,.
	\label{eq:met-geod-perp}
	\end{equation}
At $\sigma \in \Sigma^{1}$, the normal tangent space $(T_{\sigma}\Sigma)^{\perp}$ is a three three dimensional vector space. Define the magnetic $2$-form by $B^{I} = \half\, \epsilon^{IJK}\; F^{JK}$ and  observe that $\dd\left(\Phi^{I}B^{I}\right) = D\Phi^{I} \wedge B^{I}$ as a consequence of the Bianchi identity $DB^{I}=0$. If we fix $\sigma\in \Sigma^{1}$ and we apply Stoke's Theorem we find
\begin{equation}
\int_{(T_{\sigma}\Sigma)^{\perp}} D\Phi^{I} \wedge B^{I} = \int_{(T_{\sigma}\Sigma)^{\perp}} \dd\!\left(\Phi^{I}B^{I}\right) = \lim_{\nu\to\infty} \int_{S^{2}_{\nu}}\Phi^{I}B^{I} \,,
\label{eq:B0}
\end{equation}
where $S^{2}_{\nu} \hookrightarrow (T_{\sigma}\Sigma)^{\perp}$ is the $2$-sphere of radius $\nu$.  Here we used the definition $\int_{(T_{\sigma}\Sigma)^{\perp}} \cdots = \lim_{\nu\to\infty} \int_{B_{\nu}^{3}} \cdots$, where $B_{\nu}^{3} \hookrightarrow (T_{\sigma}\Sigma)^{\perp}$ is the $3$-ball of radius $\nu$ and $S_{\nu}^{2}=\partial B_{\nu}^{3}$.
As shown by 'tHooft~\cite{tHooft:1974kcl} and Polyakov~\cite{Polyakov:1974ek}, the $\SO(2)$ unbroken gauge symmetry magnetic field $2$-form is $\Phi^{I}B^{I}/\phi_{0}$ and thus we expect that, for a monopole, the abelian magnetic flux $\flux$, which is a measure of the magnetic charge, is given by:
\begin{equation}
\lim_{\nu\to\infty} \int_{S^{2}_{\nu}}\Phi^{I}B^{I} = \phi_{0}\, \flux \,.
\end{equation}
Thus, we conclude that
\begin{equation}
\int_{(T_{\sigma}\Sigma)^{\perp}} \frac{1}{\phi_{0}}\, D\Phi^{I} \wedge B^{I} = \flux \,.
\label{eq:mag-charge}
\end{equation}
The integrand in the left hand side of the equation may be interpreted as the magnetic charge density.
Note that the conventionally normalized magnetic field $B_{\text{conv}}$ is related to our magnetic field by $B = g\, B_{\text{conv}}$, and one has to be careful in analyzing the $g\downarrow 0$ limit. The topological flux quantization condition should hold for all values of $g$.

Next, we want to see how the above computation works within our ansatz for the spherically symmetric hedgehog  because this illuminates why there are no Bogomolny equations in this case. Using eq.~\eqref{eq:F} we obtain
\begin{equation}
\int_{S^{2}_{\nu}}\Phi^{I}B^{I} = \int_{S^{2}_{\nu}} \phi(\nu) f(\nu) \left( 1 - \frac{\nu^{2} f(\nu)}{4}\right) \left(\frac{\nu/\rho}{\sinh(\nu/\rho)}\right)^{2} \; \dual_{S^{2}_{\nu}}\,,
\end{equation}
where $\dual_{S^{2}_{\nu}}$ is the area $2$-form on $S^{2}_{\nu}\hookrightarrow (T_{\sigma}\Sigma)^{\perp}$. Note that that $S^{2}_{\nu}$ has area $4\pi\rho^{2} \sinh^{2}(\nu/\rho)$. Due to its spherical symmetry, the integral is easily computed and we obtain the result found by 'tHooft and Polyakov:
\begin{equation}
\int_{S^{2}_{\nu}}\Phi^{I}B^{I}\,\dual_{S^{2}_{\nu}} = 4\pi\nu^{2}\,\phi(\nu) f(\nu) \left( 1 - \frac{\nu^{2} f(\nu)}{4}\right) = 4\pi \phi(\nu) \left( 1-h(\nu)^{2}\right) \xrightarrow{\nu\to\infty\,} 4\pi\phi_{0}
\,,
\label{eq:mag-charge-val}
\end{equation}
where we used the asymptotic boundary conditions. With our normalization, the magnetic flux is $\flux =4\pi$, which is the correct topological quantization condition for the case where the $\SO(3)$ gauge group breaks down to $\SO(2)$ via a vector representation scalar field\footnote{Often in the literature, the model considered is a $\SU(2)$ gauge theory with an adjoint representation scalar field that breaks the symmetry group down to $\U(1)$. In this case, the magnetic flux quantization condition is $\flux=2\pi$.}. Note the cancellation of the hyperbolic sine factors, which was  necessary to obtain a finite non-zero magnetic flux.

Next, we discuss a couple of attempts at trying to obtain the Bogomolny bound and the associated Bogomolny equations. We implicitly assume that we are always taking the Prasad-Sommerfield limit $\lambda \downarrow 0$, and maintaining the appropriate boundary conditions. 

It is computationally useful to use the Hodge inner product. We give a brief review to establish conventions. Let $V$ be a vector space with inner product, either Euclidean or Lorentzian signature, $\dim V=n$. The normalized volume element determined by an orthonormal basis is $\vb*{\zeta}$. If $\omega$ is a $k$-form, it is convenient to define a norm by
$\hnorm{\omega} = \omega^{\mu_{1}\dotsm\mu_{k}}\omega_{\mu_{1}\dotsm\mu_{k}}/k!$. With this normalization each independent term in the $k$-form only contributes once in the summation. A pointwise bilinear product, $\innerp{\bullet}{\bullet}$, and the Hodge dual operator $\hodge$ on $k$-forms are defined by  $\innerp{\omega}{\omega}\vb*{\zeta} = \omega\wedge \hodge\omega  = \hnorm{\omega}\,\vb*{\zeta}$. If $\msign=+1$ in Euclidean signature and if $\msign=-1$ in Minkowski signature then please note that  $\hnorm{\hodge\omega} = \msign\hnorm{\omega}$, and $\hodge\!\hodge\omega = \msign (-1)^{k(n-k)}\omega$.

\subsection{From the magnetic flux to the  energy functional}

In this section, we fix $\sigma \in \Sigma$ and we restrict to the three dimensional normal tangent space $(T_{\sigma}\Sigma)^{\perp}$. In particular,  the metric on $(T_{\sigma}\Sigma)^{\perp}$ is given by \eqref{eq:met-geod-perp}, and the Hodge duality operator refers to Hodge duality with respect to this metric. The strategy in this section is to begin with the correct expression for the magnetic flux and to try to get to the energy functional whose variation gives the equations of motion. 

In this appendix, we have to use Hodge duality on the normal tangent space $(T_{\sigma}\Sigma)^{\perp}$, and on the full manifold $\AdS_{4}$. For this reason we introduce a special notation: on the normal tangent space  $(T_{\sigma}\Sigma)^{\perp}$ with metric \eqref{eq:met-geod-perp}, we denote the Hodge duality operation by $\hperp$, and the associated norm by $\hnormperp{\bullet}$.

We observe that 
\begin{align}
-2\phi_{0}\, \flux/g &= -2\int_{(T_{\sigma}\Sigma)^{\perp}} D\Phi^{I} \wedge B^{I}/g
= -2\int_{(T_{\sigma}\Sigma)^{\perp}} D\Phi^{I} \wedge (\hperp\!\hperp B^{I})/g
\nonumber\\
&= -2\int_{(T_{\sigma}\Sigma)^{\perp}} \innerp{D\Phi^{I}}{\hperp B^{I}/g} \vb*{\zeta}_{(T_\sigma \Sigma)^\perp}
\nonumber \\
&= \int_{(T_{\sigma}\Sigma)^{\perp}} \innerp{D\Phi^{I}-\hperp B^{I}/g}{D\Phi^{I}-\hperp B^{I}/g} \vb*{\zeta}_{(T_\sigma \Sigma)^\perp}
\nonumber\\
&\quad - \int_{(T_{\sigma}\Sigma)^{\perp}} \left( \hnormperp{D\Phi} + \hnormperp{B}/g^{2}\right) \vb*{\zeta}_{(T_{\sigma}\Sigma)^{\perp}} \,.
\nonumber
\end{align}
Here $\vb*{\zeta}_{(T_{\sigma}\Sigma)^{\perp}}=\hperp 1$ is the volume element on $(T_{\sigma}\Sigma)^{\perp}$.
Rearranging terms we find
\begin{align}
\half \int_{(T_{\sigma}\Sigma)^{\perp}} \left( \hnormperp{D\Phi} + \frac{1}{g^{2}}\, \hnormperp{B}\right) \vb*{\zeta}_{(T_{\sigma}\Sigma)^{\perp}} &= \half \int_{(T_{\sigma}\Sigma)^{\perp}} \hnormperp{D\Phi-\frac{1}{g}\hperp\! B} \vb*{\zeta}_{(T_{\sigma}\Sigma)^{\perp}}
\nonumber \\
&\quad +\phi_{0}\, \flux  /g\,.
\label{eq:B1}
\end{align}
This is the expression~\cite{Bogomolny:1975de,Coleman:1976uk} that is used to show in $\mathbb{M}^{4}$ that a solution to the Bogomolny equations $D\Phi^I=\hperp B^{I}/g$ is an absolute minimum of the left hand side of \eqref{eq:B1}.
Unfortunately, this is not what we need because the functional that has to be minimized to obtain the equations of motion is not the left hand side of \eqref{eq:B1} but the transverse energy functional \eqref{eq:defect-action}, which in Hodge star notation is
\begin{equation}
E_{\perp} = \half \int_{(T_{\sigma}\Sigma)^{\perp}} \cosh(\nu/\rho) \left( \hnormperp{D\Phi} + \frac{1}{g^{2}}\,\hnormperp{B}\right) \vb*{\zeta}_{(T_{\sigma}\Sigma)^{\perp}}\,.
\end{equation}
The hyperbolic cosine factor in the previous equation is necessary to obtain the correct equations of motion.  The static equations of motion are obtained by restricting the $4$-dimensional action, constructed with the $\AdS_{4}$ metric, to time translationally invariant field configurations. Note that $\partial/\partial \tau$ is a Killing vector since the metric components in \eqref{eq:met-geod-four} are independent of $\tau$.

Equation~\eqref{eq:B1} is what you would get for an instanton solution in Euclidean hyperbolic $3$-space $H^{3}$. Solutions to the BPS equations would be absolute minima of the equations of motion in the Prasad-Sommerfield limit, see also the discussion in Atiyah~\cite{Atiyah:monopoles}.

The Bogomolny argument  works in a situation where the spacetime is a product manifold $M^{4} = \mathbb{R} \times V^{3}$ with the  product metric $\dd s^{2}_{M} = -\dd\tau^{2} + \dd s^{2}_{V}$. It also works in the Euclidean signature instanton case where the manifold $M^{3}$ is Euclidean hyperbolic 3-space, \emph{ibid.} and footnote~\ref{foot:instanton}.

\subsection{From the  energy functional to the magnetic flux }

In this section, we begin with the correct functional that has to be minimized to obtain the static equations of motion and we try to see if we can find a bound for this functional that is related to the magnetic flux . From now on, the metric is the $\AdS_{4}$ Lorentzian metric \eqref{eq:met-geod-four} with associated inner product $\innerp{\bullet}{\bullet}$. The Hodge duality operation is with respect to this metric, the $\AdS_{4}$ volume element is denoted by $\vb*{\zeta} = \hodge 1$, and the normalized timelike $1$-form is $\that = \cosh(\nu/\rho)\,\dd\tau$. The action for our static configuration is
\begin{equation*}
I = - \half\int_{\AdS_{4}} \left( \hnorm{D\Phi} + \hnorm{B}/g^{2}\right) \vb*{\zeta} = -E_{\perp} \left(\int_{\Sigma} \dd\tau\right)\,.
\end{equation*}
Next we rewrite the action as
\begin{equation*}
2 I = -\int_{\AdS_{4}} \bigl( \innerp{D\Phi}{D\Phi} + \innerp{B}{B}/g^{2} \bigr)\vb*{\zeta}\,.
\end{equation*}
To try to get Bogomolny type equations, we observe that $\that\wedge D\Phi$ is a timelike $2$-form, and that $\hodge B$ is also a timelike $2$-form. Note that $\that$ is orthogonal to $D\Phi$ and thus, we can rewrite the above equation as
\begin{align*}
2 I &= -\int_{\AdS_{4}} \left( -\innerp{\that\wedge D\Phi}{\that\wedge D\Phi} - \innerp{\hodge B}{\hodge B}/g^{2} \right) \vb*{\zeta} \,,\\
&= \int_{\AdS_{4}} \innerp{\that\wedge D\Phi - \hodge B/g}{\that\wedge D\Phi - \hodge B/g}\vb*{\zeta} + 2 \int_{\AdS_{4}} \innerp{\that\wedge D\Phi}{\hodge B/g}\vb*{\zeta} \\
&= \int_{\AdS_{4}} \hnorm{\that\wedge D\Phi - \hodge B/g} \vb*{\zeta}
+ 2 \int_{\AdS_{4}} \that\wedge D\Phi \wedge (\hodge\!\hodge B/g) \\
&= \int_{\AdS_{4}} \hnorm{\that\wedge D\Phi - \hodge B/g} \vb*{\zeta}
- 2 \int_{\AdS_{4}} \that\wedge D\Phi \wedge B/g \\
&= \int_{\AdS_{4}} \hnorm{\that\wedge D\Phi - \hodge B/g} \vb*{\zeta}
- (2/g) \int_{\Sigma} \dd\tau \left( \int_{(T_{\sigma}\Sigma)^{\perp}} \cosh(\nu/\rho)\, D\Phi \wedge B \right)
\end{align*}
Using the time translational invariance we can write the above as \
\begin{equation}
E_{\perp} = \frac{-\half\int_{\AdS_{4}} \hnorm{\that\wedge D\Phi - \hodge B/g} \vb*{\zeta}}{\int_{\Sigma} \dd\tau} + \frac{1}{g} \int_{(T_{\sigma}\Sigma)^{\perp}} \cosh(\nu/\rho)\, D\Phi \wedge B \;.
\label{eq:B2}
\end{equation}
Next we simplify the integral over $\AdS_{4}$. Let $(\that,\hat{\theta}^{1} ,\hat{\theta}^{2} ,\hat{\theta}^{3})$ be the adapted orthonormal coframe for $\AdS_{4}$ we constructed early in the manuscript. On the normal tangent space  $(T_{\sigma}\Sigma)^{\perp}$ with metric \eqref{eq:met-geod-perp}, we denoted the Hodge duality operation by $\hperp$, and the associated norm by $\hnormperp{\bullet}$.
It is easy to see that $\hodge\left( \that\wedge\hat{\theta}^{i}\right) = -\hperp\hat{\theta}^{i}$ by using $\epsilon^{0123}=-1$. We observe that $-\hnorm{\that\wedge D\Phi - \hodge B/g} = \hnormperp{D\Phi - \hperp B/g}$,  $\dual = \cosh(\nu/\rho)\, \dd\tau \wedge \dual_{(T_{\sigma}\Sigma)^{\perp}}$, and we can rewrite \eqref{eq:B2} as
\begin{align}
E_{\perp}(\rho) &= \int_{(T_{\sigma}\Sigma)^{\perp}} \cosh(\nu/\rho) \hnormperp{D\Phi - \frac{1}{g} \hperp\! B} \;\dual_{(T_{\sigma}\Sigma)^{\perp}} 
\nonumber\\
&\quad + \frac{1}{g} \int_{(T_{\sigma}\Sigma)^{\perp}} \cosh(\nu/\rho)\, D\Phi \wedge B \;.
\label{eq:B3}
\end{align}
If we set $\rho=\infty$, \emph{i.e.}, $k=0$, then \eqref{eq:B3} leads to the standard Bogomolny type argument for a lower bound on the mass of $4\pi/g$. The bound is saturated by field configurations that satisfy the Bogomolny equations $D\Phi = \hperp B/g$.

If $0<\rho<\infty$ and  $\hnormperp{\bullet}$ is a positive definite norm, then this equation implies a bound
\begin{equation}
E_{\perp} \ge  \frac{1}{g} \int_{(T_{\sigma}\Sigma)^{\perp}} \cosh(\nu/\rho)\, D\Phi \wedge B \;.
\label{eq:B4}
\end{equation}
This bound does not appear very useful because the bound is field configuration dependent. You can verify the field dependency by computing the variation of the integral with respect to a variation of the scalar field. If $\delta\Phi(\nu)$ is a variation of the scalar field with compact support, then you obtain
\begin{equation*}
\delta_{\Phi}\! \int_{(T_{\sigma}\Sigma)^{\perp}} \cosh(\nu/\rho)\, D\Phi^{I} \wedge B^{I}
= - \int_{(T_{\sigma}\Sigma)^{\perp}} \frac{\sinh(\nu/\rho)}{\rho}\; \dd\nu\wedge B^{I}\; (\delta\Phi^{I})\,.
\end{equation*}
Note that the right hand side of the equation above vanishes for $\rho=\infty$, and it exemplifies the topological nature of the magnetic flux   \eqref{eq:mag-charge}, which is invariant under deformations of the field configuration.

\subsection{A bound on the mass for non-negative magnetic charge  density}
\label{sec:bound}

We previously mentioned that the left hand side of magnetic flux equation \eqref{eq:mag-charge} may be interpreted as the magnetic charge density.
Here we prove a theorem that states that if the magnetic charge density is non-negative in the Prasad-Sommerfield limit and if $0<\rho<\infty$, then the mass satisfies the strict inequality $E_{\perp}(\rho) > 4\pi\phi_{0}/g$.

A $3$-form $\alpha$ on $(T_{\sigma}\Sigma)^{\perp}$ is said to be non-negative if there exists a function $f:(T_{\sigma}\Sigma)^{\perp} \to \mathbb{R}$ with $f \ge 0$ such that $\alpha = f\, \dual_{(T_{\sigma}\Sigma)^{\perp}}$. Let $\mathcal{C}= \lbrace (\Phi,A) \rbrace$ be the space of admissible field configurations, and let $\mathcal{C}_{+} \subset \mathcal{C}$ be the subset of field configurations where the $3$-form $D\Phi^{I}\wedge B^{I} = d\left(\Phi^{I}B^{I}\right)$ is non-negative. These are the configurations with non-negative magnetic charge density. Since $\cosh(\nu/\rho) > 1$ for $\nu>0$ and $0<\rho<\infty$, the use of \eqref{eq:B4}, \eqref{eq:mag-charge} and \eqref{eq:mag-charge-val} leads  to the conclusion that if we restrict to field configurations in $\mathcal{C}_{+}$, then there is a lower bound provided by  magnetic flux quantization
\begin{equation*}
E_{\perp}(\rho)\biggr\rvert_{\mathcal{C}_{+}} > \frac{1}{g}\int_{(T_{\sigma}\Sigma)^{\perp}} D\Phi^{I} \wedge B^{I} = \frac{4\pi}{g}\;\phi_{0}\,.
\end{equation*}

\bigskip

\providecommand{\href}[2]{#2}\begingroup\raggedright\endgroup

\end{document}